\documentclass[twocolumn,aps,prx,showpacs,amsmath,superscriptaddress,longbibliography,notitlepage]{revtex4-2}
\usepackage{amssymb}
\usepackage{mathrsfs}
\usepackage{graphicx}
\usepackage{float}
\usepackage{braket}
\usepackage{xcolor}
\usepackage[colorlinks,linkcolor={red!75!black},citecolor={red!75!black},urlcolor={blue!75!black}]{hyperref}
\usepackage{verbatim}
\usepackage{subcaption}
\usepackage[normalem]{ulem}
\usepackage{bm}
\usepackage{soul}

\newcommand{\clr}{\color{red!75!black}}

\newcommand{\Rnum}[1]{\uppercase\expandafter{\romannumeral #1\relax}}

\usepackage{enumitem}
\usepackage{comment}

\begin{document}
\title{Drag-induced skin effect in a Bose-Fermi mixture}

\author{Wenjie Liu}\email{liuwenjie@dlut.edu.cn} 
\affiliation{School of General Education, Dalian University of Technology, Panjin 124221, China}

\author{Yee Sin Ang}
\email{yeesin\_ang@sutd.edu.sg} 
\affiliation{Science, Mathematics and Technology, Singapore University of Technology and Design, Singapore 487372, Singapore}

\author{Ching Hua Lee}\email{phylch@nus.edu.sg}
\affiliation{Department of Physics, National University of Singapore, Singapore 117542}

\author{Yi Qin}\email{doctor\_yi\_qin\_25@nus.edu.sg} 
\affiliation{Department of Physics, National University of Singapore, Singapore 117542}

\begin{abstract}
The non-Hermitian skin effect (NHSE) represents one of the most distinctive phenomena in non-Hermitian physics. 
Here, we uncover a new drag-induced NHSE mechanism in interacting Bose--Fermi mixtures where only bosons, but not fermions, experience asymmetric hopping.
We show that strong Bose--Fermi interactions enable fermions to inherit boundary accumulation through correlated bound states. 
The interplay of interactions, quantum statistics, and non-Hermitian dynamics gives rise to an interaction-induced blockade mechanism, leading to highly asymmetric fermionic transport. 
We demonstrate that the drag-induced NHSE is dynamically stable and propose a feasible realization in ultracold Bose--Fermi mixtures with Floquet-engineered asymmetric tunneling. 
Our results establish a qualitatively distinct route to the NHSE, where non-Hermitian localization is not generated by direct nonreciprocal hopping of the localized particles themselves, but emerges through the interaction-mediated transfer of non-Hermitian dynamics between different quantum species. This mechanism further provides a potential route toward species-selective control of localization and transport in hybrid quantum matter.
\end{abstract}

\maketitle

\emph{{\clr Introduction}.---}
Non-Hermitian physics provides a unified framework for coherent dynamics intertwined with gain, loss, or measurement backaction, and has revealed phenomena with no Hermitian counterparts~\cite{2019PhysRevX9041015,Ashida2020,Bergholtz2021RMP,Zhang2022NHSE,Ding2022NHTop,Lin2023TopNHSE,Okuma2022NHTopReview,lei2026inter,meng2025gen,lee2022exceptional,liang2025size,huang2026complex,shen2025observation,long2026sym,shen2026observation}.
A paradigmatic phenomenon is the non-Hermitian skin effect (NHSE), where nonreciprocity causes an extensive number of eigenstates to accumulate at system boundaries under open boundary conditions, leading to a breakdown of conventional bulk--boundary correspondence~\cite{Yao2018edge,Kunst2018biorthogonal,Lee2019anatomy,Song2019chiral,Ghatak2020PNAS,Xiao2020NatPhys,Helbig2020NatPhys,Weidemann2020Science,zhang2021tidal,2022PhysRevLett.129.070401,rafi2022unconventional,li2022non,ZhangPRL2023,SCPHe2023,jiang2023dimensional,XiaoLei2024PRL,Zhang2024PRB,ZhaoEntong2025,wang2025non,ammari2025non,2025bmq57tf6,yang2025conf,20254yt24rx4,2025cwwdbclc,3927n25r2025,cheng2025stochasticity,ZhangPRB2025,yang2025rev,xue2025non,2025z9m13mwb,li2025anderson,yang2025non,d5zcp1sk2025,WuPRL2025,li2025phase,gohsrich2025non,wang2025nonlinear,zhao2025magne,liu2025anom,li2025algebraic,hu2025topological,v6ht-jq3l2025,xbj1-hfyf2025,Nobuhiro2026,yi2026dir,lin2026glo,rahul2026cont,longhi2026erra,saito2026quas,
okuma2026ste,bai2026eng,deng2026conf,2026s26b8bdl,
wu2026observation,yu2026sensitivity,hu2026boundary}.
Extending the NHSE beyond single-particle physics to interacting and many-body systems has emerged as a central challenge, as interactions and quantum statistics can fundamentally reshape localization and spectral sensitivity in ways inaccessible to noninteracting models.
Recent studies have uncovered genuinely many-body manifestations of the NHSE, including interaction-modified non-Hermitian spectra~\cite{2020PhysRevB.101.121109,2020PhysRevB.102.081115,2022PhysRevResearch.4.033122,PhysRevB.106.L121102,qin2023non,shen2022non,2023PhysRevResearch.5.033173,2024hysRevLett.133.216601,2024PhysRevLett.133.136503,2024PhysRevLett.133.136502,2024PhysRevLett.132.096501,2025PhysRevB.111.035144,cpl_42_3_037301,2025wztw-l8wg,hu2025many,2026bhpz17d2,hao2025interacting,koh2025interacting,Qin2025AnyonNHSE,Ekman2026symmetry,Qin2026MBNHSE,xr3h-j9pv2026}, non-Hermitian topology~\cite{2021PhysRevB.104.195102,2021PhysRevB.103.235306,2022PhysRevLett.129.180401,2022PhysRevB.105.165137,2023PhysRevB.108.155114,2023PhysRevB.107.045131}, and entanglement transitions~\cite{2023PhysRevX.13.021007,2024PhysRevB.110.035113,liu2025non,xue2026topologically}.
These advances establish interaction-driven NHSE phenomena as correlated quantum effects governed by interactions, Hilbert-space connectivity, and quantum statistics.

Ultracold Bose--Fermi mixtures provide precise control over tunneling and interspecies interactions~\cite{Greiner2002,Lewenstein2007,Bloch2008RMP,2001PhysRevA.64.011402,2001PhysRevLett.87.080403,2002PhysRevA.65.053607,2008PhysRevA.78.013619,2010PhysRevA.82.011605,2012PhysRevA.85.063616,li2020topological,Ferrier-Barbut2014,2015PhysRevA.91.041605,Onofrio2016Cooling,2016PhysRevLett.117.145301,2016PhysRevLett.117.245302,2017PhysRevLett.118.055301,2017PhysRevA.95.043643,2018PhysRevX.8.031042,2018PhysRevLett.121.253402,DeSalvo2019,2020PhysRevLett.124.163401,qin2024kinked}. Their interspecies physics is well established~\cite{1998PhysRevLett.80.1804,2000PhysRevA.61.053605,2001PhysRevA.64.043608,2002PhysRevA.66.063604,2002PhysRevA.65.021603,2003PhysRevA.68.033605,2004PhysRevA.69.043607,2007PhysRevA.75.013623,2008PhysRevB.78.134517,2004PhysRevA.69.063603,2011hysRevA.83.041603,2013PhysRevLett.110.115303,2017PhysRevLett.119.233401,2018PhysRevB.97.020506,Modugno2002,2018PhysRevLett.120.243403,2018PhysRevA.97.033628,shen2023proposal,2021PhysRevA.103.063317,2023PhysRevLett.131.083003}, including boson-mediated interactions~\cite{PhysRevLett.132.033401}, topological pumping~\cite{Mostaan2022}, correlated-hopping effects~\cite{2025lw8k-7h6p}, and drag of fermionic impurities by a bosonic background~\cite{YanZoeZ2024}.
These advances motivate a basic yet unexplored question in the non-Hermitian regime: 
\emph{Can a species that is intrinsically impervious to the NHSE acquire non-Hermitian localization through its interaction with another species that is sensitive to the NHSE?}

In this work, we uncover a drag-induced NHSE in a one-dimensional Bose--Fermi mixture, where non-Hermitian bosons induce boundary accumulation of otherwise Hermitian fermions through interspecies interactions. In the minimal one-boson--one-fermion system, we identify a sharp bound--scattering dichotomy: only strongly correlated bound states exhibit correlated skin localization, consistent with an effective strong-coupling description. In the few-body regime, quantum statistics and interparticle interactions further give rise to band-dependent drag and an interaction-induced blockade, resulting in strongly asymmetric and directionally controllable fermionic transport. We also demonstrate the dynamical stability of the effect and propose a cold-atom implementation based on Floquet-engineered nonreciprocal tunneling and tunable Bose--Fermi interactions.

Unlike conventional NHSEs, where nonreciprocal hopping directly acts on the particles undergoing skin localization, our mechanism transfers non-Hermitian dynamics between different species through interspecies correlations. This provides a route to species-selective non-Hermitian control, whereby nonreciprocity engineered in one species can indirectly control the localization and transport of another species that remains intrinsically Hermitian.

\emph{{\clr Model and formalism}.---}  
We consider a one-dimensional Bose--Fermi mixture governed by the Hamiltonian
\begin{equation}\label{BMModel}
H = H_b + H_f + H_{bf},
\end{equation}
which comprises bosonic ($H_b$), fermionic ($H_f$), and interspecies
interaction ($H_{bf}$) terms.
The bosonic sector is described by a dimerized lattice with nonreciprocal inter-cell tunneling:
\begin{align}
H_b &= -\sum_{j=1}^{L/2}\left(t_0\, b_{2j-1}^\dagger b_{2j} + \text{H.c.}\right)
+ \frac{U_{bb}}{2}\sum_{j=1}^{L} n_j^b (n_j^b-1) \notag\\
&\quad - \sum_{j=1}^{L/2-1}\left(t_L\, b_{2j}^\dagger b_{2j+1} + t_R\, b_{2j+1}^\dagger b_{2j}\right),
\end{align}
where $t_0$ denotes the intra-cell (Hermitian) hopping amplitude,
while $t_L$ and $t_R$ introduce asymmetric (non-Hermitian) inter-cell tunneling that underlies the bosonic skin effect.
The dimerized lattice considered here is adopted as a representative nonreciprocal lattice rather than a necessary ingredient of the proposed mechanism. To demonstrate its generality, we further investigate a uniform asymmetric-hopping model in Sec.~S1 of the Supplemental Material~\cite{2026supplemental}, where the same interaction-mediated mechanism is shown to persist.
The fermionic component features uniform hopping and nearest-neighbor interactions:
\begin{equation}
H_f = -\sum_{j=1}^{L-1}\left(t_f\, c_{j+1}^\dagger c_j + \text{H.c.}  \right)
+ U_{ff}\sum_{j=1}^{L-1} n_j^f n_{j+1}^f
\end{equation}
with $U_{ff}$
representing nearest-neighbor interactions.
The fermionic component is described by a reciprocal hopping amplitude $t_f$, which is in general independent of the bosonic hopping amplitudes. Unless otherwise specified, we set $t_f=t_0$ throughout the main text to reduce the number of independent parameters.

\begin{figure}
\centering
\begin{subfigure}[b]{0.95\linewidth}
\centering
\includegraphics[width=0.95\textwidth,height=2.5cm]{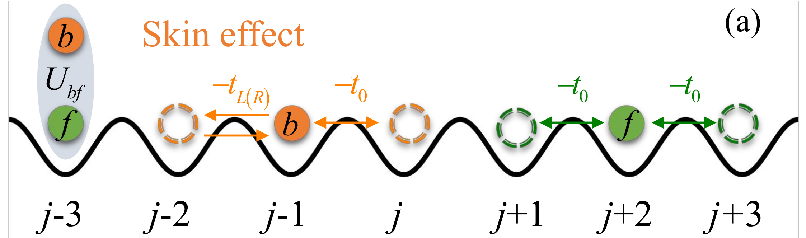}
\end{subfigure}\\
\begin{subfigure}[b]{0.95\linewidth}
\centering
\includegraphics[width=\textwidth,height=7cm]{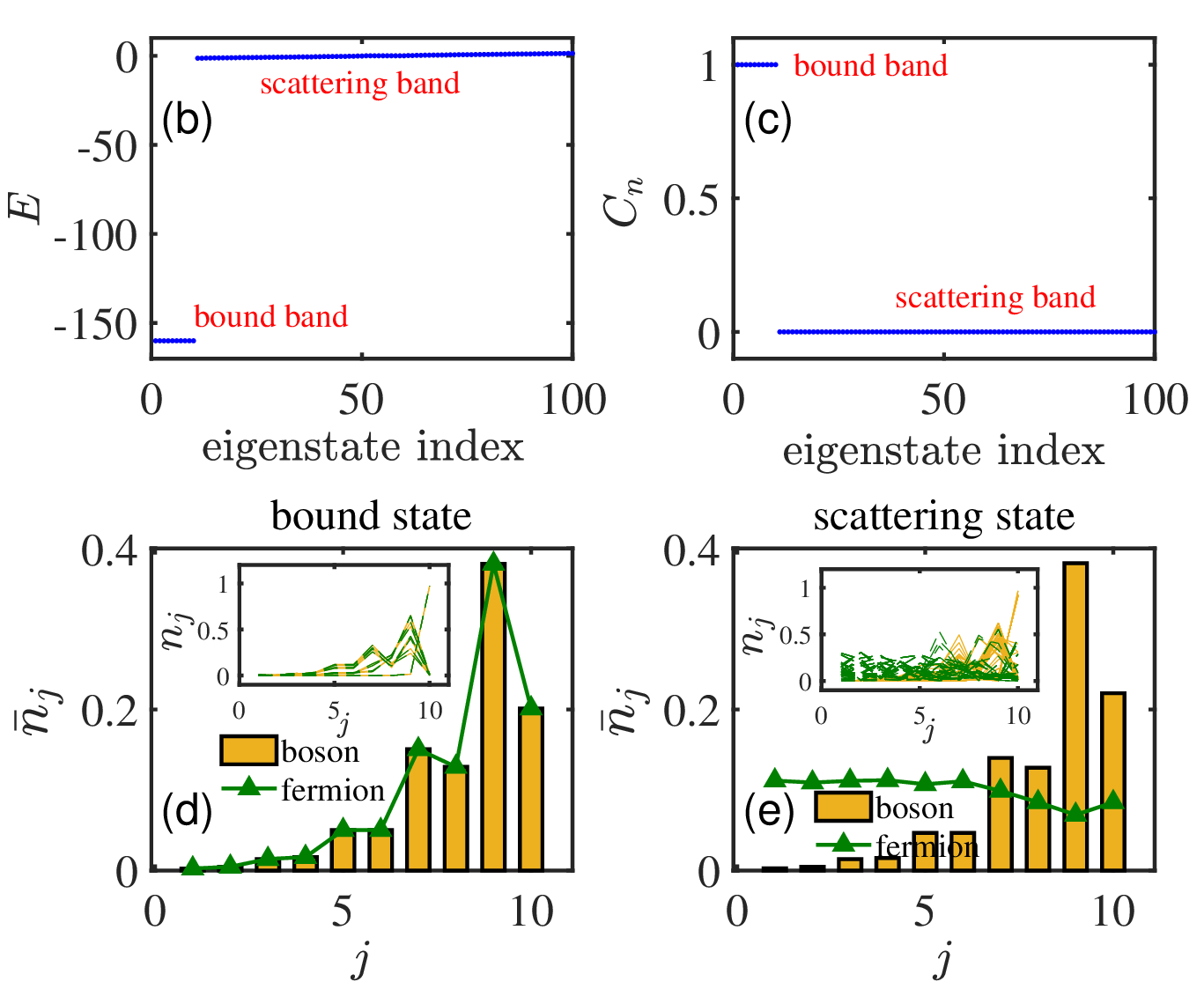}
\end{subfigure}
\caption{(a) Schematic of the Bose--Fermi mixture. Bosons (orange) experience nonreciprocal hopping ($t_L\neq t_R$), whereas fermions (green) remain Hermitian and couple to bosons via the on-site interaction $U_{bf}$.
(b) Energy spectrum and (c) boson--fermion correlation $C_n$ for the one-boson--one-fermion system.
(d),(e) Band-averaged bosonic and fermionic density
distributions in the bound and scattering sectors, respectively. The insets show the density profiles of the individual eigenstates
within the corresponding band. Only the strongly correlated bound states exhibit a drag-induced NHSE, where fermions inherit the bosonic boundary accumulation; scattering states remain largely insensitive to the bosonic skin localization. Parameters: $t_0=0.2$, $t_L=0.5$, $t_R=1.5$, and $U_{bf}=-160$.
}\label{fig:first}
\end{figure}

The two species are coupled via Bose-Fermi interactions:
\begin{equation}
H_{bf} = U_{bf}\sum_{j=1}^{L} n_j^b n_j^f,
\end{equation}
where $n_j^b=b_j^{\dagger} b_j$ and $n_j^f=c_j^{\dagger} c_j$.
As shown in Fig.~\ref{fig:first}(a), bosons experience asymmetric hopping and exhibit the NHSE, whereas fermions remain intrinsically Hermitian. We show below that NHSE transfer occurs in strongly bound Bose--Fermi composites but is absent in scattering states, and can be further reshaped by quantum statistics and kinetic constraints.

\emph{{\clr One boson and one fermion}.---} 
We begin with one boson and one fermion in the strong-coupling regime, taking $t_0=0.2$, $t_L=0.5$, $t_R=1.5$, and $U_{bf}=-160$, where bound and scattering states are well separated.

The energy spectrum of this hybrid system is shown in Fig.~\ref{fig:first}(b). Despite the non-Hermitian nature of the bosonic sector, the full spectrum remains real because the Hamiltonian is related to a Hermitian counterpart by a non-unitary similarity transformation~\cite{2026supplemental}. Although this transformation preserves the energy spectrum, it changes the eigenstates and physical observables, so the system remains genuinely non-Hermitian.
A sharply isolated band appears at $E \approx -160$, signaling a nearly stable bound pair formed by the strong boson–fermion interaction. In contrast, a continuum of scattering states occupies the energy window $E$ around zero.
The degree of spatial correlation between the two particles is quantified by the correlation function
$ C_n = \sum_{j} \langle \psi_n | n_j^b n_j^f | \psi_n \rangle $ 
where $C_n = 1$ indicates perfect co-localization on the same site, while $C_n = 0$ indicates the absence of on-site Bose--Fermi co-localization. As shown in Fig.~\ref{fig:first}(c), the deep-bound state ($E_n \approx -160$) exhibits $C_n \approx 1$, confirming a tightly bound composite, whereas states near $E_n \approx 0$ yield $C_n \approx 0$, reflecting weak correlations.

This dichotomy is directly visualized in the density profiles of Figs.~\ref{fig:first}(d) and (e).
We define the band-averaged density as
\begin{equation}
    \overline{n}_j
    =
    \frac{1}{N_{\mathrm{band}}}
    \sum_{n\in \mathrm{band}}
    \left\langle
    \psi_n
    \left|
    n_j
    \right|
    \psi_n
    \right\rangle,
    \label{eq:uniform_band_density}
\end{equation}
where $N_{\mathrm{band}}$ is the number of eigenstates included in the corresponding band.
In the bound regime [Fig.~\ref{fig:first}(d)], the boson displays a pronounced NHSE---its density accumulates sharply at the right edge due to asymmetric tunneling. Remarkably, the fermion, though governed by Hermitian dynamics, inherits this localization through strong $U_{bf}$ coupling, resulting in a correlated “drag-induced” skin effect. The inset highlights the near-perfect spatial overlap of both densities for representative bound states.
Conversely, in the scattering band [Fig.~\ref{fig:first}(e)], the bosonic density remains right-localized, but the fermionic density is nearly uniform across the lattice. This decoupling arises because scattering states involve weak effective interactions: the fermion retains its intrinsic delocalized character and does not respond to the bosonic skin accumulation.
Importantly, the drag-induced NHSE persists beyond the strong-coupling regime and continuously evolves into a correlation-assisted localization effect at weaker interactions~\cite{2026supplemental}.

The effective Hamiltonian in the bound-state subspace can be obtained via a strong-coupling expansion~\cite{2026supplemental}. Up to an overall energy shift and weak boundary corrections, the dynamics is described by
\begin{equation}
\begin{aligned}
\hat H_{\rm eff}
&=
2\frac{t_0t_f}{U_{bf}}
\sum_{j}
\left(
\hat S_{2j-1}^{\dagger}\hat S_{2j}
+\mathrm{H.c.}
\right) \\
&\quad+
2\frac{t_f}{U_{bf}}
\sum_{j}
\left(
t_L\,\hat S_{2j}^{\dagger}\hat S_{2j+1}
+t_R\,\hat S_{2j+1}^{\dagger}\hat S_{2j}
\right)
\end{aligned}
\end{equation}
where $\hat S_j^\dagger$ creates a bound boson--fermion composite at site $j$.
The effective inter-cell hoppings remain asymmetric, being proportional to $t_L t_f/U_{bf}$ and $t_R t_f/U_{bf}$, respectively. Their non-reciprocity ratio is therefore independent of $t_f$, which only sets the overall dynamical scale. Consequently, the bound composite inherits the non-reciprocity of the bosonic sector and exhibits the same skin accumulation under open boundary conditions.
We have further verified that the same bound-state-selective drag-induced NHSE occurs in a uniform asymmetric-hopping model without dimerization; see Sec.~S1 of the Supplemental Material.

\begin{figure}[htp]
\center
\includegraphics[width=0.47\textwidth]{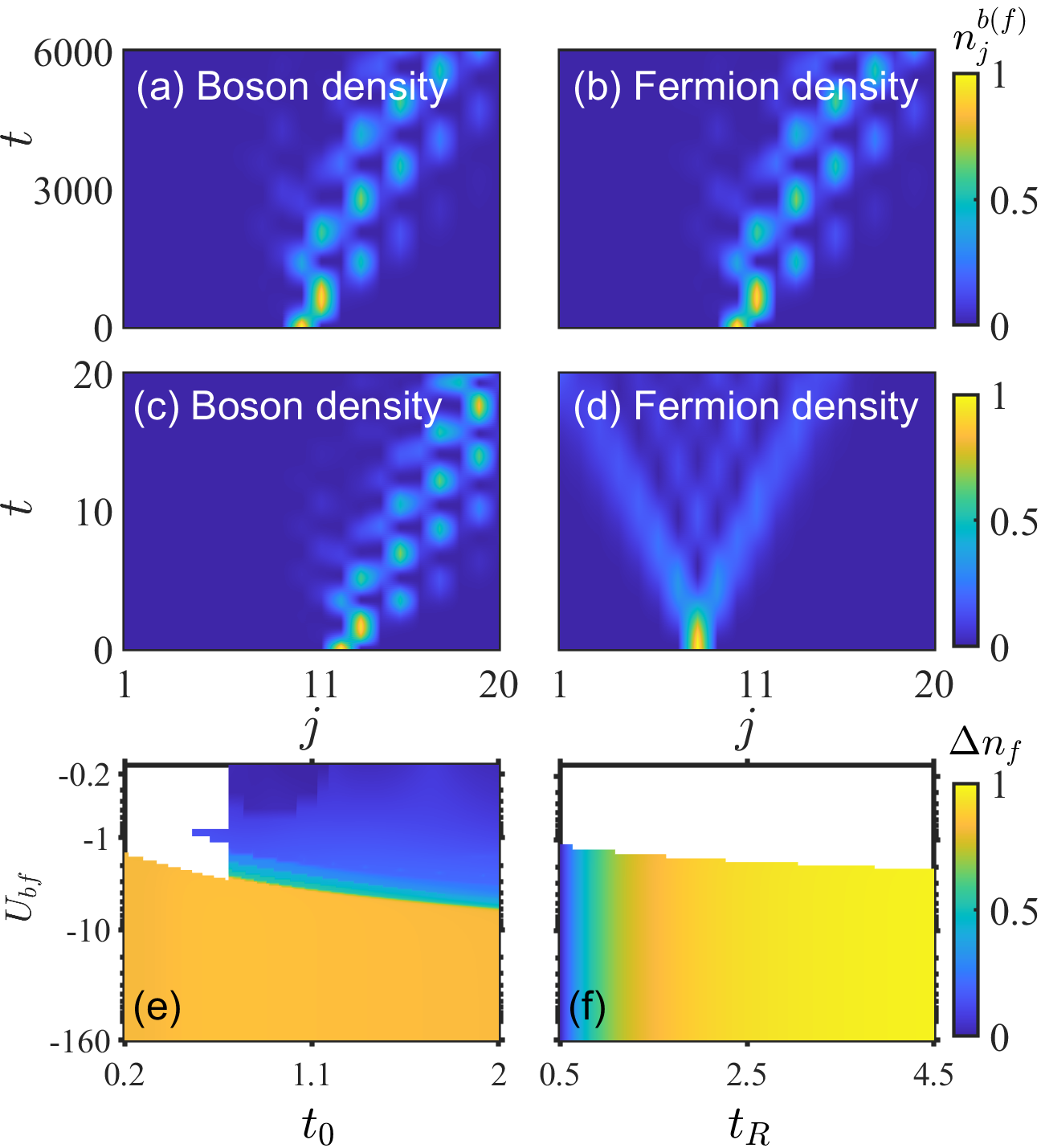}
\caption{
Dynamical stability and parameter dependence of the drag-induced NHSE. (a),(b) Time evolution of bosonic and fermionic densities for an initial Bose--Fermi bound pair, showing correlated rightward accumulation of both species. (c),(d) Corresponding dynamics for an initial scattering state, where bosons remain skin-localized while fermions exhibit nearly symmetric ballistic transport. 
(e),(f) Average fermion density imbalance $\Delta n_f$ in the parameter spaces $(t_0,U_{bf})$ and $(t_R,U_{bf})$, respectively. 
For (e), $L=10$, $t_L=0.5$ and $t_R=1.5$; for (f), $L=10$, $t_0=0.2$ and $t_L=0.5$. In both panels, $t_f=t_0$, while $U_{bb}$ and $U_{ff}$ do not contribute to the one-boson--one-fermion system. The drag-induced skin effect strengthens with increasing $|U_{bf}|$ and bosonic nonreciprocity, and saturates in the strong-coupling regime. 
Blank regions indicate parameter regimes in which the spectral gap $\Delta E_{L,L+1}=E_{L+1}-E_L$ between the nominal bound and scattering sectors is smaller than the numerical threshold $0.1$, so that an isolated bound band and its averaged imbalance cannot be reliably defined.
Parameters for (a)--(d): $L=20$, $t_0=0.2$, $t_L=0.5$, $t_R=1.5$, and $U_{bf}=-160$.
\label{fig:5}}
\end{figure}

\emph{{\clr Dynamical stability of the drag-induced skin effect}.---} 
Having established the emergence of a drag-induced skin effect through boson–fermion correlations, we now examine its dynamics. Figure.~\ref{fig:5} shows the time evolution of bosonic and fermionic densities for two distinct initial conditions: a bound pair [Figs.~\ref{fig:5}(a),(b)] and a scattering state [Figs.~\ref{fig:5}(c),(d)]. In the bound case, both species exhibit coherent rightward localization---evidencing a dynamically stable drag-induced skin effect. By contrast, for scattering initial states, while bosons still accumulate at the right edge due to their intrinsic non-Hermitian dynamics, fermions spread symmetrically in a light-cone pattern characteristic of Hermitian ballistic transport. This dichotomy mirrors the static spectral structure shown in Fig.~\ref{fig:first}, confirming that the drag-induced skin effect persists under time evolution and is thus dynamically stable.

\begin{figure*}[htp]
\center
\includegraphics[width=0.95\textwidth]{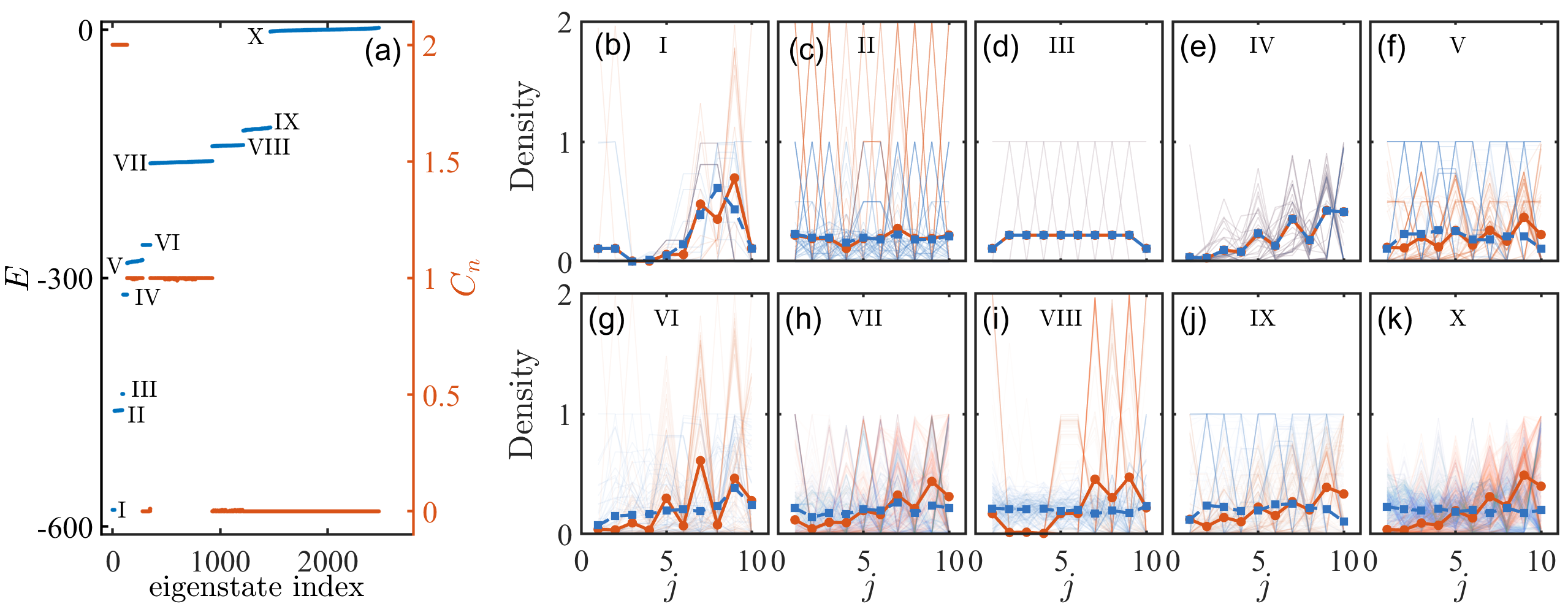}
\caption{
Band-resolved spectrum and density structures of a two-boson--two-fermion system.
(a) Energy spectrum and boson--fermion correlation $C_n$ as functions of the eigenstate index. 
The spectrum separates into ten clusters, labeled I--X.
(b)--(k) Density distributions for the ten corresponding bands. 
The faint lines show the density profiles of individual eigenstates within each band, while the thick lines with markers denote the band-averaged densities. 
Orange and blue curves correspond to bosons and fermions, respectively.
This representation reveals not only the average boundary accumulation, but also the intra-band structure of the correlated few-body states. 
In several bands, individual eigenstates display strongly configuration-dependent density patterns that are partially washed out after band averaging, indicating that the bands are composed of distinct correlated few-body configurations rather than simple single-particle-like density profiles.
The parameters are $L = 10$, $t_0 = 0.2$, $t_L = 0.5$, $t_R = 1.5$, $U_{ff} = -120$, $U_{bb} = -140$, and $U_{bf} = -160$. 
}
\label{fig:3}
\end{figure*}

To quantify the fermionic skin response, we introduce the mean density imbalance
\begin{equation}
    \Delta n_f = \frac{1}{N_f} \sum_{j=1}^{L/2} \left( \bar n^f_{L-j+1} - \bar n^f_j \right),
    \label{eq:density_imbalance}
\end{equation}
where $N_f$ denotes the total number of fermions in the system, and $\bar n_j^f$ is the average fermion density at site $j$, evaluated over eigenstates within the bound band. A large positive $\Delta n_f$ indicates strong right-edge accumulation, while $\Delta n_f \approx 0$ corresponds to delocalized fermions.
The parameter dependence of $\Delta n_f$ is shown in Figs.~\ref{fig:5}(e) and (f). 
Figure~\ref{fig:5}(e) shows that the drag-induced skin effect strongly depends on the Bose--Fermi interaction strength $U_{bf}$. 
For weak interactions, $\Delta n_f$ remains small, indicating that the fermions largely retain their intrinsic Hermitian character. As $|U_{bf}|$ increases, $\Delta n_f$ rapidly grows and saturates in the strong-coupling regime, reflecting the formation of tightly bound boson--fermion composites that efficiently inherit the bosonic skin localization. The weak dependence on $t_0$ further indicates that, once the bound states are established, the drag-induced skin effect is governed primarily by the Bose--Fermi interaction strength and the bosonic non-reciprocity.

By contrast, Fig.~\ref{fig:5}(f) shows that the nonreciprocal hopping amplitude $t_R$ acts as the primary driving mechanism for the drag-induced skin effect. Increasing $t_R$ strongly enhances $\Delta n_f$, leading to pronounced fermionic boundary accumulation inherited from the bosonic NHSE. Together, these results show that the drag-induced skin effect requires both sufficiently strong Bose--Fermi correlations and finite nonreciprocal bosonic hopping.
The gap criterion, extended interaction regime, and robustness against variations of $t_f$ are detailed in the Supplemental Material~\cite{2026supplemental}.

\emph{{\clr Interaction-organized band hierarchy in the few-body regime}---} 
Having established the interaction-induced skin effect in the minimal one-boson–one-fermion setting, we now explore richer physics emerging in a few-body regime where quantum statistics become decisive. 
We consider a system with two bosons and two fermions. 
Here the interaction parameters are chosen to clearly resolve the interaction-organized few-body bands, thereby facilitating the discussion of the band hierarchy and the interaction-induced blockade.
Figure~\ref{fig:3}(a) displays the full energy spectrum---ordered by increasing $E$---alongside the eigenstate-resolved boson–fermion correlation function.
The spectrum organizes into ten distinct bands, each exhibiting a characteristic value of $C_n$. 
High-$C_n$ bands correspond to states where strong interspecies interaction enforces local boson–fermion binding, while low-$C_n$ bands reflect spatial segregation driven by interparticle interactions and statistics.
While the band hierarchy is largely dictated by interactions and quantum statistics, non-Hermiticity controls how different correlated configurations respond to the bosonic skin drive. This interplay leads to band-dependent drag-induced localization and, in certain configurations, interaction-induced blockade effects.

Figures~\ref{fig:3}(b)–(k) show the corresponding bosonic and fermionic density distributions for Bands I--X, revealing pronounced band-dependent localization and distinct correlated occupation patterns.
In the lowest-energy bands (I–II), all particles tend to cluster: typically, both bosons and one fermion occupy the same site, while the second fermion occupies either a nearest-neighbor site (Band I) or a more distant site (Band II). 
Middle-energy bands (III–IV) feature one tightly bound boson–fermion pair, with the remaining particles forming secondary pairs at adjacent or farther sites. 
Higher bands (V–VIII) display more intricate arrangements---for instance, a localized boson–fermion dimer coexisting with unpaired particles on non-overlapping sites (Bands V, VII), or two bosons co-localized with two fermions on neighboring sites (Bands VI, VIII).
In the highest-energy bands (IX–X), fermions localize on adjacent or separated sites, while bosons avoid fermion-occupied regions entirely. 

Notably, in Band VI, the density distribution is such that two bosons localize at the same lattice site, while the two fermions occupy the two nearest-neighbor sites excluding the one occupied by the bosons.
This configuration becomes particularly transparent from the individual eigenstate density profiles, where the spatial avoidance between bosons and fermions can be directly identified.
An interaction-induced blockade emerges from the interplay among boson--boson, fermion--fermion, and boson--fermion interactions, which collectively stabilize a localized few-body configuration. 
This correlated configuration effectively forms a few-body domain wall separating distinct occupation sectors. Due to fermionic statistics and kinetic constraints, certain transport channels become strongly suppressed, leading to highly asymmetric fermionic dynamics and contributing to the drag-induced skin effect.
The resulting band-dependent localization requires a correlated few-body description; eigenstate-resolved densities are provided in the Supplemental Material~\cite{2026supplemental}.

\begin{figure}[htp]
\center
\includegraphics[width=0.45\textwidth]{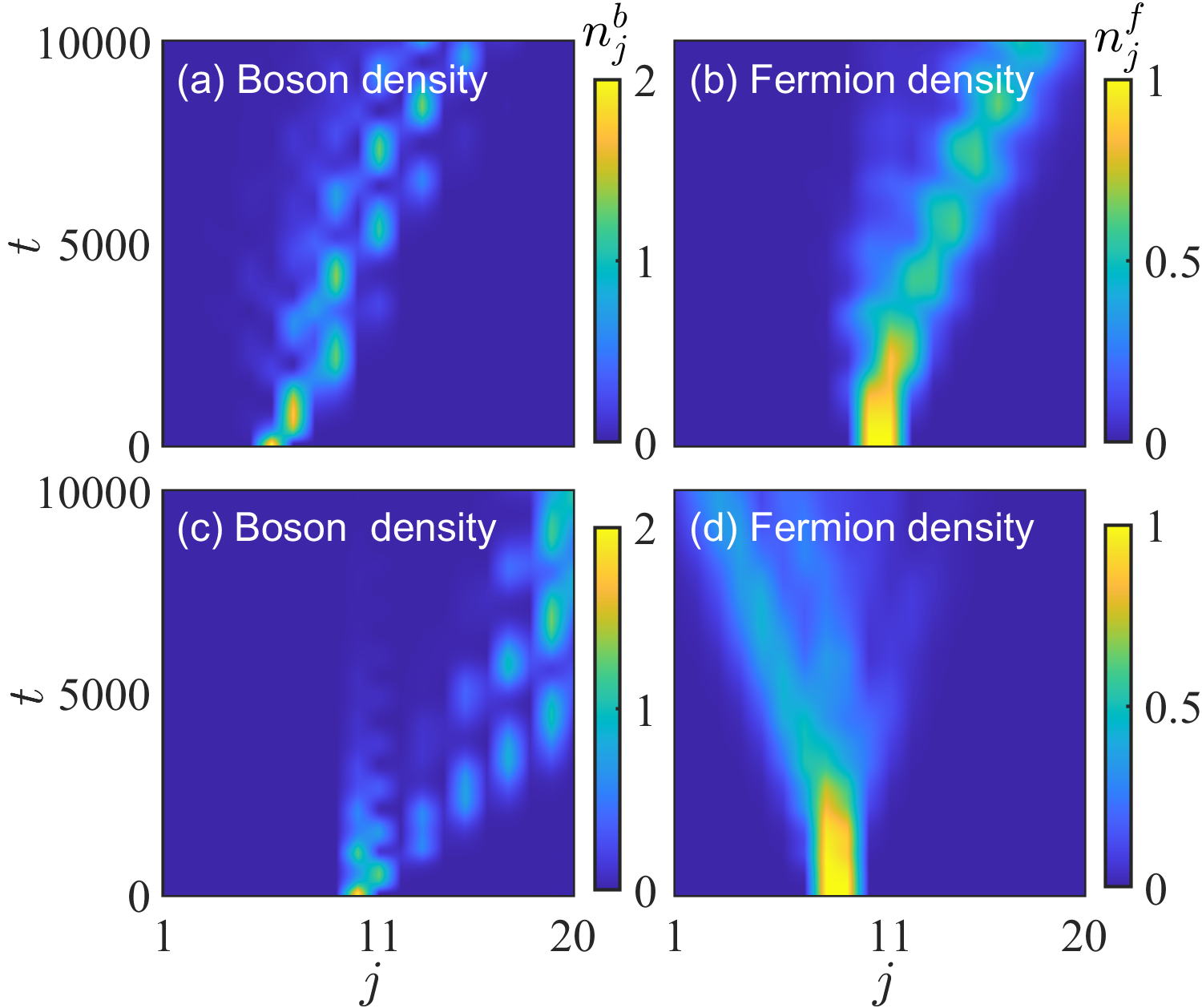}
\caption{Dynamical manifestation of the interaction blockade effect in a two-boson--two-fermion system ($L=20$).
(a),(b) Dynamics for an initial configuration in which the bosons are located to the left of the fermions.
(c),(d) Corresponding dynamics when the bosons are initially located to the right of the fermions.
The blockade originates from an interaction-stabilized few-body occupation configuration, generated by the combined effects of interparticle interactions and fermionic statistics, which kinetically suppresses selected fermionic transport channels.
Parameters: $t_0 = 0.2$, $t_L = 0.2$, $t_R = 0.5$, $U_{ff} = -120$, $U_{bb} = -140$, $U_{bf} = -160$.
\label{fig:6}}
\end{figure}

\emph{{\clr Blockade Effect and Its Dynamical Consequences}.---} 
Our few-body analysis reveals a striking interaction-induced \textit{blockade effect}. To illustrate this, Fig.~\ref{fig:6} shows the time evolution of bosonic and fermionic densities for two representative product-state configurations motivated by the occupation ordering associated with Band VI in Fig.~\ref{fig:3}. We set $t_L = 0.2$ and $t_R = 0.5$ (all other parameters identical to Fig.~\ref{fig:3}) to activate the non-Hermitian skin drive.

Due to the asymmetric tunneling, the bosons are driven toward the right edge, resulting in a pronounced non-Hermitian skin effect, as shown in Figs.~\ref{fig:6}(a) and (c).
Specifically, Figs.~\ref{fig:6}(a) and (b) show the case in which the bosons are initially positioned to the left of the fermions $\left|\psi_0\right\rangle=\left|2_b\right\rangle_j\left|1_f\right\rangle_{j+4}\left|1_f\right\rangle_{j+5}$. In this configuration, both species exhibit a rightward skin effect: the bosons are driven to the right due to asymmetric hopping, while the fermions, though Hermitian, are pulled along via the strong boson-fermion interaction $U_{bf} = -160$. This co-propagation reflects the drag-induced localization mechanism discussed above.
In contrast, when the initial state has the bosons positioned to the right of the fermions $\left|\psi_0\right\rangle=\left|1_f\right\rangle_j\left|1_f\right\rangle_{j+1}\left|2_b\right\rangle_{j+2}$, the bosons continue to exhibit rightward skin propagation, while the fermions display asymmetric behavior, primarily propagating to the left [Figs.~\ref{fig:6}(c) and (d)]. 
This striking reversal arises from the interplay between the interaction blockade and the directionality of the non-Hermitian drive. The effect of $U_{bf}$ dynamically stabilizes a localized few-body domain that separates distinct occupation configurations. Due to fermionic statistics and the associated kinetic constraints, certain right-moving fermionic channels cannot efficiently penetrate or reconstruct this moving domain and therefore become strongly suppressed. As a result, the fermions predominantly propagate in the opposite direction.
The corresponding blockade behavior also survives in a uniform asymmetric-hopping model, confirming that it originates from interaction-stabilized few-body configurations and kinetic constraints rather than from dimerization.

For $U_{bf}=0$, the fermions recover symmetric light-cone spreading~\cite{2026supplemental}, confirming that the asymmetry is interaction driven. Thus, the bosonic spatial configuration can control the propagation direction of otherwise Hermitian fermions.

\emph{{\clr Possible experimental route in a cold-atom system}.---}
Our model Eq.~\eqref{BMModel} can be realized using ultracold $^{87}$Rb bosons and $^{40}$K fermions in a one-dimensional optical lattice, a well-established platform for quantum simulation~\cite{Greiner2002,Lewenstein2007,Bloch2008RMP}. After cooling both species to quantum degeneracy, a dilute Bose--Fermi mixture can be loaded into a sufficiently long optical lattice, with the lattice depths controlling the tunneling amplitudes of the two species. The asymmetric inter-cell hopping ($t_L\neq t_R$) can be engineered using a bichromatic lattice combined with periodic driving or phase modulation through Floquet engineering~\cite{2026supplemental}.

The Bose--Fermi interaction $U_{bf}$ can be tuned via a magnetic Feshbach resonance~\cite{PhysRevLett.93.183201,Wang_2011}, while the boson--boson interaction $U_{bb}$ is likewise tunable through Feshbach resonances~\cite{Chin2010}. For identical fermions, on-site interactions vanish due to Pauli exclusion, whereas effective nearest-neighbor interactions $U_{ff}$ can be engineered through density-assisted tunneling or Raman dressing~\cite{Ozawa2019}.

The drag-induced NHSE can be directly probed using species-resolved quantum gas microscopy~\cite{BakrWS2009,Sherson2010}, which provides single-site measurements of $\langle n_j^b\rangle$ and $\langle n_j^f\rangle$. The dilute few-particle regime can be accessed through sparse loading and post-selection of the desired particle-number sector. The measured density profiles and boundary imbalance then provide direct signatures of the drag-induced skin effect.

\emph{{\clr Conclusion and discussion}.---} We have uncovered a drag-induced non-Hermitian skin effect in a one-dimensional Bose--Fermi mixture, where fermions inherit boundary localization from non-Hermitian bosons through strong interspecies correlations. Starting from the minimal one-boson--one-fermion system, we showed that only Bose--Fermi bound states exhibit correlated skin localization, while scattering states remain insensitive to the bosonic NHSE. Extending to the few-body regime, we demonstrated that the interplay of interspecies interactions and quantum statistics gives rise to an interaction-induced blockade, leading to strongly asymmetric and directionally controllable fermionic transport. Dynamical simulations further confirm the robustness of the drag-induced NHSE, highlighting the rich interplay between non-Hermitian dynamics, interactions, and quantum statistics in hybrid quantum systems.

More broadly, our work establishes an interaction-mediated mechanism for transferring non-Hermitian dynamics between different quantum species. Unlike conventional NHSEs, where nonreciprocity must be engineered directly in the particles exhibiting boundary accumulation, the present mechanism allows non-Hermitian dynamics engineered in one species to be transferred to another that remains intrinsically Hermitian through interspecies correlations. The drag-induced NHSE therefore provides a potential route toward species-selective control of localization and transport in hybrid quantum systems, particularly when direct non-Hermitian engineering is more accessible for one component than for another.
The drag-induced skin effect also persists at higher particle numbers, with its observability governed by particle number, system size, filling, and few-body correlations~\cite{2026supplemental}.

\begin{acknowledgments}
W.L. acknowledges support from the Doctoral Scientific Research Start-up Fund of Liaoning Province (Grant No. 2026-BS-0068) and the Central Universities, Dalian University of Technology [DUT25RC(3)084]. 
C.H.L. and Q.Y. acknowledge support from the Singapore Ministry of Education Academic Research Fund Tier 2 under Grant Nos. MOE-T2EP50224-0007  and MOE-T2EP50224-0021.
\end{acknowledgments}


%


\clearpage
\onecolumngrid

\section*{Supplemental Material: Drag-induced skin effect in a Bose-Fermi mixture}
\addcontentsline{toc}{section}{Supplemental Materials} 
\renewcommand{\theequation}{S\arabic{equation}}
\renewcommand{\thefigure}{S\arabic{figure}}
\renewcommand{\thetable}{S\arabic{table}}
\setcounter{equation}{0}
\setcounter{figure}{0}
\setcounter{table}{0}

\renewcommand{\thesection}{S\arabic{section}}
\renewcommand{\thesubsection}{\thesection.\arabic{subsection}}
\renewcommand{\theequation}{S\arabic{equation}}
\renewcommand{\thefigure}{S\arabic{figure}}
\renewcommand{\thetable}{S\arabic{table}}

\section{Persistence of the drag-induced skin effect and interaction blockade in a uniform asymmetric-hopping model}
\label{uniform_model}

In the main text, we consider a dimerized nonreciprocal bosonic lattice as a representative setting for studying the interaction-mediated transfer of non-Hermitian dynamics. Here, we examine a uniform asymmetric-hopping model without dimerization to determine whether the drag-induced skin effect and the interaction-induced blockade depend on the dimerized lattice structure.

The Hamiltonian of the uniform Bose--Fermi mixture is written as
\begin{equation}
    H^{\mathrm{uni}}
    =
    H_b^{\mathrm{uni}}
    +
    H_f
    +
    H_{bf},
    \label{eq:uniform_total_H}
\end{equation}
where the bosonic part is replaced by
\begin{equation}
\begin{aligned}
    H_b^{\mathrm{uni}}
    ={}&
    -\sum_{j=1}^{L-1}
    \left(
    t_L b_j^\dagger b_{j+1}
    +
    t_R b_{j+1}^\dagger b_j
    \right)
    \\
    &+
    \frac{U_{bb}}{2}
    \sum_{j=1}^{L}
    n_j^b
    \left(
    n_j^b-1
    \right).
\end{aligned}
\label{eq:uniform_boson_H}
\end{equation}
Here, $t_L$ and $t_R$ denote the left- and right-moving bosonic hopping amplitudes, respectively. For $t_L\neq t_R$, the bosonic hopping is nonreciprocal and supports the non-Hermitian
skin effect (NHSE) under open boundary conditions. Unlike the model considered in the main text, Eq.~\eqref{eq:uniform_boson_H} contains no distinction between intra-cell and inter-cell bonds.

The fermionic Hamiltonian remains reciprocal,
\begin{equation}
    H_f
    =
    -t_f
    \sum_{j=1}^{L-1}
    \left(
    c_{j+1}^{\dagger}c_j
    +
    c_j^{\dagger}c_{j+1}
    \right)
    +
    U_{ff}
    \sum_{j=1}^{L-1}
    n_j^f n_{j+1}^f,
    \label{eq:uniform_fermion_H}
\end{equation}
and the two species are coupled through the on-site Bose--Fermi interaction
\begin{equation}
    H_{bf}
    =
    U_{bf}
    \sum_{j=1}^{L}
    n_j^b n_j^f.
    \label{eq:uniform_bf_H}
\end{equation}
The uniform model therefore retains the asymmetric bosonic hopping, reciprocal fermionic hopping, and interspecies interaction of the main-text model, while completely removing the dimerized lattice structure.

\subsection{One-boson--one-fermion sector}
\label{subsec:uniform_two_particle}

\begin{figure}[t]
    \centering
    \includegraphics[width=0.5\linewidth]{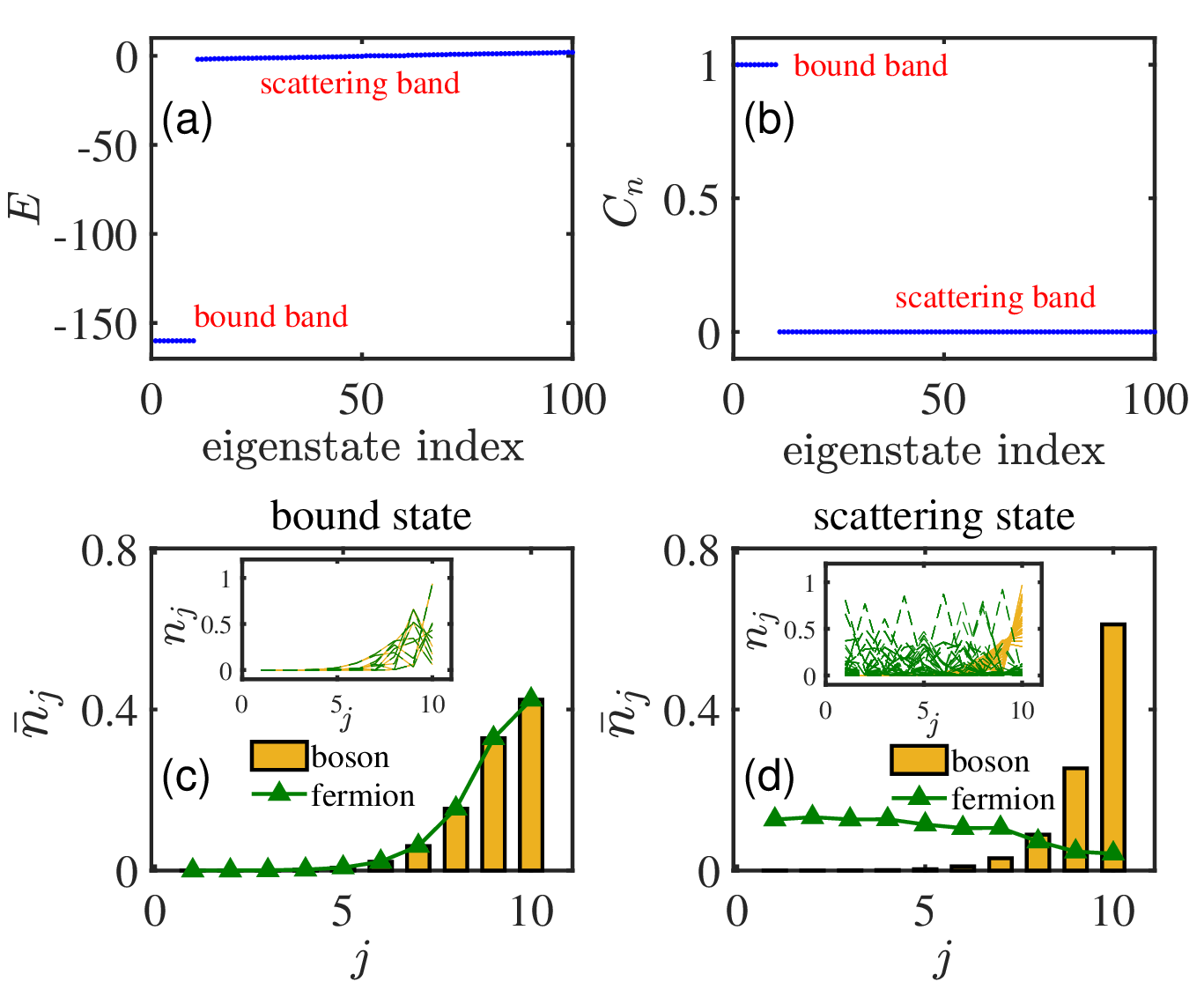}
    \caption{
    Drag-induced NHSE in the uniform asymmetric-hopping model.
    (a) Energy spectrum and
    (b) eigenstate-resolved Bose--Fermi correlation $C_n$ for the one-boson--one-fermion system.
    The low-energy bound band near $E\simeq U_{bf}$ is characterized by $C_n\simeq 1$, whereas the scattering band near zero energy has $C_n\simeq 0$.
    (c),(d) Band-averaged bosonic and fermionic density distributions in the bound and scattering sectors, respectively.
    The insets show the density profiles of the individual eigenstates within the corresponding band.
    In the bound sector, the fermions accumulate at the same boundary as the nonreciprocal bosons.
    In the scattering sector, the bosons remain skin localized, while the fermionic density is nearly extended.
    Parameters are identical to those used in Fig.~1 of the main text, except that the bosonic dimerization is removed and replaced by uniform asymmetric hopping.
    }
    \label{fig:uniform_static}
\end{figure}

We first consider the minimal system containing one boson and one fermion. Figure~\ref{fig:uniform_static} shows the energy spectrum, the eigenstate-resolved Bose--Fermi correlation, and the density distributions in the bound and scattering sectors.
The Bose--Fermi correlation is defined as
\begin{equation}
    C_n
    =
    \sum_{j=1}^{L}
    \left\langle
    \psi_n
    \left|
    n_j^b n_j^f
    \right|
    \psi_n
    \right\rangle,
    \label{eq:uniform_correlation}
\end{equation}
where $\left|\psi_n\right\rangle$ denotes the $n$th eigenstate of the Hamiltonian. In the one-boson--one-fermion sector, $C_n\simeq 1$ indicates that the two particles predominantly occupy the same lattice site, whereas $C_n\simeq 0$ characterizes weakly correlated scattering states.

As shown in Fig.~\ref{fig:uniform_static}(a), the energy spectrum separates into a low-energy bound band near $E\simeq U_{bf}$ and a scattering band located near zero energy. The corresponding correlation function in Fig.~\ref{fig:uniform_static}(b) approaches unity throughout the bound band, while it remains close to zero in the scattering sector. The clear separation between the two sectors therefore originates from the strong Bose--Fermi interaction rather than from the dimerized structure of the bosonic lattice.
Figures~\ref{fig:uniform_static}(c) and \ref{fig:uniform_static}(d) show the density distributions averaged over the bound and scattering bands, respectively. We define the band-averaged density as
\begin{equation}
    \overline{n}_j^{\,\alpha}
    =
    \frac{1}{N_{\mathrm{band}}}
    \sum_{n\in \mathrm{band}}
    \left\langle
    \psi_n
    \left|
    n_j^\alpha
    \right|
    \psi_n
    \right\rangle,
    \qquad
    \alpha=b,f,
    \label{eq:uniform_band_density}
\end{equation}
where $N_{\mathrm{band}}$ is the number of eigenstates included in the corresponding band.
In the bound sector, both the bosonic and fermionic densities accumulate toward the same boundary, as shown in Fig.~\ref{fig:uniform_static}(c). Since only the bosons experience asymmetric hopping, the fermionic boundary accumulation is induced entirely through the strong Bose--Fermi correlation. The boson and fermion form an effective composite object that inherits the nonreciprocal motion of the bosonic component.
By contrast, in the scattering sector shown in Fig.~\ref{fig:uniform_static}(d), the bosonic density remains strongly localized near the right boundary, whereas the fermionic density is nearly extended over the lattice. The fermion therefore does not automatically inherit the bosonic skin effect. Instead, the transfer of non-Hermitian localization requires the formation of a strongly correlated Bose--Fermi bound state.
The same bound--scattering dichotomy obtained for the dimerized model thus persists in the uniform asymmetric-hopping model. This confirms that the drag-induced NHSE is controlled by interspecies correlations and bosonic nonreciprocity, rather than by dimerization.

The persistence of the drag-induced NHSE can also be understood from the strong-coupling effective description. As shown by the derivation presented in Sec.~\ref{Effectiveness}, projecting onto the Bose--Fermi bound-state subspace yields an effective nonreciprocal hopping of the composite particle. Repeating the same procedure for the uniform asymmetric-hopping model leads to effective left- and right-moving hoppings proportional to $t_Lt_f/U_{bf}$ and $t_Rt_f/U_{bf}$, respectively. Their nonreciprocity ratio therefore remains $t_R/t_L$, demonstrating that the drag mechanism is independent of the dimerized lattice structure.

\subsection{Dynamics of the bound and scattering states}
\label{subsec:uniform_two_particle_dynamics}

\begin{figure}[t]
    \centering
    \includegraphics[width=0.5\linewidth]
    {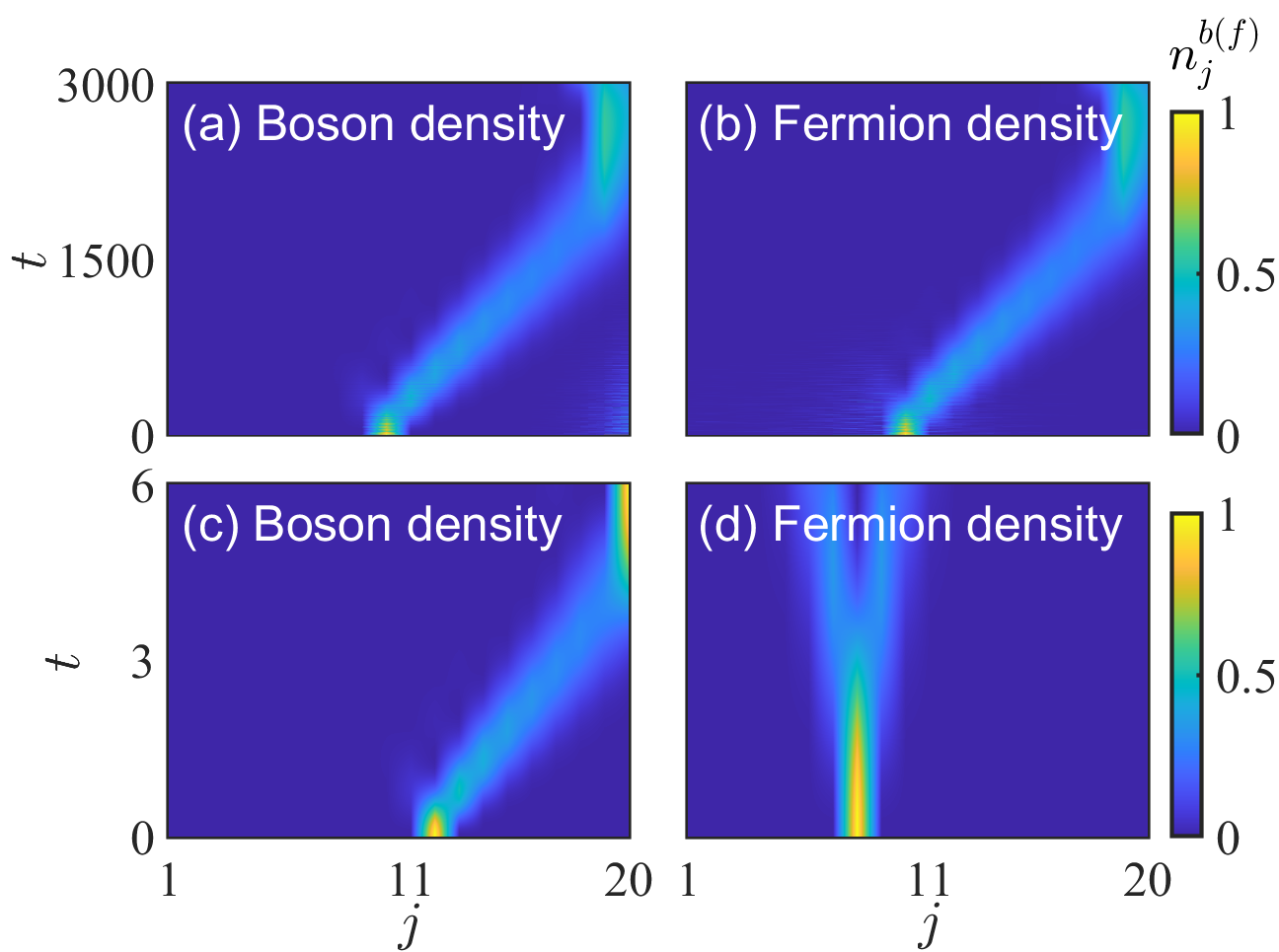}
    \caption{
    Dynamics of the drag-induced NHSE in the uniform asymmetric-hopping model.
    (a),(b) Time evolution of the bosonic and fermionic densities for an initially bound Bose--Fermi pair.
    The two species remain spatially correlated and propagate together toward the right boundary.
    (c),(d) Corresponding dynamics for an initial scattering configuration.
    The boson undergoes directional non-Hermitian transport, whereas the fermion spreads nearly symmetrically around its initial position.
    The longer time interval used in (a),(b) reflects the reduced effective hopping of the strongly bound composite.
    Parameters and initial states are identical to those used in Figs.~2(a)--(d) of the main text.
    }
    \label{fig:uniform_two_particle_dynamics}
\end{figure}

We next examine the real-time dynamics of the uniform model. Figure~\ref{fig:uniform_two_particle_dynamics} compares the evolution from an initially bound Bose--Fermi pair with that from an initial scattering configuration.

For the bound initial state, the bosonic and fermionic wavepackets remain spatially correlated throughout the evolution and propagate together toward the right boundary, as shown in Figs.~\ref{fig:uniform_two_particle_dynamics}(a) and \ref{fig:uniform_two_particle_dynamics}(b). The fermion therefore follows the directional motion of the nonreciprocal boson, despite having reciprocal hopping in isolation. Upon reaching the boundary, both components exhibit pronounced accumulation.
The long time scale of the bound-state dynamics is a consequence of the reduced effective hopping of the composite particle, as derived in Sec.~\ref{Effectiveness}.

For the scattering initial state, the boson still propagates preferentially toward the right due to its asymmetric hopping, as shown in Fig.~\ref{fig:uniform_two_particle_dynamics}(c). The fermion, however, exhibits nearly symmetric spreading around its initial position, as shown in Fig.~\ref{fig:uniform_two_particle_dynamics}(d). The fermionic dynamics remains essentially reciprocal because the weakly correlated scattering configuration does not support an effective bound composite.

These results reproduce the dynamical bound--scattering distinction found in the dimerized model and demonstrate that the dynamical drag-induced NHSE does not require alternating bonds.

\subsection{Band hierarchy in the two-boson--two-fermion sector}
\label{subsec:uniform_four_particle_spectrum}

We now consider a system containing two bosons and two fermions. The purpose is to determine whether the few-body band hierarchy and the associated configuration-dependent localization discussed in the main text originate from the dimerized lattice structure.
For multiple particles, the Bose--Fermi correlation is again defined by
\begin{equation}
    C_n
    =
    \sum_{j=1}^{L}
    \left\langle
    \psi_n
    \left|
    n_j^b n_j^f
    \right|
    \psi_n
    \right\rangle.
    \label{eq:uniform_four_particle_correlation}
\end{equation}
Unlike the one-boson--one-fermion sector, $C_n$ can exceed unity because more than one Bose--Fermi pair may occupy the same lattice site.

\begin{figure*}[t]
    \centering
    \includegraphics[width=0.98\textwidth]
    {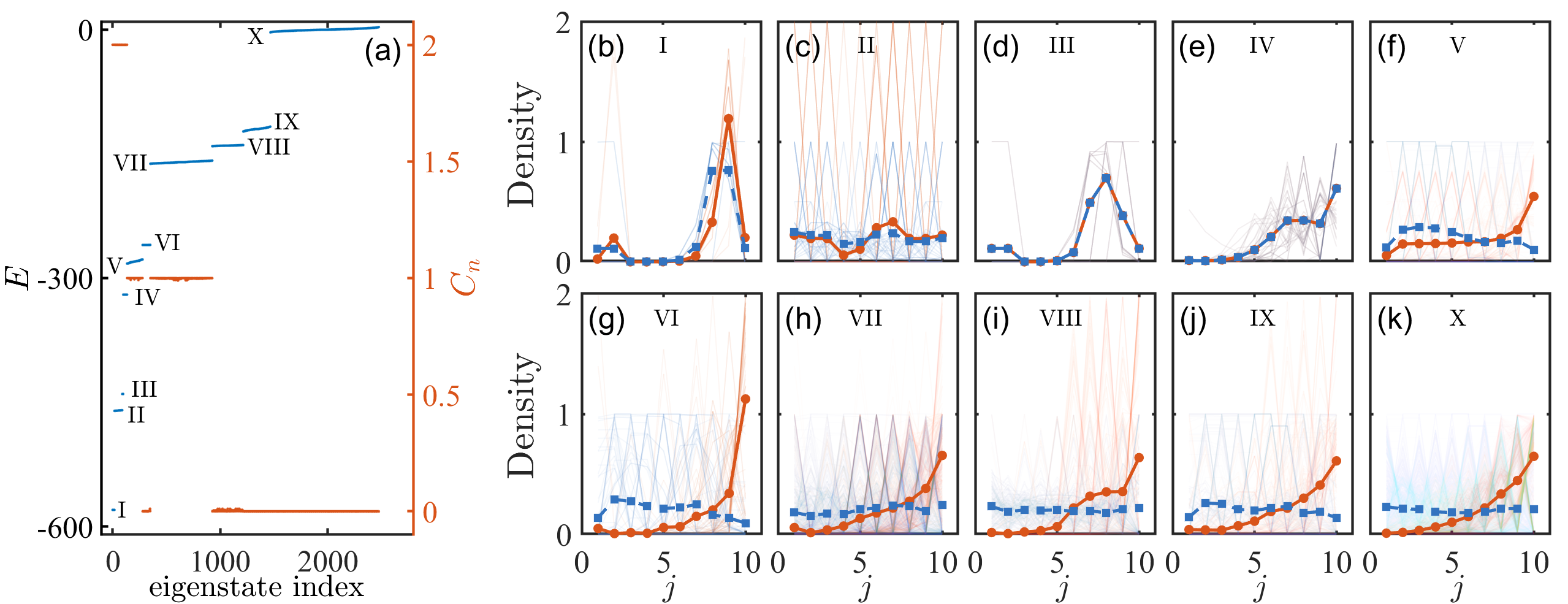}
    \caption{
    Band hierarchy and density structures of the uniform two-boson--two-fermion model.
    (a) Ordered energy spectrum and eigenstate-resolved Bose--Fermi correlation $C_n$.
    The spectrum separates into ten interaction-organized clusters labeled I--X.
    (b)--(k) Bosonic and fermionic density profiles in the corresponding bands.
    The faint curves represent individual eigenstates, while the thick curves with markers denote the band-averaged densities.
    The orange curves correspond to bosons and the blue curves correspond to fermions.
    The persistence of the band hierarchy in the absence of alternating bonds demonstrates that it is primarily generated by interactions, quantum statistics, and the allowed few-body occupation configurations.
    Parameters are identical to those used in Fig.~3 of the main text.
    }
    \label{fig:uniform_four_particle_spectrum}
\end{figure*}

Figure~\ref{fig:uniform_four_particle_spectrum}(a) shows the full energy spectrum, ordered by increasing energy, together with the corresponding eigenstate-resolved correlation $C_n$. Even in the absence of alternating bonds, the spectrum separates into ten well-resolved clusters, labeled I--X. Different bands exhibit distinct correlation values, indicating different microscopic occupation structures.

The persistence of this band hierarchy in the uniform model demonstrates that its main origin is the interaction energy and quantum statistics rather than dimerization. In the strong-interaction regime, different few-body configurations carry different combinations of interaction energies. These include bosonic double occupancy, nearest-neighbor fermionic occupation, and on-site Bose--Fermi pairing. Schematically, the interaction energy of a given configuration can be expressed as
\begin{equation}
    E_{\mathrm{int}}
    =
    N_{bb}U_{bb}
    +
    N_{ff}U_{ff}
    +
    N_{bf}U_{bf},
    \label{eq:uniform_interaction_energy}
\end{equation}
where $N_{bb}$, $N_{ff}$, and $N_{bf}$ count the corresponding interaction channels. Different allowed occupation configurations therefore organize into distinct energy sectors, which are broadened into bands by particle hopping.

Figures~\ref{fig:uniform_four_particle_spectrum}(b)--\ref{fig:uniform_four_particle_spectrum}(k) show the bosonic and fermionic density profiles in the ten corresponding bands. The faint curves denote the densities of individual eigenstates, while the thick curves with markers show the band-averaged densities. As in the dimerized model, different bands exhibit markedly different localization behavior.

In some bands, the bosonic and fermionic components accumulate toward the same boundary, indicating a strong interaction-mediated drag response. In other bands, the bosons remain strongly skin localized while the fermions are more extended or weakly localized. These differences reflect the distinct correlated occupation structures and the different kinetic constraints associated with each band.
The interaction-organized band hierarchy and the band-dependent transfer of non-Hermitian localization remain clearly visible. The dimerized structure is therefore not responsible for the existence of the few-body hierarchy, although it can enrich its detailed spectral and spatial features.

\subsection{Interaction-induced blockade without dimerization}
\label{subsec:uniform_blockade}

\begin{figure}[t]
    \centering
    \includegraphics[width=0.5\linewidth]
    {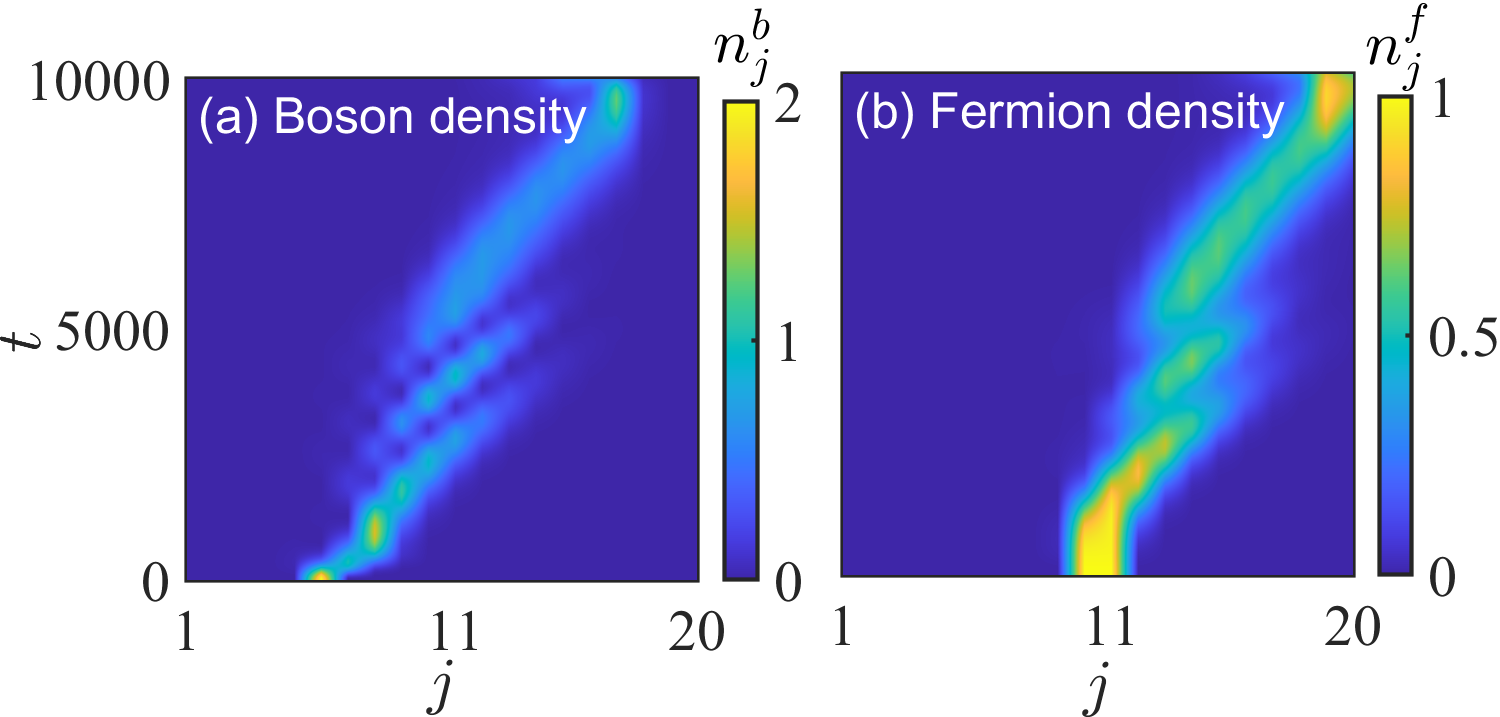}
    \caption{
    Interaction-induced blockade in the uniform asymmetric-hopping model.
    Time evolution of the
    (a) bosonic and
    (b) fermionic densities for the initial state
$\left|\psi_0\right\rangle=\left|2_b\right\rangle_j\left|1_f\right\rangle_{j+4}\left|1_f\right\rangle_{j+5}$.
    Although the bosons are driven toward the right by the asymmetric hopping, the fermionic propagation is strongly constrained and becomes highly asymmetric.
    The persistence of this behavior in the absence of alternating bonds confirms that the blockade originates from the interaction-stabilized few-body configuration, fermionic statistics, and the associated kinetic constraints rather than from dimerization.
    Parameters are identical to those used in Fig.~4 of the main text.
    }
    \label{fig:uniform_blockade}
\end{figure}

Finally, we examine whether the interaction-induced blockade discussed in the main text survives in the uniform asymmetric-hopping model. Figure~\ref{fig:uniform_blockade} shows the time evolution of the bosonic and fermionic densities for a representative interaction-stabilized few-body configuration.

The initial state $\left|\psi_0\right\rangle=\left|2_b\right\rangle_j\left|1_f\right\rangle_{j+4}\left|1_f\right\rangle_{j+5}$ is identical to that used in Figs.~4(a) and 4(b) of the main text.
As shown in Fig.~\ref{fig:uniform_blockade}(a), the bosons are driven preferentially toward the right by the asymmetric hopping. Their propagation is substantially slower than that of isolated particles because the initial configuration is stabilized by strong interactions and its motion requires higher-order rearrangement processes.
The fermionic dynamics in Fig.~\ref{fig:uniform_blockade}(b) is strongly asymmetric and differs qualitatively from the symmetric light-cone spreading expected for freely propagating reciprocal fermions. The fermions cannot move independently without modifying the strongly interacting Bose--Fermi occupation structure. Their propagation therefore requires a sequence of intermediate configurations that may carry a large interaction-energy cost.

In addition, Pauli exclusion prevents the two identical fermions from occupying the same lattice site. Together with the nearest-neighbor fermionic interaction and strong Bose--Fermi coupling, this statistical constraint eliminates or suppresses selected rearrangement pathways. The resulting configuration-dependent kinetic constraints produce an interaction-induced blockade of fermionic transport.

Importantly, the uniform model contains no alternating strong and weak bonds. The persistence of the strongly constrained and asymmetric fermionic motion therefore demonstrates that the blockade does not originate from the dimerized hopping structure. Its essential ingredients are instead the interaction-stabilized few-body configuration, fermionic statistics, and the directional drive supplied by the nonreciprocal bosons.

In summary, the uniform asymmetric-hopping model reproduces all central qualitative phenomena obtained in the dimerized model. In the one-boson--one-fermion sector, strongly correlated bound states exhibit a drag-induced NHSE, whereas weakly correlated scattering states do not transfer the bosonic skin localization to the fermions. In the two-boson--two-fermion sector, the interaction-organized band hierarchy and the interaction-induced blockade also remain present.

These results demonstrate that neither the drag-induced NHSE nor the blockade mechanism requires a dimerized bosonic lattice. The dimerized model considered in the main text was chosen as a representative nonreciprocal lattice to demonstrate that the proposed mechanism is not restricted to the simplest uniform asymmetric-hopping model. The results presented here further establish the robustness and generality of the interaction-mediated transfer of non-Hermitian dynamics in Bose--Fermi mixtures.

\section{Real spectrum from a non-unitary similarity transformation}
\label{sec:hermitian_mapping}

The non-Hermitian nature of the Hamiltonian arises from the asymmetric
inter-cell hopping amplitudes $t_L\neq t_R$ in the bosonic sector. Under open
boundary conditions, and provided that $t_L/t_R>0$, this non-Hermitian
Hamiltonian can be mapped to a Hermitian counterpart by a non-unitary
similarity transformation. This implies that the Hamiltonian is
pseudo-Hermitian and hence possesses a real spectrum~\cite{MostafazadehA2002}.
However, pseudo-Hermiticity and spectral reality do not imply that the system becomes physically Hermitian. The left and right eigenstates remain distinct, and the non-unitary similarity transformation modifies the eigenstates, physical observables, and nonequilibrium dynamics.
Importantly, the existence of such a Hermitian counterpart does not make the
skin effect trivial. The similarity transformation is non-unitary and therefore
changes the spatial structure of eigenstates and physical observables such as
density profiles and non-equilibrium dynamics. 
The drag-induced NHSE discussed in the main text is therefore encoded in the non-unitarily transformed eigenstates and observables, rather than in the energy spectrum itself.

We introduce the cell-dependent similarity transformation
\begin{equation}
    \tilde{H}=\eta H\eta^{-1},
\end{equation}
with
\begin{equation}
    \eta =
    \exp\left[
    \kappa \sum_{m=1}^{L/2}
    m\left(n^b_{2m-1}+n^b_{2m}\right)
    \right],
    \qquad
    \kappa=\frac{1}{2}\ln\left(\frac{t_L}{t_R}\right).
    \label{eq:eta_cell}
\end{equation}
This transformation is well-defined for $t_L/t_R>0$, for which $\kappa$ is real
and $\eta$ is Hermitian, positive definite, and invertible. It acts on the
bosonic operators as
\begin{align}
    \eta b_{2m-1}\eta^{-1} &= e^{-\kappa m} b_{2m-1},&
    \eta b^\dagger_{2m-1}\eta^{-1} &= e^{\kappa m} b^\dagger_{2m-1},\\
    \eta b_{2m}\eta^{-1} &= e^{-\kappa m} b_{2m},&
    \eta b^\dagger_{2m}\eta^{-1} &= e^{\kappa m} b^\dagger_{2m}.
\end{align}
The bosonic number operators are invariant,
\begin{equation}
    \eta n^b_j \eta^{-1}=n^b_j .
\end{equation}
The fermionic operators are also unchanged because $\eta$ acts only on the
bosonic sector.

We now transform the bosonic hopping terms. The intra-cell hopping is unchanged:
\begin{align}
&\eta\left[
-t_0\sum_{m=1}^{L/2}
\left(b^\dagger_{2m-1}b_{2m}+b^\dagger_{2m}b_{2m-1}\right)
\right]\eta^{-1}
\nonumber\\
&=
-t_0\sum_{m=1}^{L/2}
\left(b^\dagger_{2m-1}b_{2m}+b^\dagger_{2m}b_{2m-1}\right),
\end{align}
because the two sites within the same unit cell carry the same scaling factor.
For the inter-cell hopping, one obtains
\begin{align}
&\eta\left[
-\sum_{m=1}^{L/2-1}
\left(t_L b^\dagger_{2m}b_{2m+1}
+t_R b^\dagger_{2m+1}b_{2m}\right)
\right]\eta^{-1}
\nonumber\\
&=
-\sum_{m=1}^{L/2-1}
\left(
t_L e^{-\kappa} b^\dagger_{2m}b_{2m+1}
+
t_R e^{\kappa} b^\dagger_{2m+1}b_{2m}
\right).
\end{align}
Using $e^\kappa=\sqrt{t_L/t_R}$, we have
\begin{equation}
    t_L e^{-\kappa}=t_R e^{\kappa}=\sqrt{t_Lt_R}.
\end{equation}
Therefore, the transformed inter-cell hopping becomes Hermitian:
\begin{equation}
-\sqrt{t_Lt_R}
\sum_{m=1}^{L/2-1}
\left(
b^\dagger_{2m}b_{2m+1}
+b^\dagger_{2m+1}b_{2m}
\right).
\end{equation}
Since all density operators are invariant under $\eta$, the boson--boson
interaction and the Bose--Fermi interaction remain unchanged:
\begin{align}
\eta\left[
\frac{U_{bb}}{2}\sum_{j=1}^L n^b_j(n^b_j-1)
\right]\eta^{-1}
&=
\frac{U_{bb}}{2}\sum_{j=1}^L n^b_j(n^b_j-1),\\
\eta\left[
U_{bf}\sum_{j=1}^L n^b_j n^f_j
\right]\eta^{-1}
&=
U_{bf}\sum_{j=1}^L n^b_j n^f_j .
\end{align}
The fermionic Hamiltonian is also invariant because the transformation acts
only on bosons.

Combining these terms, the transformed Hamiltonian reads
\begin{align}
\tilde{H}_b
&=
-t_0\sum_{m=1}^{L/2}
\left(b^\dagger_{2m-1}b_{2m}+b^\dagger_{2m}b_{2m-1}\right)
\nonumber\\
&\quad
-\sqrt{t_Lt_R}\sum_{m=1}^{L/2-1}
\left(b^\dagger_{2m}b_{2m+1}+b^\dagger_{2m+1}b_{2m}\right)
\nonumber\\
&\quad
+\frac{U_{bb}}{2}\sum_{j=1}^L n^b_j(n^b_j-1),
\\
\tilde{H}_f&=H_f,\\
\tilde{H}_{bf}&=U_{bf}\sum_{j=1}^L n^b_j n^f_j .
\end{align}
Thus,
\begin{equation}
    \tilde{H}=\tilde{H}_b+\tilde{H}_f+\tilde{H}_{bf}
\end{equation}
is Hermitian for real $t_0$, $t_f$, $U_{bb}$, $U_{ff}$, $U_{bf}$, and $t_L/t_R>0$.
Consequently, the original non-Hermitian Hamiltonian has the same real spectrum as its Hermitian counterpart under open boundary conditions. Nevertheless, the two Hamiltonians are not physically equivalent because the non-unitary similarity transformation changes the eigenstates and observables. The resulting spatial reshaping of the right eigenstates is responsible for the non-Hermitian skin effect.

\section{Drag-induced skin effect beyond the strong-coupling limit}

In the main text, we have focused on the strong Bose–Fermi interaction regime (e.g., $U_{bf}=-160$), where the bound and scattering sectors are well separated in energy. This separation allows for a particularly transparent identification of the drag-induced NHSE: fermions inherit the boundary localization of bosons only in the bound sector, while remaining delocalized in the scattering sector.
In this section, we examine the fate of this phenomenon when the interaction strength is reduced to values comparable to the single-particle energy scales. Specifically, we consider intermediate and weak interaction regimes with $U_{bf}=-16$ and $U_{bf}=-1.6$, while keeping all other parameters identical to those in Fig.~1 of the main text.
The parameters used in the main text are selected to clearly separate the bound and scattering sectors and highlight the drag mechanism. However, the phenomenon is not limited to this parameter regime.

\begin{figure}[t]
\centering
\includegraphics[width=0.5\linewidth]{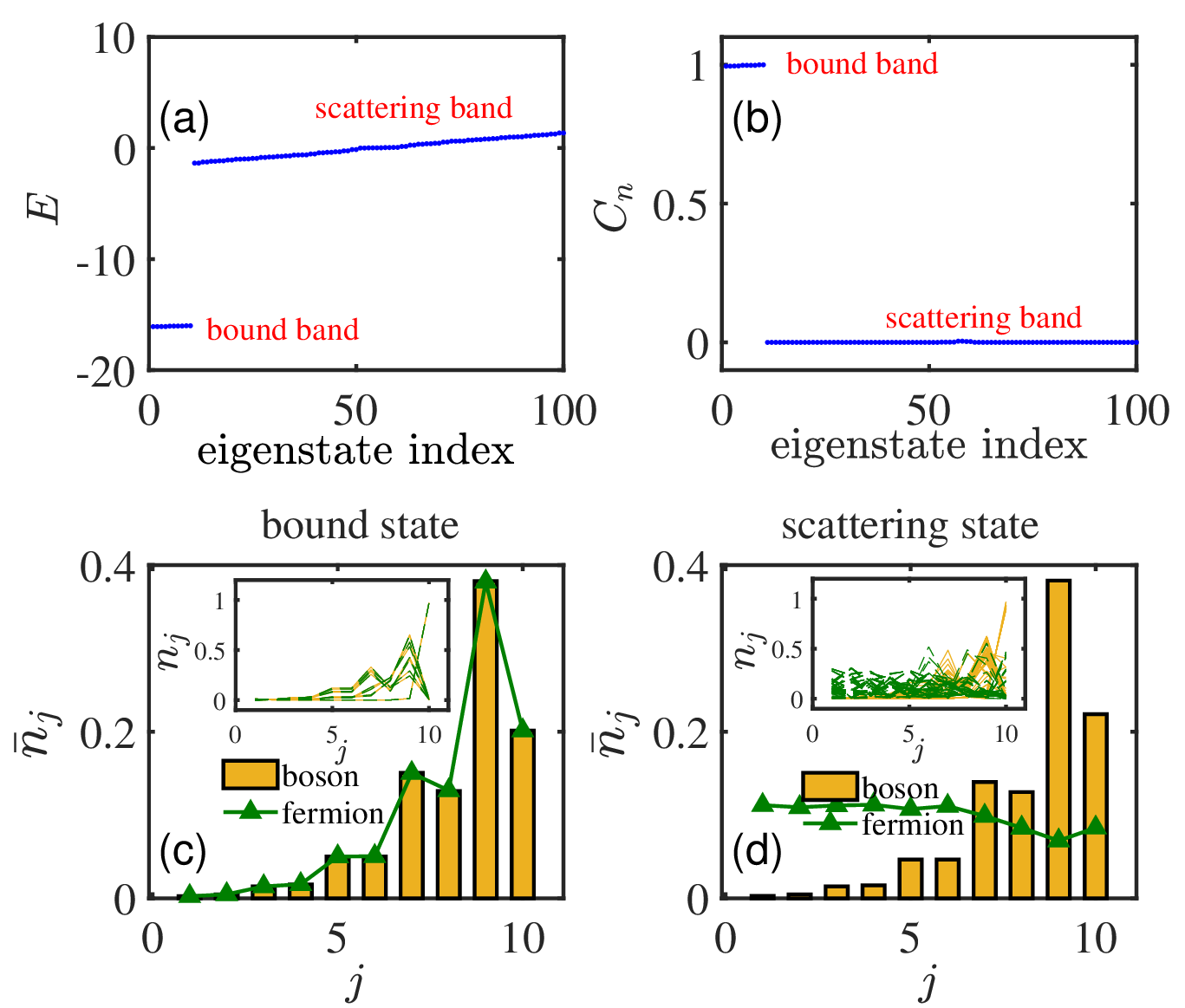}
\caption{
Energy spectrum, boson–fermion correlation, and density distributions for the intermediate interaction strength $U_{bf}=-16$, with all other parameters identical to Fig.~1 of the main text. 
(a) Energy spectrum as a function of eigenstate index, showing a reduced but still discernible separation between the bound and scattering sectors. 
(b) Boson–fermion correlation function $C_n$, where low-energy states retain strong correlations close to unity, indicating the persistence of tightly bound composites. 
(c),(d) Average density distributions for representative low-energy and high-energy states, respectively. 
In the low-energy sector, both bosons and fermions exhibit pronounced boundary accumulation with nearly identical spatial profiles, demonstrating that the drag-induced skin effect remains essentially unchanged compared to the strong-coupling case. 
In higher-energy states, bosons retain their intrinsic skin localization, while fermions remain extended, consistent with the scattering sector behavior.
}
\label{fig:16}
\end{figure}

\subsection{Intermediate interaction regime}

We first consider the case $U_{bf}=-16$, shown in Fig.~\ref{fig:16}. Compared to the strong-coupling case in Fig.~1 of the main text, the energy separation between the bound band and the scattering band is reduced [Fig.~\ref{fig:16}(a)]. However, the overall band structure and correlation characteristics remain qualitatively unchanged. In particular, the low-energy states still exhibit strong boson–fermion correlations, with $C_n$ remaining close to unity [Fig.~\ref{fig:16}(b)], indicating the persistence of tightly bound composite states.

The corresponding density distributions further confirm that the drag-induced skin effect remains essentially intact in this regime. As shown in Fig.~\ref{fig:16}(c), both bosons and fermions are strongly localized near the right boundary in the low-energy sector, demonstrating that fermions continue to be efficiently dragged by the bosonic skin accumulation. For higher-energy states [Fig.~\ref{fig:16}(d)], the bosons retain their intrinsic skin localization, while the fermions remain largely extended, consistent with the behavior observed in the strong-coupling limit.
These results indicate that the drag-induced skin effect is robust against a substantial reduction of the interaction strength, and does not rely on an extreme separation of energy scales.

\subsection{Weak interaction regime}

\begin{figure}[t]
\centering
\includegraphics[width=0.5\linewidth]{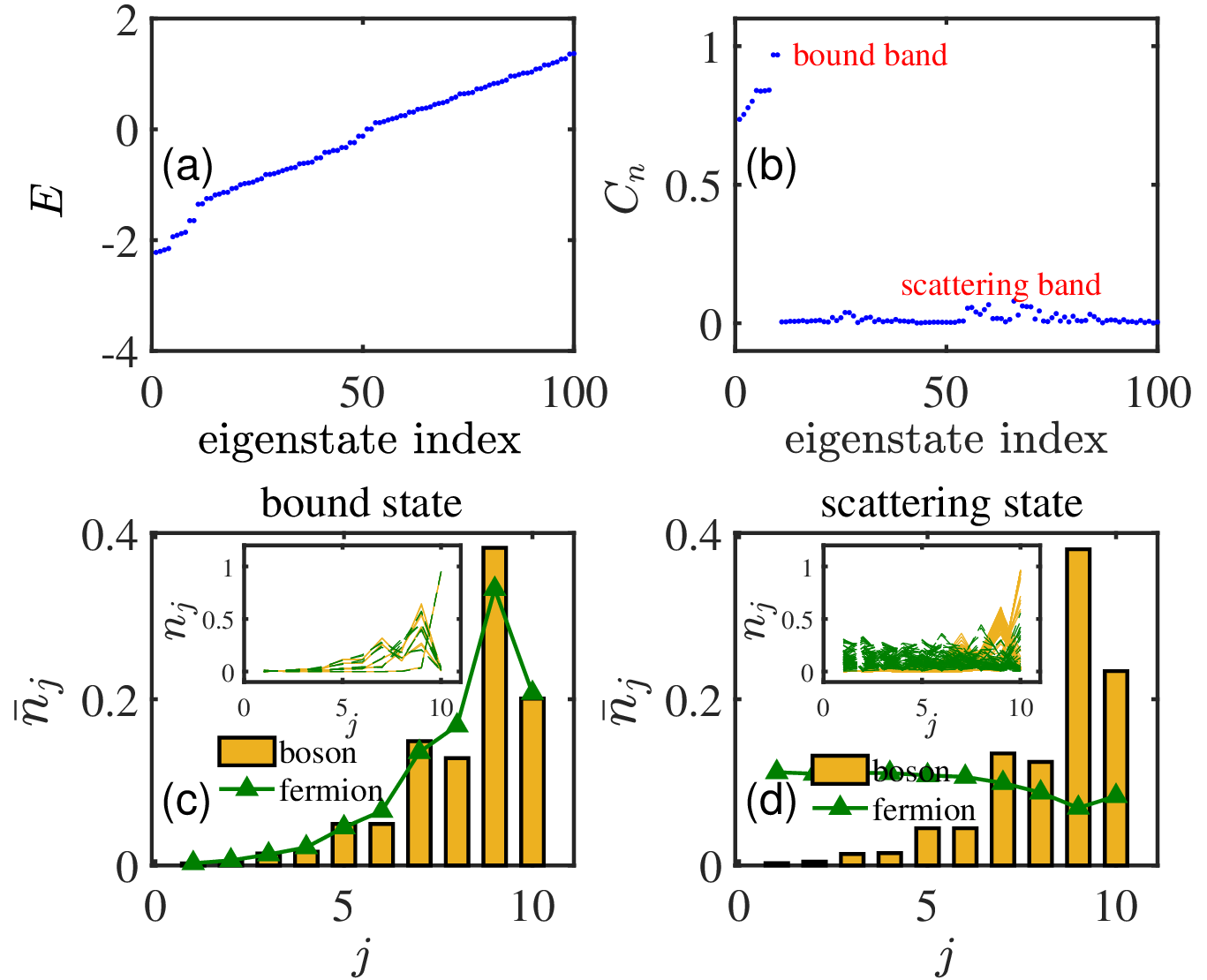}
\caption{
Same as Fig.~S5, but for weak interaction strength $U_{bf}=-1.6$. 
(a) The energy spectrum appears as a nearly continuous band without a clearly isolated bound sector. 
(b) The boson–fermion correlation $C_n$ nevertheless reveals a distinction between weakly and strongly correlated states, with a subset of low-energy states retaining relatively large correlation values. 
(c),(d) Corresponding density distributions for representative states. 
In the correlated low-energy sector, both bosons and fermions exhibit boundary accumulation, indicating that a drag-induced effect persists even in the weak interaction regime. However, the fermionic localization no longer perfectly follows that of the bosons, and their density profiles show reduced overlap. 
For higher-energy states, bosons remain strongly localized due to the NHSE, while fermions are largely delocalized, indicating the suppression of the drag mechanism in weakly correlated states.
}
\label{fig:1dot6}
\end{figure}

We now turn to the weak interaction case $U_{bf}=-1.6$, shown in Fig.~\ref{fig:1dot6}. In this regime, the energy spectrum appears as a nearly continuous band without a clearly isolated bound sector [Fig.~\ref{fig:1dot6}(a)]. Nevertheless, the boson–fermion correlation function still reveals a distinction between different classes of states: a subset of low-energy eigenstates retains relatively large values of $C_n$, indicating residual boson–fermion binding [Fig.~\ref{fig:1dot6}(b)].

This residual correlation is reflected in the density distributions. In the correlated low-energy sector, bosons and fermions both exhibit boundary accumulation, signaling that a drag-induced effect is still present [Fig.~\ref{fig:1dot6}(c)]. However, in contrast to the strong- and intermediate-coupling regimes, the fermionic localization no longer perfectly follows that of the bosons, and the overlap between their density profiles is reduced. 
For higher-energy states [Fig.~\ref{fig:1dot6}(d)], the bosons continue to display a clear NHSE, while the fermions are largely delocalized, indicating that the drag mechanism becomes ineffective in weakly correlated states.

Combining these results with those in the main text, we obtain a unified picture of the drag-induced skin effect across different interaction regimes. In both the strong and intermediate interaction regimes, well-defined boson–fermion bound states enable an efficient transfer of non-Hermitian localization from bosons to fermions, leading to nearly identical skin accumulation for the two species.
As the interaction strength is further reduced, the system enters a crossover regime where the distinction between bound and scattering sectors becomes less apparent in the energy spectrum, but remains visible in correlation measures. In this regime, the drag-induced skin effect persists in correlated states, although the fermionic localization gradually deviates from that of the bosons.

Importantly, the drag-induced skin effect does not disappear abruptly. Instead, it evolves continuously from a strongly correlated, bound-state-driven mechanism to a weaker, correlation-assisted effect that survives even when the interaction strength is comparable to the single-particle energy scales.
These results demonstrate that the phenomenon is robust over a broad parameter regime and does not rely on extreme interaction strengths, thereby reinforcing the experimental feasibility of observing drag-induced skin effects in Bose–Fermi mixtures.


\section{Effective single-particle model of a boson-fermion pair} \label{Effectiveness}

We separate the hybrid system into two parts with the dominant part $H_{b f}$
and the perturbation one $H_{b}$+$H_{f}$.
The domination $H_{b f}$ can be divided into two degenerate subspaces $\mathcal{U}$ and $\mathcal{V}$.
The subspace $\mathcal{U}\equiv\left\{|2\rangle_{j}\right\}$ contains bound states which two particles occupy the same site, and its degenerate energy is $E_{j}=U_{bf}$.
The subspace $\mathcal{V}\equiv\left\{|1\rangle_{j}|1\rangle_{k}\right\} $ contains scattering states which two particles populate on different sites, and its degenerate energy is $E_{j k}=0$ with $j\neq k$.
The projection operators of subspaces $\mathcal{U}$ and $\mathcal{V}$ are written as
\begin{equation}
\hat{P}=\sum_{j}|2\rangle_{j}\langle 2|_{j}
\end{equation}
and
\begin{equation}
\hat{S}=\sum_{j \neq k} \frac{1}{E_{j}-E_{jk}}|1\rangle_{j}|1\rangle_{k}\langle 1|_{k}\langle 1|_{j}.
\end{equation}
Perturbation theory gives the effective single-particle model in the subspace $\mathcal{U}$ by
\begin{equation}
\hat{H}_{\mathrm{eff}}=\hat{h}_{0}+\hat{h}_{1}+\hat{h}_{2}.
\end{equation}
The zeroth-order perturbation calculation is
\begin{equation}\label{ZeroTerm}
\hat{h}_{0}=E_{j} \hat{P}=U_{bf}\sum_{j}|2\rangle_{j}\langle 2|_{j}.
\end{equation}
The first-order perturbation calculation is zero.
The second-order perturbation calculation is
\begin{equation}
\begin{aligned}
\hat{h}_{2} &= \hat{P}\left(H_{b}+H_{f}\right)\hat{S}\left(H_{b}+H_{f}\right)\hat{P} \\
&= \frac{t_{0}^{2}+t_{f}^{2}}{U_{bf}}(|2\rangle_{1}\langle 2|_{1} + |2\rangle_{L}\langle 2|_{L} )+ \left(\frac{t_{0}^{2}+2t_{f}^{2}+t_{L} t_{R}}{U_{bf}}\right) \sum_{j=2}^{L-1}|2\rangle_{j}\langle 2|_{j} \\
&\quad + 2\frac{t_{0}t_f}{U_{bf}} \sum_{j=1}^{L / 2}|2\rangle_{2 j-1}\langle 2|_{2 j} + 2\frac{t_{0}t_f}{U_{bf}} \sum_{j=1}^{L / 2}|2\rangle_{2 j}\langle 2|_{2 j-1} \\
&\quad + 2\frac{t_{L} t_{f}}{U_{bf}} \sum_{j=1}^{L / 2-1}|2\rangle_{2 j}\langle 2|_{2 j+1} + 2\frac{t_{f} t_{R}}{U_{bf}} \sum_{j=1}^{L / 2-1}|2\rangle_{2 j+1}\langle 2|_{2 j}.
\end{aligned}
\end{equation}
Then we obtain the effective single-particle model as
\begin{equation}
\begin{aligned}
\hat{H}_{\text{eff}} &= \mathcal{C} \sum_{j=1}^{L} \hat{n}_{j} - \left(\frac{t_{f}^{2}}{U_{bf}} + \frac{t_{L} t_{R}}{U_{bf}}\right)\left(\hat{n}_{1} + \hat{n}_{L}\right) \\
&\quad + 2 \frac{t_{0}t_{f}}{U_{bf}} \sum_{j=1}^{L / 2} \hat{S}_{2 j-1}^{\dagger} \hat{S}_{2 j} + 2 \frac{t_{0}t_{f}}{U_{bf}} \sum_{j=1}^{L / 2} \hat{S}_{2 j}^{\dagger} \hat{S}_{2 j-1} \\
&\quad + 2 \frac{t_{L} t_{f}}{U_{bf}} \sum_{j=1}^{L / 2 - 1} \hat{S}_{2 j}^{\dagger} \hat{S}_{2 j+1} + 2 \frac{t_{f} t_{R}}{U_{bf}} \sum_{j=1}^{L / 2 - 1} \hat{S}_{2 j+1}^{\dagger} \hat{S}_{2 j}.
\end{aligned}
\end{equation}

Here, $\hat{S}_{j}^{\dagger}$ creates a Bose--Fermi bound composite at
the $j$th site,
\begin{equation}
    \hat{S}_{j}^{\dagger}|\mathbf{0}\rangle
    =
    |2\rangle_{j},
\end{equation}
and $\hat{n}_{j}=\hat{S}_{j}^{\dagger}\hat{S}_{j}$ denotes the
corresponding composite-particle number operator. The constant
\begin{equation}
    \mathcal{C}
    =
    U_{bf}
    +
    \frac{t_{0}^{2}+2t_{f}^{2}+t_{L}t_{R}}{U_{bf}}
\end{equation}
produces only an overall energy shift and therefore does not affect the
bulk localization properties. The second term in
$\hat{H}_{\mathrm{eff}}$ represents a weak boundary energy correction,
which likewise does not modify the bulk nonreciprocity of the effective
composite model.

The effective Hamiltonian shows that a tightly bound Bose--Fermi pair
behaves as a single composite particle. Its intra-cell hopping is
reciprocal and proportional to $t_{0}t_f/U_{bf}$, whereas its effective
inter-cell hoppings are nonreciprocal and proportional to
$t_Lt_f/U_{bf}$ and $t_Rt_f/U_{bf}$. The corresponding nonreciprocity
ratio is therefore
\begin{equation}
    \frac{J_R^{\mathrm{eff}}}{J_L^{\mathrm{eff}}}
    =
    \frac{t_R}{t_L},
\end{equation}
which is identical to that of the nonreciprocal bosonic hopping and is
independent of the reciprocal fermionic hopping amplitude $t_f$.
Consequently, the Bose--Fermi bound composite inherits the bosonic
nonreciprocity and exhibits the NHSE under open boundary conditions.

For completeness, the same perturbative construction can be directly
applied to the uniform asymmetric-hopping model introduced in
Sec.~S1. In that case, the bosonic Hamiltonian contains no alternating
intra-cell and inter-cell bonds, and the bound composite moves between
all neighboring sites through the effective Hamiltonian
\begin{equation}
\begin{aligned}
    \hat{H}_{\mathrm{eff}}^{\mathrm{uni}}
    ={}&
    \mathcal{C}_{\mathrm{uni}}
    \sum_{j=1}^{L}\hat{n}_{j}
    -
    \frac{t_f^{2}+t_Lt_R}{U_{bf}}
    \left(
    \hat{n}_{1}+\hat{n}_{L}
    \right)
    \\
    &+
    2\frac{t_Lt_f}{U_{bf}}
    \sum_{j=1}^{L-1}
    \hat{S}_{j}^{\dagger}\hat{S}_{j+1}
    +
    2\frac{t_Rt_f}{U_{bf}}
    \sum_{j=1}^{L-1}
    \hat{S}_{j+1}^{\dagger}\hat{S}_{j},
\end{aligned}
\label{eq:uniform_effective_model}
\end{equation}
where
\begin{equation}
    \mathcal{C}_{\mathrm{uni}}
    =
    U_{bf}
    +
    2\frac{t_f^{2}+t_Lt_R}{U_{bf}}.
\end{equation}
The effective left- and right-moving hopping amplitudes are thus
\begin{equation}
    J_L^{\mathrm{eff}}
    =
    2\frac{t_Lt_f}{U_{bf}},
    \qquad
    J_R^{\mathrm{eff}}
    =
    2\frac{t_Rt_f}{U_{bf}},
\end{equation}
and satisfy
\begin{equation}
    \frac{J_R^{\mathrm{eff}}}{J_L^{\mathrm{eff}}}
    =
    \frac{t_R}{t_L}.
\end{equation}
Therefore, removing the dimerization changes the detailed band
structure of the effective composite model but leaves its
nonreciprocity unchanged. This result provides an analytical
explanation for the numerical results in Sec.~S1: the drag-induced
NHSE originates from the interaction-mediated inheritance of the
bosonic nonreciprocity and does not require a dimerized lattice.

\section{Criterion for distinguishing the bound and scattering sectors}

In the strong-coupling limit, the bound-state subspace of the
one-boson--one-fermion system on an $L$-site lattice contains $L$
basis states in which the boson and fermion occupy the same site.
We therefore sort the eigenenergies in ascending order,
$E_1\leq E_2\leq\cdots$, and define the spectral gap
\begin{equation}
    \Delta E_{L,L+1}=E_{L+1}-E_L.
\end{equation}
When the bound-state band is well separated from the scattering
continuum, $E_L$ and $E_{L+1}$ correspond to the upper edge of the
bound band and the lower edge of the scattering sector, respectively.
Energy gaps between appropriately indexed eigenstates have similarly
been used to determine whether isolated states remain separated from
continuum bands and to identify the associated spectral transitions~\cite{Liu2023NJP}.

\begin{figure}[htp]
\centering
\includegraphics[width=0.5\linewidth]{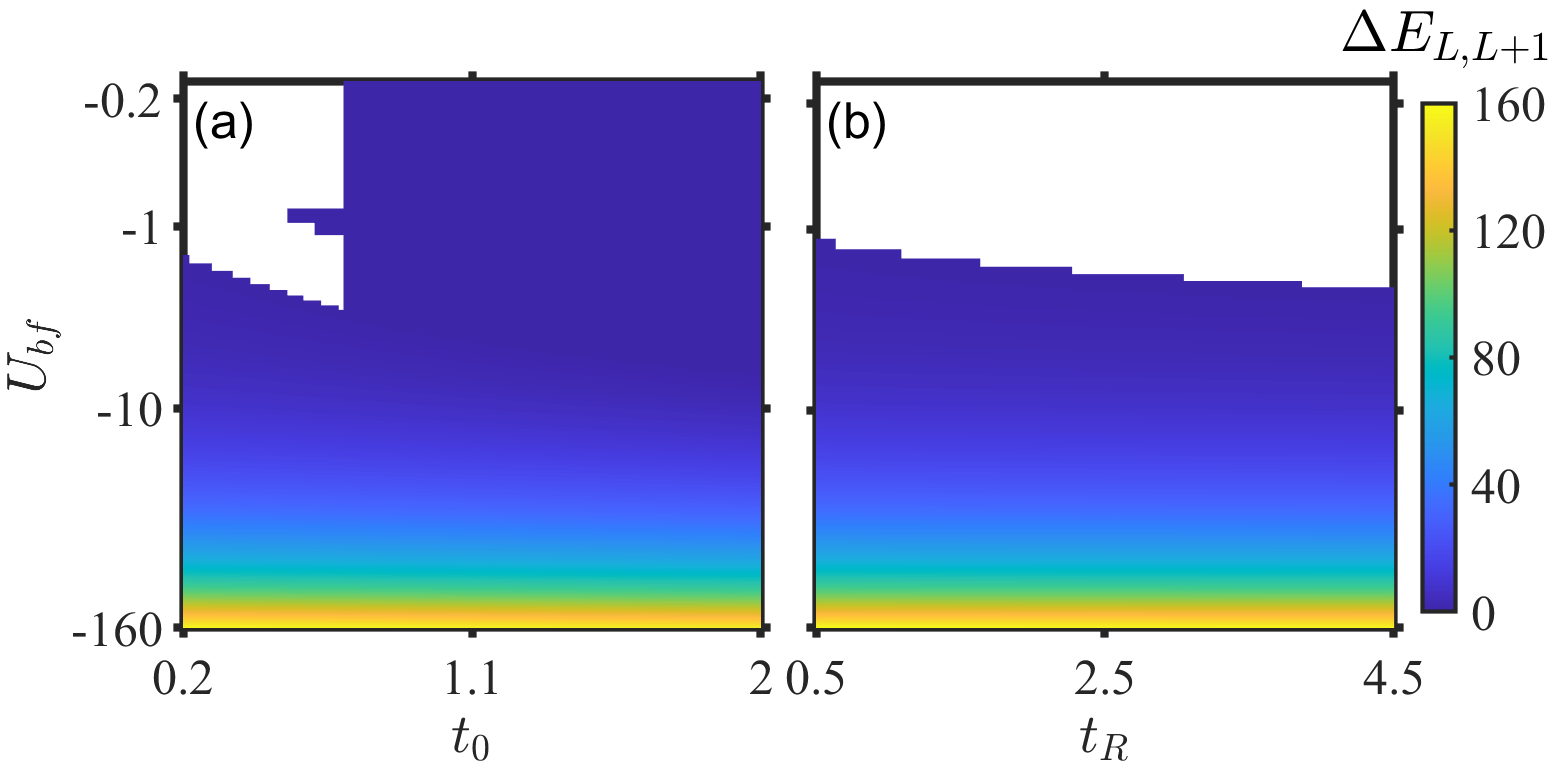}
\caption{
Spectral criterion for distinguishing the bound and scattering
sectors. The energy gap
$\Delta E_{L,L+1}=E_{L+1}-E_L$ is shown as a function of
(a) $(t_0,U_{bf})$ and (b) $(t_R,U_{bf})$, where the eigenenergies are
ordered in ascending order. The lowest $L$ eigenstates constitute the
nominal bound-state sector, while the $(L+1)$th state marks the onset
of the scattering sector. 
Blank regions correspond to
$\Delta E_{L,L+1}\leq0.1$, for which the two sectors are regarded as
insufficiently separated and the bound-band-averaged fermionic
imbalance is not evaluated. For (a), $L=10$, $t_L=0.5$,
$t_R=1.5$, and $t_f=t_0$; for (b), $L=10$, $t_0=0.2$,
$t_L=0.5$, and $t_f=t_0$. The interactions $U_{bb}$ and $U_{ff}$ do
not enter the one-boson--one-fermion calculation.
\label{fig:bandgapcriterion}}
\end{figure}

Figure~\ref{fig:bandgapcriterion} shows $\Delta E_{L,L+1}$ in the same
parameter spaces as Figs.~2(e) and 2(f) of the main text. In our
numerical calculations, we adopt the threshold
$\Delta E_{L,L+1}=0.1$. For
$\Delta E_{L,L+1}>0.1$, the lowest $L$ eigenstates form a
spectrally distinguishable bound-state band, and the fermionic density
imbalance is evaluated by averaging over these states. For
$\Delta E_{L,L+1}\leq0.1$, the spectral separation is too small for an
isolated bound-state band to be reliably identified. The
bound-band-averaged imbalance is therefore not evaluated in this
regime, giving rise to the blank regions in the corresponding
parameter-space plots.

\section{Interaction dependence of the drag-induced skin effect}

\begin{figure}[htp]
\centering
\includegraphics[width=0.5\linewidth]{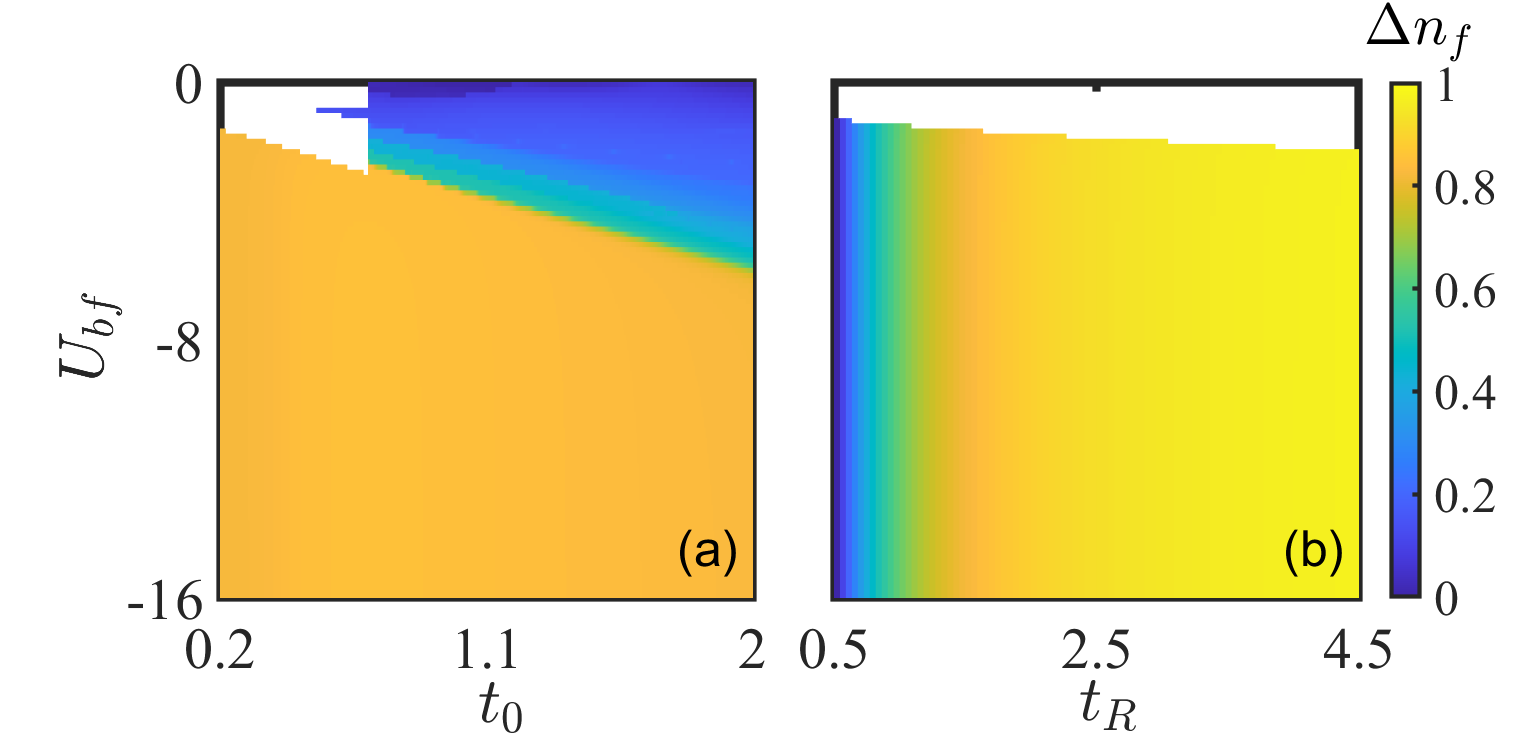}
\caption{
Average fermion density imbalance $\Delta n_f$ as a function of (a) $(t_0,U_{bf})$ and (b) $(t_R,U_{bf})$. 
Panel (a) shows that $\Delta n_f$ increases rapidly with increasing $|U_{bf}|$ and saturates in the strong-coupling regime, while remaining weakly dependent on $t_0$. 
Panel (b) demonstrates a strong enhancement of $\Delta n_f$ with increasing $t_R$, indicating that the nonreciprocal bosonic hopping drives the drag-induced skin effect. 
Blank regions indicate parameter regimes in which the spectral gap $\Delta E_{L,L+1}=E_{L+1}-E_L$ between the nominal bound and scattering sectors is smaller than the numerical threshold $0.1$, so that an isolated bound band and its averaged imbalance cannot be reliably defined.
\label{fig:additionalphase}}
\end{figure}

To further clarify the parameter dependence of the drag-induced skin effect discussed in Figs.~2(e) and (f) of the main text, we examine the fermionic density imbalance $\Delta n_f$ over an extended interaction regime, as shown in Fig.~\ref{fig:additionalphase}. 
Here, $\Delta n_f$ is evaluated using the bound-band eigenstates defined in the main text.

Figure~\ref{fig:additionalphase}(a) shows the dependence of $\Delta n_f$ on the symmetric hopping scale $t_0$ and the Bose--Fermi interaction strength $U_{bf}$. For weak interactions, $\Delta n_f$ remains small, indicating that the fermions largely retain their intrinsic Hermitian character and do not significantly inherit the bosonic skin localization. 
As $|U_{bf}|$ increases, $\Delta n_f$ rapidly grows and eventually saturates, reflecting the formation of tightly bound boson--fermion composites that efficiently transfer the non-Hermitian skin accumulation from bosons to fermions. The relatively weak dependence on $t_0$ further indicates that, once the bound states are established, the drag-induced skin effect is governed primarily by the Bose--Fermi interaction strength and the bosonic nonreciprocity.

Figure~\ref{fig:additionalphase}(b) shows the corresponding dependence on the nonreciprocal hopping amplitude $t_R$ and $U_{bf}$. 
Increasing $t_R$ strongly enhances $\Delta n_f$, demonstrating that the asymmetric bosonic hopping acts as the primary driving mechanism for the drag-induced skin effect. 
For sufficiently large $t_R$ and $|U_{bf}|$, $\Delta n_f$ approaches saturation, indicating robust fermionic boundary accumulation inherited from the bosonic NHSE.

The blank regions in Fig.~\ref{fig:additionalphase} correspond to parameter regimes where the bound and scattering sectors strongly hybridize and no sufficiently isolated bound band exists. In these regimes, the quantity $\Delta n_f$, defined through the bound-band eigenstates, is no longer well defined.
The relatively large values of $U_{bb}$ and $U_{ff}$ adopted in the multi-particle calculations of the main text are chosen only to make the interaction-organized few-body bands more clearly distinguishable. They are not required for the drag-induced skin effect itself, which is already established in the one-boson--one-fermion system where $U_{bb}$ and $U_{ff}$ do not contribute.

\section{Robustness of the drag-induced skin effect against fermionic hopping}

\begin{figure}[htp]
\centering
\includegraphics[width=0.5\linewidth]{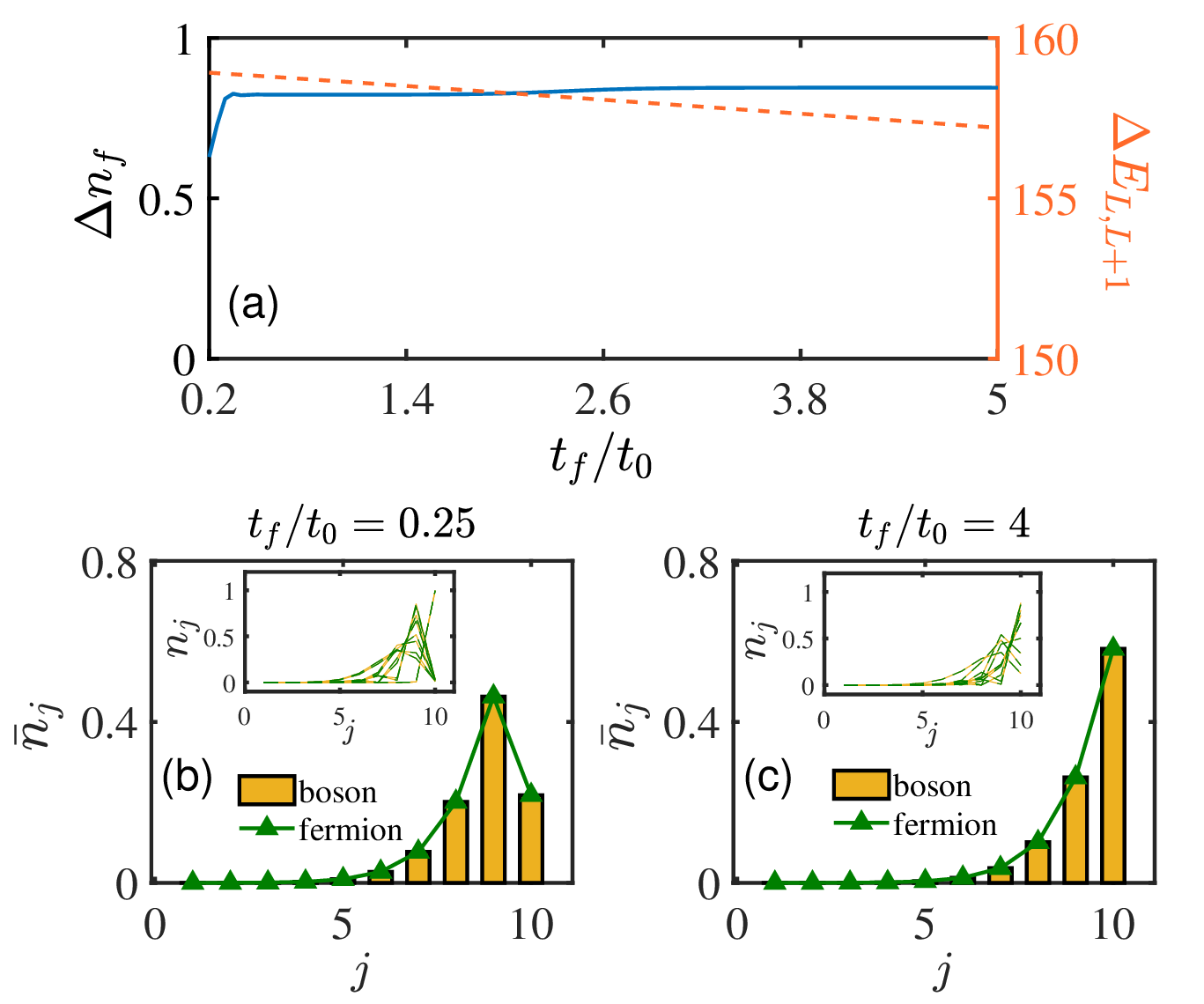}
\caption{
Robustness of the drag-induced skin effect against variations of the
fermionic hopping amplitude $t_f$.
(a) Fermionic density imbalance $\Delta n_f$ within the bound-state band
(blue solid line, left axis) and the spectral gap
$\Delta E_{L,L+1}=E_{L+1}-E_L$
(orange dashed line, right axis), which serves as the criterion for
distinguishing the bound and scattering sectors, as functions of
$t_f/t_0$.
(b,c) Bosonic and fermionic density profiles for
$t_f/t_0=0.25$ and $4$, respectively.
The other parameters are fixed at
$t_0=0.2$, $t_L=0.5$, $t_R=1.5$, and $U_{bf}=-160$.
\label{fig:tfresults}}
\end{figure}

To examine whether the drag-induced skin effect relies on the specific choice
$t_f=t_0$, we independently vary the reciprocal fermionic hopping amplitude
$t_f$ while keeping the other parameters fixed at
$t_0=0.2$, $t_L=0.5$, $t_R=1.5$, and $U_{bf}=-160$.
Figure~\ref{fig:tfresults}(a) shows the fermionic density imbalance $\Delta n_f$
(blue solid line, left axis) within the bound-state band as a function of
$t_f/t_0$. 
Over the broad range $0.2\lesssim t_f/t_0\lesssim5$,
$\Delta n_f$ remains positive and sizable, indicating that the fermionic
component continues to exhibit boundary accumulation even when its hopping
amplitude differs substantially from the bosonic hopping scale. 
For comparison, the spectral gap
$\Delta E_{L,L+1}=E_{L+1}-E_L$
(orange dashed line, right axis), defined in the main text as the criterion
for distinguishing the bound and scattering sectors, is also shown.
Throughout the considered parameter range,
$\Delta E_{L,L+1}$ remains above the numerical threshold
$\Delta E_{L,L+1}=0.1$ adopted in this work,
indicating that the bound-state band can be reliably distinguished
from the scattering sector and then the bound-band-averaged imbalance
$\Delta n_f$ is well defined.
The persistence of both a finite $\Delta E_{L,L+1}$ and a sizable
$\Delta n_f$ demonstrates that the drag-induced skin effect is robust
against variations of the reciprocal fermionic hopping.

Figure~\ref{fig:tfresults}(b) and Figure~\ref{fig:tfresults}(c) further show the bosonic and fermionic density
profiles for $t_f/t_0=0.25$ and $4$, respectively. 
In both cases, the
fermions accumulate toward the same boundary as the bosons despite the
substantial difference between their hopping amplitudes.
These results demonstrate that the choice $t_f=t_0$ is not essential for
the drag-induced skin effect, but merely serves to reduce the number of
independent parameters in the main text.

\section{Eigenstate-resolved density distributions within each energy band}
\label{sec:eigenstate_density}

\begin{figure*}[t]
\centering
\includegraphics[width=\textwidth]{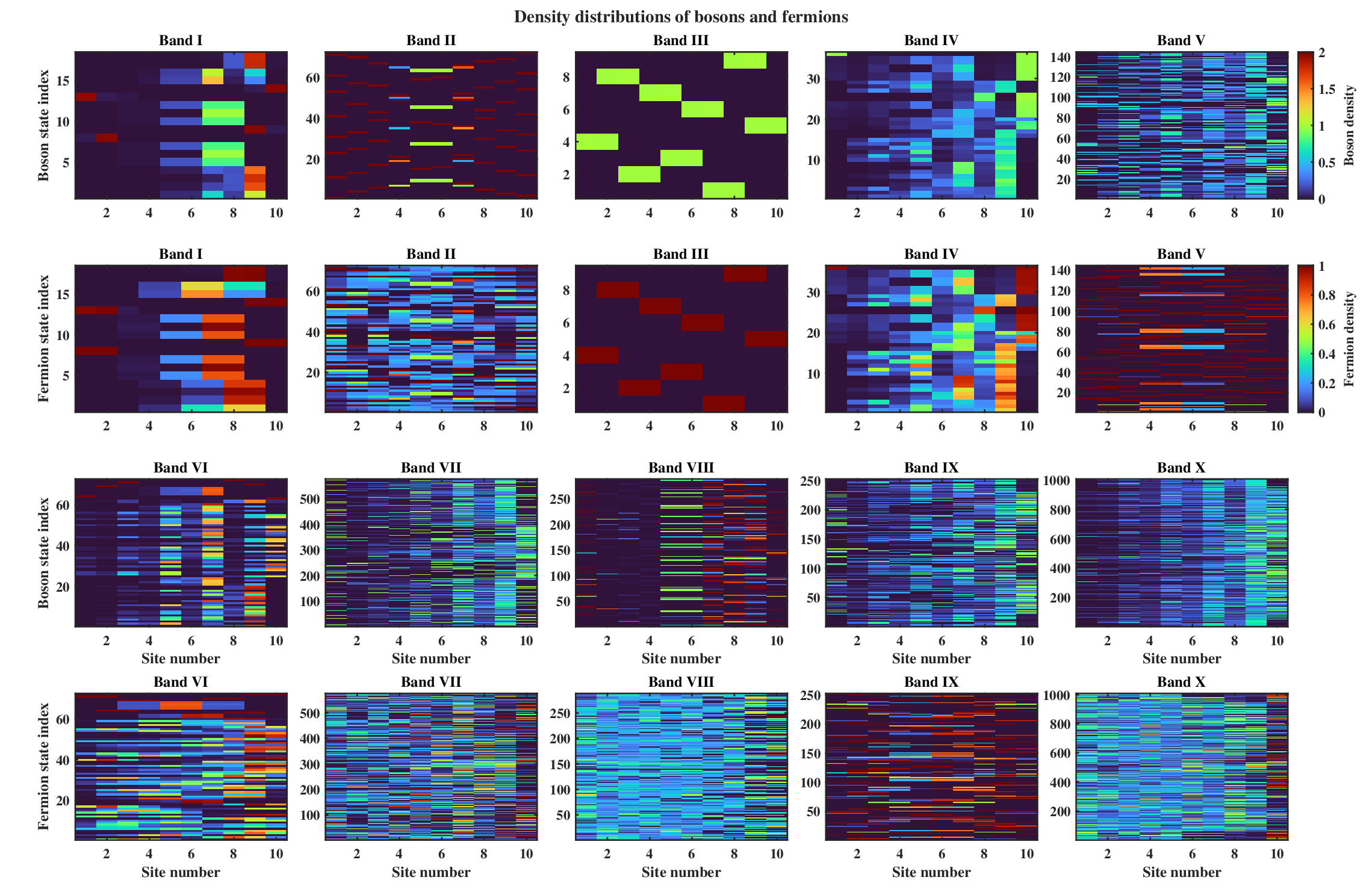}
\caption{
Eigenstate-resolved bosonic and fermionic density distributions for all ten energy bands of the two-boson--two-fermion system. Each horizontal row corresponds to a single many-body eigenstate within the corresponding band, and the horizontal axis denotes the lattice site index. The color scale indicates the local particle density. These density maps provide the microscopic details underlying the band-averaged density distributions shown in Fig.3 of the main text. Parameters are identical to those used in Fig.3.
}
\label{fig:allstates_density}
\end{figure*}

In the main text, Fig.~3 presents the band-averaged bosonic and fermionic density distributions for the two-boson--two-fermion system. To complement that analysis, Fig.~\ref{fig:allstates_density} shows the bosonic and fermionic density distributions of all eigenstates belonging to Bands I--X.

Each horizontal row corresponds to a single many-body eigenstate within the corresponding band, while the horizontal axis denotes the lattice site index. The color scale represents the local particle density. Compared with the band-averaged densities shown in the main text, these eigenstate-resolved density maps provide a more detailed view of the microscopic occupation configurations and state-to-state variations within each band.
The results show that, although the band-averaged densities capture the overall localization characteristics of each band, individual eigenstates can still exhibit distinct density profiles and occupation patterns. These eigenstate-resolved distributions therefore provide additional microscopic information underlying the averaged results presented in Fig.3.

\section{Dynamics in the absence of boson-fermion interaction}

To further clarify the role of the boson-fermion interaction $U_{bf}$ in shaping the dynamical behavior observed in Fig.~4, we present here the spatiotemporal evolution of the system under identical initial conditions but with $U_{bf} = 0$. The remaining parameters are kept consistent with those used in Fig.4: $L = 20$, $t_0 = 0.2$, $t_L = 0.2$, $t_R = 0.5$, $U_{ff} = -120$, and $U_{bb} = -140$.

\begin{figure}[htp]
\center
\includegraphics[width=0.45\textwidth]{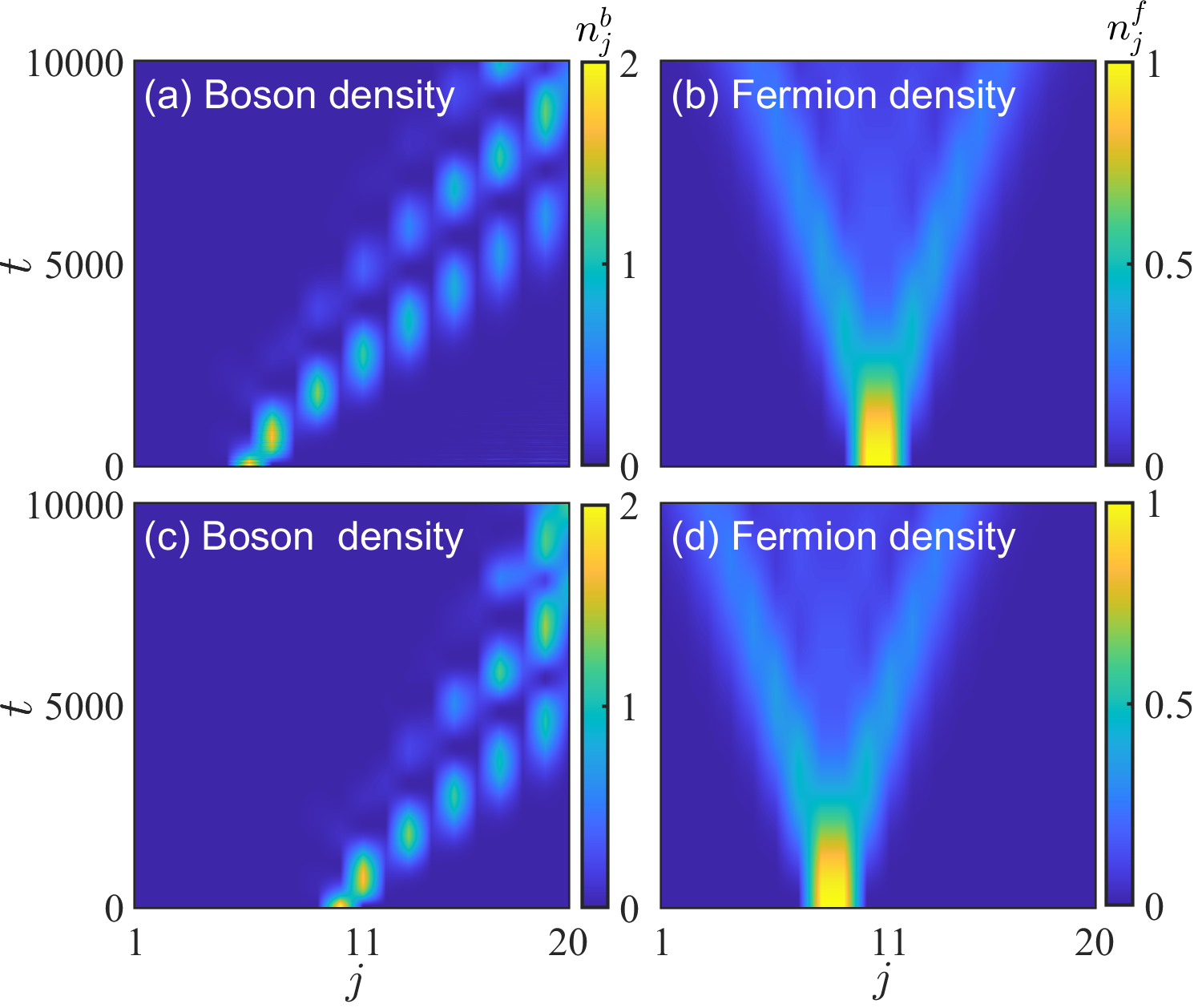}
\caption{Spatiotemporal evolution of boson and fermion densities for $U_{bf} = 0$.
(a),(b) Initial state with bosons to the left of fermions; (c),(d) bosons to the right of fermions.
The bosons (left column) still exhibit rightward skin localization due to $t_R > t_L$, while the fermions (right column) display symmetric light-cone spreading in both cases.
Parameters: $L = 20$, $t_0 = 0.2$, $t_L = 0.2$, $t_R = 0.5$, $U_{ff} = -120$, $U_{bb} = -140$, $U_{bf} = 0$.
\label{fig:noninteracting}}
\end{figure}

As shown in Fig.~\ref{fig:noninteracting}, both configurations-(a,b) with bosons initially to the left of fermions, and (c,d) with bosons to the right-exhibit symmetric light-cone spreading for the fermions [panels (b) and (d)]. This indicates that in the absence of $U_{bf}$, the fermions propagate ballistically without directional bias, regardless of the bosonic configuration.
Meanwhile, the bosons still exhibit a rightward skin effect due to asymmetric tunneling ($t_R > t_L$), as seen in panels (a) and (c). However, since there is no coupling between species, the fermions remain unaffected by the boson dynamics and evolve independently.
This contrast with the case of $U_{bf} = -160$ (Fig.4) unambiguously demonstrates that the asymmetric propagation of fermions is entirely mediated by the strong interspecies interaction. In the absence of $U_{bf}$, no such blockade or directional control emerges.
Therefore, the dynamic blockade effect is not an intrinsic property of the non-Hermitian lattice alone, but rather an emergent phenomenon arising from the interplay between non-Hermitian driving and strong boson-fermion coupling.

\section{Floquet engineering of nonreciprocal tunneling}
\label{sec:floquet}

In static Hermitian systems, microscopic reversibility enforces symmetric tunneling amplitudes ($t_L = t_R$). To realize the asymmetric hopping $t_L \neq t_R$ required for the NHSE within a conservative cold-atom platform, we employ \textit{Floquet engineering}---a technique that uses periodic driving to generate effective Hamiltonians with tailored complex-valued or nonreciprocal couplings.

A concrete implementation utilizes a bichromatic optical lattice subjected to a time-periodic phase modulation. Consider a 1D lattice potential of the form
\begin{equation}
V(x,t) = V_0 \cos^2(kx) + V_1 \cos^2\big(kx + \phi(t)\big),
\end{equation}
where the relative phase is modulated as $\phi(t) = \omega t$, with driving frequency $\omega$. In the high-frequency limit ($\omega \gg t_0, U$), the system is described by an effective time-independent Floquet Hamiltonian obtained via the Magnus expansion~\cite{Bukov2015}. To leading order, the nearest-neighbor tunneling acquires a complex Peierls-like phase:
\begin{equation}
t_{j,j+1}^{\rm eff} = t_0 e^{i \theta}, \quad
t_{j+1,j}^{\rm eff} = t_0 e^{-i \theta},
\end{equation}
with $\theta \propto (V_1/V_0) \sin(\omega t)$ averaged over one period. While this yields Hermitian dynamics with reciprocal magnitude ($|t_{L}| = |t_{R}|$), true nonreciprocity in amplitude can be achieved by combining phase modulation with an additional energy offset (e.g., a linear potential gradient) or by driving multiple internal states. Through the Magnus expansion, Floquet engineering has also been proposed to generate effective many-body interactions~\cite{lee2018floquet}.

The few-body low-filling regime considered here is experimentally accessible in optical lattices using controlled preparation of a small number of bosons and fermions, where the correlated dynamics can be directly probed with single-site-resolved measurements.

\section{Drag-induced skin effect at higher particle numbers}

\begin{figure}[htp]
\center
\includegraphics[width=0.45\textwidth]{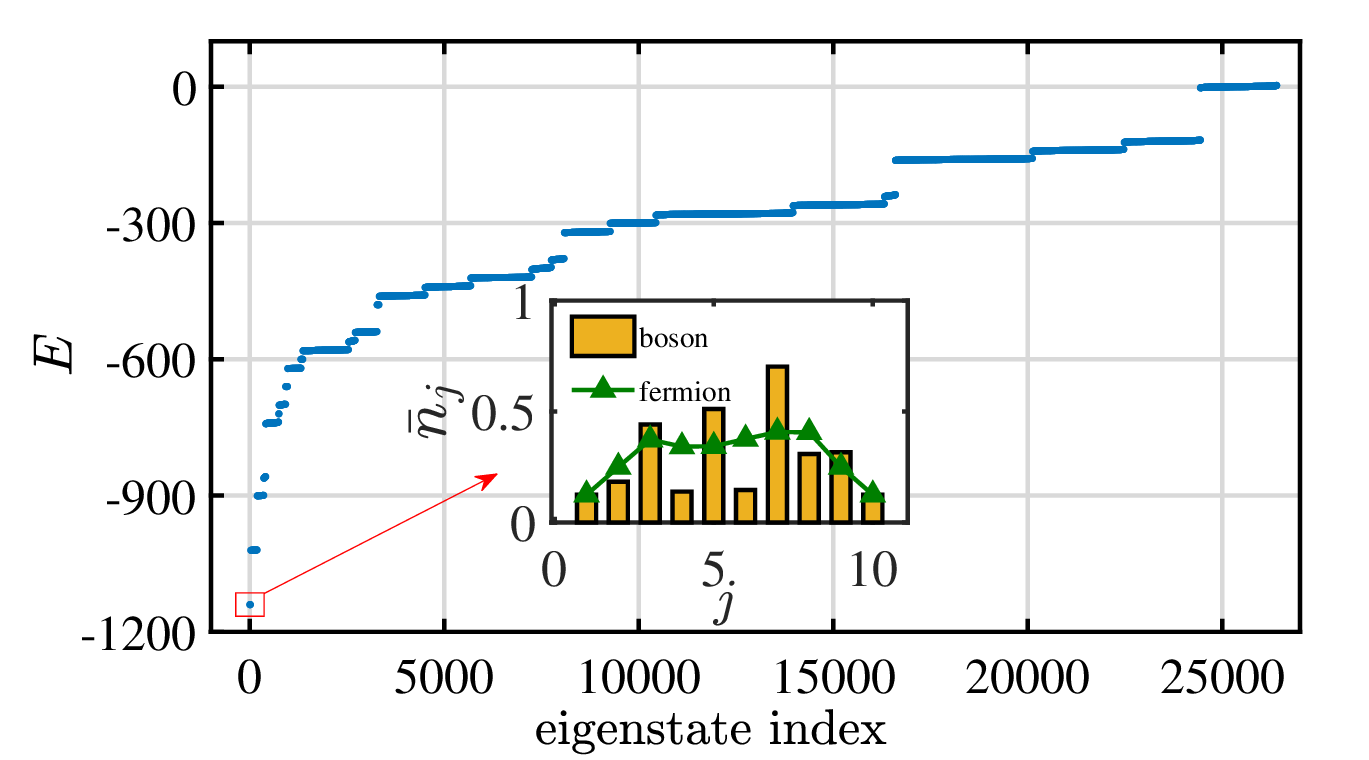}
\caption{Energy spectrum of the system with three bosons and three fermions. The inset shows the average density distribution of bosons (yellow bars) and fermions (green triangles) for the lowest energy band. The lowest-energy configuration consists of three bosons and one fermion occupying the same site, with the remaining two fermions located on the nearest and next-nearest neighboring sites. The parameters are $L=10$, $t_0=0.2$, $t_L=0.5$, $t_R=1.5$, $U_{ff}=-120$, $U_{bb}=-140$, and $U_{bf}=-160$. The relatively weak localization reflects the reduced effective hopping of the composite cluster in a finite system, rather than a suppression of the drag-induced skin effect.
\label{fig:threeparticlecase}}
\end{figure}

To investigate the behavior of the drag-induced skin effect at larger particle numbers, we analyze a system with three bosons and three fermions ($N_b=3, N_f=3$). The calculated energy spectrum and the density distribution of the lowest band are presented in Fig.~\ref{fig:threeparticlecase}. 

In the strong-coupling regime ($|U_{bf}| \gg t_{0}, t_{L,R}$), the lowest-energy states exhibit a clustered structure in which the three bosons and one fermion predominantly occupy the same lattice site, while the remaining two fermions reside on the nearest and next-nearest neighboring sites. This configuration reflects the interplay between strong boson–fermion attraction and fermionic statistics, which prevents all fermions from occupying the same site.
The motion of such a composite object arises from higher-order virtual processes; for instance, the tunneling of the full six-particle cluster corresponds to a sixth-order process in perturbation theory. As a result, the effective hopping amplitude is parametrically reduced compared to the bare single-particle scales. Importantly, however, the nonreciprocal mechanism inherited from the bosonic sector remains present in the effective dynamics.

The relatively weak localization observed in this case should therefore be understood as a finite-size effect. For small systems (e.g., $L=10$), the reduced effective hopping competes with the limited spatial extent, making the boundary accumulation less pronounced. In larger systems, or at lower filling fractions, the directional accumulation is expected to become more visible. These results demonstrate that the drag-induced skin effect is not intrinsically suppressed with increasing particle number, but rather that its observability is governed by the interplay among particle number, system size, filling, and few-body correlations.

\begin{thebibliography}{160}%
\makeatletter
\providecommand \@ifxundefined [1]{%
 \@ifx{#1\undefined}
}%
\providecommand \@ifnum [1]{%
 \ifnum #1\expandafter \@firstoftwo
 \else \expandafter \@secondoftwo
 \fi
}%
\providecommand \@ifx [1]{%
 \ifx #1\expandafter \@firstoftwo
 \else \expandafter \@secondoftwo
 \fi
}%
\providecommand \natexlab [1]{#1}%
\providecommand \enquote  [1]{``#1''}%
\providecommand \bibnamefont  [1]{#1}%
\providecommand \bibfnamefont [1]{#1}%
\providecommand \citenamefont [1]{#1}%
\providecommand \href@noop [0]{\@secondoftwo}%
\providecommand \href [0]{\begingroup \@sanitize@url \@href}%
\providecommand \@href[1]{\@@startlink{#1}\@@href}%
\providecommand \@@href[1]{\endgroup#1\@@endlink}%
\providecommand \@sanitize@url [0]{\catcode `\\12\catcode `\$12\catcode
  `\&12\catcode `\#12\catcode `\^12\catcode `\_12\catcode `\%12\relax}%
\providecommand \@@startlink[1]{}%
\providecommand \@@endlink[0]{}%
\providecommand \url  [0]{\begingroup\@sanitize@url \@url }%
\providecommand \@url [1]{\endgroup\@href {#1}{\urlprefix }}%
\providecommand \urlprefix  [0]{URL }%
\providecommand \Eprint [0]{\href }%
\providecommand \doibase [0]{https://doi.org/}%
\providecommand \selectlanguage [0]{\@gobble}%
\providecommand \bibinfo  [0]{\@secondoftwo}%
\providecommand \bibfield  [0]{\@secondoftwo}%
\providecommand \translation [1]{[#1]}%
\providecommand \BibitemOpen [0]{}%
\providecommand \bibitemStop [0]{}%
\providecommand \bibitemNoStop [0]{.\EOS\space}%
\providecommand \EOS [0]{\spacefactor3000\relax}%
\providecommand \BibitemShut  [1]{\csname bibitem#1\endcsname}%
\let\auto@bib@innerbib\@empty
\bibitem [{\citenamefont {Kawabata}\ \emph {et~al.}(2019)\citenamefont
  {Kawabata}, \citenamefont {Shiozaki}, \citenamefont {Ueda},\ and\
  \citenamefont {Sato}}]{2019PhysRevX9041015}%
  \BibitemOpen
  \bibfield  {author} {\bibinfo {author} {\bibfnamefont {K.}~\bibnamefont
  {Kawabata}}, \bibinfo {author} {\bibfnamefont {K.}~\bibnamefont {Shiozaki}},
  \bibinfo {author} {\bibfnamefont {M.}~\bibnamefont {Ueda}},\ and\ \bibinfo
  {author} {\bibfnamefont {M.}~\bibnamefont {Sato}},\ }\bibfield  {title}
  {\bibinfo {title} {Symmetry and topology in non-hermitian physics},\ }\href
  {https://doi.org/10.1103/PhysRevX.9.041015} {\bibfield  {journal} {\bibinfo
  {journal} {Phys. Rev. X}\ }\textbf {\bibinfo {volume} {9}},\ \bibinfo {pages}
  {041015} (\bibinfo {year} {2019})}\BibitemShut {NoStop}%
\bibitem [{\citenamefont {Ashida}\ \emph {et~al.}(2020)\citenamefont {Ashida},
  \citenamefont {Gong},\ and\ \citenamefont {Ueda}}]{Ashida2020}%
  \BibitemOpen
  \bibfield  {author} {\bibinfo {author} {\bibfnamefont {Y.}~\bibnamefont
  {Ashida}}, \bibinfo {author} {\bibfnamefont {Z.}~\bibnamefont {Gong}},\ and\
  \bibinfo {author} {\bibfnamefont {M.}~\bibnamefont {Ueda}},\ }\bibfield
  {title} {\bibinfo {title} {Non-hermitian physics},\ }\href
  {https://doi.org/10.1080/00018732.2021.1876991} {\bibfield  {journal}
  {\bibinfo  {journal} {Adv. Phys.}\ }\textbf {\bibinfo {volume} {69}},\
  \bibinfo {pages} {249} (\bibinfo {year} {2020})}\BibitemShut {NoStop}%
\bibitem [{\citenamefont {Bergholtz}\ \emph {et~al.}(2021)\citenamefont
  {Bergholtz}, \citenamefont {Budich},\ and\ \citenamefont
  {Kunst}}]{Bergholtz2021RMP}%
  \BibitemOpen
  \bibfield  {author} {\bibinfo {author} {\bibfnamefont {E.~J.}\ \bibnamefont
  {Bergholtz}}, \bibinfo {author} {\bibfnamefont {J.~C.}\ \bibnamefont
  {Budich}},\ and\ \bibinfo {author} {\bibfnamefont {F.~K.}\ \bibnamefont
  {Kunst}},\ }\bibfield  {title} {\bibinfo {title} {Exceptional topology of
  non-hermitian systems},\ }\href
  {https://doi.org/10.1103/RevModPhys.93.015005} {\bibfield  {journal}
  {\bibinfo  {journal} {Rev. Mod. Phys.}\ }\textbf {\bibinfo {volume} {93}},\
  \bibinfo {pages} {015005} (\bibinfo {year} {2021})}\BibitemShut {NoStop}%
\bibitem [{\citenamefont {Zhang}\ \emph
  {et~al.}(2022{\natexlab{a}})\citenamefont {Zhang}, \citenamefont {Zhang},
  \citenamefont {Lu},\ and\ \citenamefont {Chen}}]{Zhang2022NHSE}%
  \BibitemOpen
  \bibfield  {author} {\bibinfo {author} {\bibfnamefont {X.}~\bibnamefont
  {Zhang}}, \bibinfo {author} {\bibfnamefont {T.}~\bibnamefont {Zhang}},
  \bibinfo {author} {\bibfnamefont {M.-H.}\ \bibnamefont {Lu}},\ and\ \bibinfo
  {author} {\bibfnamefont {Y.-F.}\ \bibnamefont {Chen}},\ }\bibfield  {title}
  {\bibinfo {title} {A review on non-hermitian skin effect},\ }\href
  {https://doi.org/10.1080/23746149.2022.2109431} {\bibfield  {journal}
  {\bibinfo  {journal} {Advances in Physics: X}\ }\textbf {\bibinfo {volume}
  {7}},\ \bibinfo {pages} {2109431} (\bibinfo {year}
  {2022}{\natexlab{a}})}\BibitemShut {NoStop}%
\bibitem [{\citenamefont {Ding}\ \emph {et~al.}(2022)\citenamefont {Ding},
  \citenamefont {Fang},\ and\ \citenamefont {Ma}}]{Ding2022NHTop}%
  \BibitemOpen
  \bibfield  {author} {\bibinfo {author} {\bibfnamefont {K.}~\bibnamefont
  {Ding}}, \bibinfo {author} {\bibfnamefont {C.}~\bibnamefont {Fang}},\ and\
  \bibinfo {author} {\bibfnamefont {G.}~\bibnamefont {Ma}},\ }\bibfield
  {title} {\bibinfo {title} {Non-hermitian topology and exceptional-point
  geometries},\ }\href {https://doi.org/10.1038/s42254-022-00516-5} {\bibfield
  {journal} {\bibinfo  {journal} {Nature Reviews Physics}\ }\textbf {\bibinfo
  {volume} {4}},\ \bibinfo {pages} {745} (\bibinfo {year} {2022})}\BibitemShut
  {NoStop}%
\bibitem [{\citenamefont {Lin}\ \emph {et~al.}(2023)\citenamefont {Lin},
  \citenamefont {Tai}, \citenamefont {Li},\ and\ \citenamefont
  {Lee}}]{Lin2023TopNHSE}%
  \BibitemOpen
  \bibfield  {author} {\bibinfo {author} {\bibfnamefont {R.}~\bibnamefont
  {Lin}}, \bibinfo {author} {\bibfnamefont {T.}~\bibnamefont {Tai}}, \bibinfo
  {author} {\bibfnamefont {L.}~\bibnamefont {Li}},\ and\ \bibinfo {author}
  {\bibfnamefont {C.~H.}\ \bibnamefont {Lee}},\ }\bibfield  {title} {\bibinfo
  {title} {Topological non-hermitian skin effect},\ }\href
  {https://doi.org/10.1007/s11467-023-1309-z} {\bibfield  {journal} {\bibinfo
  {journal} {Frontiers of Physics}\ }\textbf {\bibinfo {volume} {18}},\
  \bibinfo {pages} {53605} (\bibinfo {year} {2023})}\BibitemShut {NoStop}%
\bibitem [{\citenamefont {Okuma}\ and\ \citenamefont
  {Sato}(2023)}]{Okuma2022NHTopReview}%
  \BibitemOpen
  \bibfield  {author} {\bibinfo {author} {\bibfnamefont {N.}~\bibnamefont
  {Okuma}}\ and\ \bibinfo {author} {\bibfnamefont {M.}~\bibnamefont {Sato}},\
  }\bibfield  {title} {\bibinfo {title} {Non-hermitian topological phenomena: A
  review},\ }\href {https://doi.org/10.1146/annurev-conmatphys-040521-033133}
  {\bibfield  {journal} {\bibinfo  {journal} {Annual Review of Condensed Matter
  Physics}\ }\textbf {\bibinfo {volume} {14}},\ \bibinfo {pages} {83} (\bibinfo
  {year} {2023})}\BibitemShut {NoStop}%
\bibitem [{\citenamefont {Lei}\ and\ \citenamefont {Li}(2026)}]{lei2026inter}%
  \BibitemOpen
  \bibfield  {author} {\bibinfo {author} {\bibfnamefont {Z.}~\bibnamefont
  {Lei}}\ and\ \bibinfo {author} {\bibfnamefont {L.}~\bibnamefont {Li}},\
  }\bibfield  {title} {\bibinfo {title} {Inter-species topological phases via a
  dynamical gauge field},\ }\href@noop {} {\bibfield  {journal} {\bibinfo
  {journal} {Science China Physics, Mechanics \& Astronomy}\ }\textbf {\bibinfo
  {volume} {69}},\ \bibinfo {pages} {257811} (\bibinfo {year}
  {2026})}\BibitemShut {NoStop}%
\bibitem [{\citenamefont {Meng}\ \emph {et~al.}(2025)\citenamefont {Meng},
  \citenamefont {Ang},\ and\ \citenamefont {Lee}}]{meng2025gen}%
  \BibitemOpen
  \bibfield  {author} {\bibinfo {author} {\bibfnamefont {H.}~\bibnamefont
  {Meng}}, \bibinfo {author} {\bibfnamefont {Y.~S.}\ \bibnamefont {Ang}},\ and\
  \bibinfo {author} {\bibfnamefont {C.~H.}\ \bibnamefont {Lee}},\ }\bibfield
  {title} {\bibinfo {title} {Generalized brillouin zone fragmentation},\
  }\href@noop {} {\bibfield  {journal} {\bibinfo  {journal} {arXiv preprint
  arXiv:2508.13275}\ } (\bibinfo {year} {2025})}\BibitemShut {NoStop}%
\bibitem [{\citenamefont {Lee}(2022)}]{lee2022exceptional}%
  \BibitemOpen
  \bibfield  {author} {\bibinfo {author} {\bibfnamefont {C.~H.}\ \bibnamefont
  {Lee}},\ }\bibfield  {title} {\bibinfo {title} {Exceptional bound states and
  negative entanglement entropy},\ }\href@noop {} {\bibfield  {journal}
  {\bibinfo  {journal} {Physical Review Letters}\ }\textbf {\bibinfo {volume}
  {128}},\ \bibinfo {pages} {010402} (\bibinfo {year} {2022})}\BibitemShut
  {NoStop}%
\bibitem [{\citenamefont {Liang}\ \emph {et~al.}(2025)\citenamefont {Liang},
  \citenamefont {Li},\ and\ \citenamefont {Xu}}]{liang2025size}%
  \BibitemOpen
  \bibfield  {author} {\bibinfo {author} {\bibfnamefont {H.-Q.}\ \bibnamefont
  {Liang}}, \bibinfo {author} {\bibfnamefont {L.}~\bibnamefont {Li}},\ and\
  \bibinfo {author} {\bibfnamefont {G.-F.}\ \bibnamefont {Xu}},\ }\bibfield
  {title} {\bibinfo {title} {Size-dependent critical localization},\
  }\href@noop {} {\bibfield  {journal} {\bibinfo  {journal} {arXiv preprint
  arXiv:2509.18943}\ } (\bibinfo {year} {2025})}\BibitemShut {NoStop}%
\bibitem [{\citenamefont {Huang}\ \emph {et~al.}(2026)\citenamefont {Huang},
  \citenamefont {Hu},\ and\ \citenamefont {Yang}}]{huang2026complex}%
  \BibitemOpen
  \bibfield  {author} {\bibinfo {author} {\bibfnamefont {J.}~\bibnamefont
  {Huang}}, \bibinfo {author} {\bibfnamefont {J.}~\bibnamefont {Hu}},\ and\
  \bibinfo {author} {\bibfnamefont {Z.}~\bibnamefont {Yang}},\ }\bibfield
  {title} {\bibinfo {title} {Complex frequency detection in a subsystem},\
  }\href@noop {} {\bibfield  {journal} {\bibinfo  {journal} {Communications
  Physics}\ }\textbf {\bibinfo {volume} {9}},\ \bibinfo {pages} {84} (\bibinfo
  {year} {2026})}\BibitemShut {NoStop}%
\bibitem [{\citenamefont {Shen}\ \emph {et~al.}(2025)\citenamefont {Shen},
  \citenamefont {Chen}, \citenamefont {Yang},\ and\ \citenamefont
  {Lee}}]{shen2025observation}%
  \BibitemOpen
  \bibfield  {author} {\bibinfo {author} {\bibfnamefont {R.}~\bibnamefont
  {Shen}}, \bibinfo {author} {\bibfnamefont {T.}~\bibnamefont {Chen}}, \bibinfo
  {author} {\bibfnamefont {B.}~\bibnamefont {Yang}},\ and\ \bibinfo {author}
  {\bibfnamefont {C.~H.}\ \bibnamefont {Lee}},\ }\bibfield  {title} {\bibinfo
  {title} {Observation of the non-hermitian skin effect and fermi skin on a
  digital quantum computer},\ }\href@noop {} {\bibfield  {journal} {\bibinfo
  {journal} {Nature Communications}\ }\textbf {\bibinfo {volume} {16}},\
  \bibinfo {pages} {1340} (\bibinfo {year} {2025})}\BibitemShut {NoStop}%
\bibitem [{\citenamefont {Long}\ \emph {et~al.}(2026)\citenamefont {Long},
  \citenamefont {Yang}, \citenamefont {Mu},\ and\ \citenamefont
  {Li}}]{long2026sym}%
  \BibitemOpen
  \bibfield  {author} {\bibinfo {author} {\bibfnamefont {S.}~\bibnamefont
  {Long}}, \bibinfo {author} {\bibfnamefont {C.}~\bibnamefont {Yang}}, \bibinfo
  {author} {\bibfnamefont {S.}~\bibnamefont {Mu}},\ and\ \bibinfo {author}
  {\bibfnamefont {L.}~\bibnamefont {Li}},\ }\bibfield  {title} {\bibinfo
  {title} {Symmetry-protected control of liouvillian topological phases via
  hamiltonian band topology},\ }\href@noop {} {\bibfield  {journal} {\bibinfo
  {journal} {arXiv preprint arXiv:2602.22323}\ } (\bibinfo {year}
  {2026})}\BibitemShut {NoStop}%
\bibitem [{\citenamefont {Shen}\ and\ \citenamefont
  {Lee}(2026)}]{shen2026observation}%
  \BibitemOpen
  \bibfield  {author} {\bibinfo {author} {\bibfnamefont {R.}~\bibnamefont
  {Shen}}\ and\ \bibinfo {author} {\bibfnamefont {C.~H.}\ \bibnamefont {Lee}},\
  }\bibfield  {title} {\bibinfo {title} {Observation of feedback-directed
  quantum dynamics in large-scale quantum processors},\ }\href@noop {}
  {\bibfield  {journal} {\bibinfo  {journal} {arXiv preprint arXiv:2604.11900}\
  } (\bibinfo {year} {2026})}\BibitemShut {NoStop}%
\bibitem [{\citenamefont {Yao}\ and\ \citenamefont {Wang}(2018)}]{Yao2018edge}%
  \BibitemOpen
  \bibfield  {author} {\bibinfo {author} {\bibfnamefont {S.}~\bibnamefont
  {Yao}}\ and\ \bibinfo {author} {\bibfnamefont {Z.}~\bibnamefont {Wang}},\
  }\bibfield  {title} {\bibinfo {title} {Edge states and topological invariants
  of non-hermitian systems},\ }\href
  {https://doi.org/10.1103/PhysRevLett.121.086803} {\bibfield  {journal}
  {\bibinfo  {journal} {Phys. Rev. Lett.}\ }\textbf {\bibinfo {volume} {121}},\
  \bibinfo {pages} {086803} (\bibinfo {year} {2018})}\BibitemShut {NoStop}%
\bibitem [{\citenamefont {Kunst}\ \emph {et~al.}(2018)\citenamefont {Kunst},
  \citenamefont {Edvardsson}, \citenamefont {Budich},\ and\ \citenamefont
  {Bergholtz}}]{Kunst2018biorthogonal}%
  \BibitemOpen
  \bibfield  {author} {\bibinfo {author} {\bibfnamefont {F.~K.}\ \bibnamefont
  {Kunst}}, \bibinfo {author} {\bibfnamefont {E.}~\bibnamefont {Edvardsson}},
  \bibinfo {author} {\bibfnamefont {J.~C.}\ \bibnamefont {Budich}},\ and\
  \bibinfo {author} {\bibfnamefont {E.~J.}\ \bibnamefont {Bergholtz}},\
  }\bibfield  {title} {\bibinfo {title} {Biorthogonal bulk-boundary
  correspondence in non-hermitian systems},\ }\href
  {https://doi.org/10.1103/PhysRevLett.121.026808} {\bibfield  {journal}
  {\bibinfo  {journal} {Phys. Rev. Lett.}\ }\textbf {\bibinfo {volume} {121}},\
  \bibinfo {pages} {026808} (\bibinfo {year} {2018})}\BibitemShut {NoStop}%
\bibitem [{\citenamefont {Lee}\ and\ \citenamefont
  {Thomale}(2019)}]{Lee2019anatomy}%
  \BibitemOpen
  \bibfield  {author} {\bibinfo {author} {\bibfnamefont {C.~H.}\ \bibnamefont
  {Lee}}\ and\ \bibinfo {author} {\bibfnamefont {R.}~\bibnamefont {Thomale}},\
  }\bibfield  {title} {\bibinfo {title} {Anatomy of skin modes and topology in
  non-hermitian systems},\ }\href {https://doi.org/10.1103/PhysRevB.99.201103}
  {\bibfield  {journal} {\bibinfo  {journal} {Phys. Rev. B}\ }\textbf {\bibinfo
  {volume} {99}},\ \bibinfo {pages} {201103(R)} (\bibinfo {year}
  {2019})}\BibitemShut {NoStop}%
\bibitem [{\citenamefont {Song}\ \emph {et~al.}(2019)\citenamefont {Song},
  \citenamefont {Yao},\ and\ \citenamefont {Wang}}]{Song2019chiral}%
  \BibitemOpen
  \bibfield  {author} {\bibinfo {author} {\bibfnamefont {F.}~\bibnamefont
  {Song}}, \bibinfo {author} {\bibfnamefont {S.}~\bibnamefont {Yao}},\ and\
  \bibinfo {author} {\bibfnamefont {Z.}~\bibnamefont {Wang}},\ }\bibfield
  {title} {\bibinfo {title} {Non-hermitian skin effect and chiral damping in
  open quantum systems},\ }\href
  {https://doi.org/10.1103/PhysRevLett.123.170401} {\bibfield  {journal}
  {\bibinfo  {journal} {Phys. Rev. Lett.}\ }\textbf {\bibinfo {volume} {123}},\
  \bibinfo {pages} {170401} (\bibinfo {year} {2019})}\BibitemShut {NoStop}%
\bibitem [{\citenamefont {Ghatak}\ \emph {et~al.}(2020)\citenamefont {Ghatak},
  \citenamefont {Brandenbourger}, \citenamefont {van Wezel},\ and\
  \citenamefont {Coulais}}]{Ghatak2020PNAS}%
  \BibitemOpen
  \bibfield  {author} {\bibinfo {author} {\bibfnamefont {A.}~\bibnamefont
  {Ghatak}}, \bibinfo {author} {\bibfnamefont {M.}~\bibnamefont
  {Brandenbourger}}, \bibinfo {author} {\bibfnamefont {J.}~\bibnamefont {van
  Wezel}},\ and\ \bibinfo {author} {\bibfnamefont {C.}~\bibnamefont
  {Coulais}},\ }\bibfield  {title} {\bibinfo {title} {Observation of
  non-hermitian topology and its bulk-edge correspondence in an active
  mechanical metamaterial},\ }\href {https://doi.org/10.1073/pnas.2010580117}
  {\bibfield  {journal} {\bibinfo  {journal} {Proc. Natl. Acad. Sci. U.S.A.}\
  }\textbf {\bibinfo {volume} {117}},\ \bibinfo {pages} {29561} (\bibinfo
  {year} {2020})}\BibitemShut {NoStop}%
\bibitem [{\citenamefont {Xiao}\ \emph {et~al.}(2020)\citenamefont {Xiao},
  \citenamefont {Deng}, \citenamefont {Wang}, \citenamefont {Zhu},
  \citenamefont {Wang}, \citenamefont {Yi},\ and\ \citenamefont
  {Xue}}]{Xiao2020NatPhys}%
  \BibitemOpen
  \bibfield  {author} {\bibinfo {author} {\bibfnamefont {L.}~\bibnamefont
  {Xiao}}, \bibinfo {author} {\bibfnamefont {T.}~\bibnamefont {Deng}}, \bibinfo
  {author} {\bibfnamefont {K.}~\bibnamefont {Wang}}, \bibinfo {author}
  {\bibfnamefont {G.}~\bibnamefont {Zhu}}, \bibinfo {author} {\bibfnamefont
  {Z.}~\bibnamefont {Wang}}, \bibinfo {author} {\bibfnamefont {W.}~\bibnamefont
  {Yi}},\ and\ \bibinfo {author} {\bibfnamefont {P.}~\bibnamefont {Xue}},\
  }\bibfield  {title} {\bibinfo {title} {Non-hermitian bulk--boundary
  correspondence in quantum dynamics},\ }\href
  {https://doi.org/10.1038/s41567-020-0836-6} {\bibfield  {journal} {\bibinfo
  {journal} {Nature Physics}\ }\textbf {\bibinfo {volume} {16}},\ \bibinfo
  {pages} {761} (\bibinfo {year} {2020})}\BibitemShut {NoStop}%
\bibitem [{\citenamefont {Helbig}\ \emph {et~al.}(2020)\citenamefont {Helbig},
  \citenamefont {Hofmann}, \citenamefont {Imhof}, \citenamefont {Abdelghany},
  \citenamefont {Kiessling}, \citenamefont {Molenkamp}, \citenamefont {Lee},
  \citenamefont {Szameit}, \citenamefont {Greiter},\ and\ \citenamefont
  {Thomale}}]{Helbig2020NatPhys}%
  \BibitemOpen
  \bibfield  {author} {\bibinfo {author} {\bibfnamefont {T.}~\bibnamefont
  {Helbig}}, \bibinfo {author} {\bibfnamefont {T.}~\bibnamefont {Hofmann}},
  \bibinfo {author} {\bibfnamefont {S.}~\bibnamefont {Imhof}}, \bibinfo
  {author} {\bibfnamefont {M.}~\bibnamefont {Abdelghany}}, \bibinfo {author}
  {\bibfnamefont {T.}~\bibnamefont {Kiessling}}, \bibinfo {author}
  {\bibfnamefont {L.~W.}\ \bibnamefont {Molenkamp}}, \bibinfo {author}
  {\bibfnamefont {C.~H.}\ \bibnamefont {Lee}}, \bibinfo {author} {\bibfnamefont
  {A.}~\bibnamefont {Szameit}}, \bibinfo {author} {\bibfnamefont
  {M.}~\bibnamefont {Greiter}},\ and\ \bibinfo {author} {\bibfnamefont
  {R.}~\bibnamefont {Thomale}},\ }\bibfield  {title} {\bibinfo {title}
  {Generalized bulk--boundary correspondence in non-hermitian topolectrical
  circuits},\ }\href {https://doi.org/10.1038/s41567-020-0922-9} {\bibfield
  {journal} {\bibinfo  {journal} {Nature Physics}\ }\textbf {\bibinfo {volume}
  {16}},\ \bibinfo {pages} {747} (\bibinfo {year} {2020})}\BibitemShut
  {NoStop}%
\bibitem [{\citenamefont {Weidemann}\ \emph {et~al.}(2020)\citenamefont
  {Weidemann}, \citenamefont {Kremer}, \citenamefont {Helbig}, \citenamefont
  {Hofmann}, \citenamefont {Stegmaier}, \citenamefont {Greiter}, \citenamefont
  {Thomale},\ and\ \citenamefont {Szameit}}]{Weidemann2020Science}%
  \BibitemOpen
  \bibfield  {author} {\bibinfo {author} {\bibfnamefont {S.}~\bibnamefont
  {Weidemann}}, \bibinfo {author} {\bibfnamefont {M.}~\bibnamefont {Kremer}},
  \bibinfo {author} {\bibfnamefont {T.}~\bibnamefont {Helbig}}, \bibinfo
  {author} {\bibfnamefont {T.}~\bibnamefont {Hofmann}}, \bibinfo {author}
  {\bibfnamefont {A.}~\bibnamefont {Stegmaier}}, \bibinfo {author}
  {\bibfnamefont {M.}~\bibnamefont {Greiter}}, \bibinfo {author} {\bibfnamefont
  {R.}~\bibnamefont {Thomale}},\ and\ \bibinfo {author} {\bibfnamefont
  {A.}~\bibnamefont {Szameit}},\ }\bibfield  {title} {\bibinfo {title}
  {Topological funneling of light},\ }\href
  {https://doi.org/10.1126/science.aaz8727} {\bibfield  {journal} {\bibinfo
  {journal} {Science}\ }\textbf {\bibinfo {volume} {368}},\ \bibinfo {pages}
  {311} (\bibinfo {year} {2020})}\BibitemShut {NoStop}%
\bibitem [{\citenamefont {Zhang}\ \emph {et~al.}(2021)\citenamefont {Zhang},
  \citenamefont {Li}, \citenamefont {Liu}, \citenamefont {Tai}, \citenamefont
  {Thomale},\ and\ \citenamefont {Lee}}]{zhang2021tidal}%
  \BibitemOpen
  \bibfield  {author} {\bibinfo {author} {\bibfnamefont {X.}~\bibnamefont
  {Zhang}}, \bibinfo {author} {\bibfnamefont {G.}~\bibnamefont {Li}}, \bibinfo
  {author} {\bibfnamefont {Y.}~\bibnamefont {Liu}}, \bibinfo {author}
  {\bibfnamefont {T.}~\bibnamefont {Tai}}, \bibinfo {author} {\bibfnamefont
  {R.}~\bibnamefont {Thomale}},\ and\ \bibinfo {author} {\bibfnamefont {C.~H.}\
  \bibnamefont {Lee}},\ }\bibfield  {title} {\bibinfo {title} {Tidal surface
  states as fingerprints of non-hermitian nodal knot metals},\ }\href@noop {}
  {\bibfield  {journal} {\bibinfo  {journal} {Communications Physics}\ }\textbf
  {\bibinfo {volume} {4}},\ \bibinfo {pages} {47} (\bibinfo {year}
  {2021})}\BibitemShut {NoStop}%
\bibitem [{\citenamefont {Liang}\ \emph {et~al.}(2022)\citenamefont {Liang},
  \citenamefont {Xie}, \citenamefont {Dong}, \citenamefont {Li}, \citenamefont
  {Li}, \citenamefont {Gadway}, \citenamefont {Yi},\ and\ \citenamefont
  {Yan}}]{2022PhysRevLett.129.070401}%
  \BibitemOpen
  \bibfield  {author} {\bibinfo {author} {\bibfnamefont {Q.}~\bibnamefont
  {Liang}}, \bibinfo {author} {\bibfnamefont {D.}~\bibnamefont {Xie}}, \bibinfo
  {author} {\bibfnamefont {Z.}~\bibnamefont {Dong}}, \bibinfo {author}
  {\bibfnamefont {H.}~\bibnamefont {Li}}, \bibinfo {author} {\bibfnamefont
  {H.}~\bibnamefont {Li}}, \bibinfo {author} {\bibfnamefont {B.}~\bibnamefont
  {Gadway}}, \bibinfo {author} {\bibfnamefont {W.}~\bibnamefont {Yi}},\ and\
  \bibinfo {author} {\bibfnamefont {B.}~\bibnamefont {Yan}},\ }\bibfield
  {title} {\bibinfo {title} {Dynamic signatures of non-hermitian skin effect
  and topology in ultracold atoms},\ }\href
  {https://doi.org/10.1103/PhysRevLett.129.070401} {\bibfield  {journal}
  {\bibinfo  {journal} {Phys. Rev. Lett.}\ }\textbf {\bibinfo {volume} {129}},\
  \bibinfo {pages} {070401} (\bibinfo {year} {2022})}\BibitemShut {NoStop}%
\bibitem [{\citenamefont {Rafi-Ul-Islam}\ \emph {et~al.}(2022)\citenamefont
  {Rafi-Ul-Islam}, \citenamefont {Siu}, \citenamefont {Sahin}, \citenamefont
  {Lee},\ and\ \citenamefont {Jalil}}]{rafi2022unconventional}%
  \BibitemOpen
  \bibfield  {author} {\bibinfo {author} {\bibfnamefont {S.}~\bibnamefont
  {Rafi-Ul-Islam}}, \bibinfo {author} {\bibfnamefont {Z.~B.}\ \bibnamefont
  {Siu}}, \bibinfo {author} {\bibfnamefont {H.}~\bibnamefont {Sahin}}, \bibinfo
  {author} {\bibfnamefont {C.~H.}\ \bibnamefont {Lee}},\ and\ \bibinfo {author}
  {\bibfnamefont {M.~B.}\ \bibnamefont {Jalil}},\ }\bibfield  {title} {\bibinfo
  {title} {Unconventional skin modes in generalized topolectrical circuits with
  multiple asymmetric couplings},\ }\href@noop {} {\bibfield  {journal}
  {\bibinfo  {journal} {Physical Review Research}\ }\textbf {\bibinfo {volume}
  {4}},\ \bibinfo {pages} {043108} (\bibinfo {year} {2022})}\BibitemShut
  {NoStop}%
\bibitem [{\citenamefont {Li}\ and\ \citenamefont {Lee}(2022)}]{li2022non}%
  \BibitemOpen
  \bibfield  {author} {\bibinfo {author} {\bibfnamefont {L.}~\bibnamefont
  {Li}}\ and\ \bibinfo {author} {\bibfnamefont {C.~H.}\ \bibnamefont {Lee}},\
  }\bibfield  {title} {\bibinfo {title} {Non-hermitian pseudo-gaps},\
  }\href@noop {} {\bibfield  {journal} {\bibinfo  {journal} {Science Bulletin}\
  }\textbf {\bibinfo {volume} {67}},\ \bibinfo {pages} {685} (\bibinfo {year}
  {2022})}\BibitemShut {NoStop}%
\bibitem [{\citenamefont {Zhang}\ \emph {et~al.}(2023)\citenamefont {Zhang},
  \citenamefont {Fang},\ and\ \citenamefont {Yang}}]{ZhangPRL2023}%
  \BibitemOpen
  \bibfield  {author} {\bibinfo {author} {\bibfnamefont {K.}~\bibnamefont
  {Zhang}}, \bibinfo {author} {\bibfnamefont {C.}~\bibnamefont {Fang}},\ and\
  \bibinfo {author} {\bibfnamefont {Z.}~\bibnamefont {Yang}},\ }\bibfield
  {title} {\bibinfo {title} {Dynamical degeneracy splitting and directional
  invisibility in non-hermitian systems},\ }\href
  {https://doi.org/10.1103/PhysRevLett.131.036402} {\bibfield  {journal}
  {\bibinfo  {journal} {Phys. Rev. Lett.}\ }\textbf {\bibinfo {volume} {131}},\
  \bibinfo {pages} {036402} (\bibinfo {year} {2023})}\BibitemShut {NoStop}%
\bibitem [{\citenamefont {Wang}\ \emph {et~al.}(2023)\citenamefont {Wang},
  \citenamefont {Li}, \citenamefont {Song},\ and\ \citenamefont
  {Wang}}]{SCPHe2023}%
  \BibitemOpen
  \bibfield  {author} {\bibinfo {author} {\bibfnamefont {H.-R.}\ \bibnamefont
  {Wang}}, \bibinfo {author} {\bibfnamefont {B.}~\bibnamefont {Li}}, \bibinfo
  {author} {\bibfnamefont {F.}~\bibnamefont {Song}},\ and\ \bibinfo {author}
  {\bibfnamefont {Z.}~\bibnamefont {Wang}},\ }\bibfield  {title} {\bibinfo
  {title} {{Scale-free non-Hermitian skin effect in a boundary-dissipated spin
  chain}},\ }\href {https://doi.org/10.21468/SciPostPhys.15.5.191} {\bibfield
  {journal} {\bibinfo  {journal} {SciPost Phys.}\ }\textbf {\bibinfo {volume}
  {15}},\ \bibinfo {pages} {191} (\bibinfo {year} {2023})}\BibitemShut
  {NoStop}%
\bibitem [{\citenamefont {Jiang}\ and\ \citenamefont
  {Lee}(2023)}]{jiang2023dimensional}%
  \BibitemOpen
  \bibfield  {author} {\bibinfo {author} {\bibfnamefont {H.}~\bibnamefont
  {Jiang}}\ and\ \bibinfo {author} {\bibfnamefont {C.~H.}\ \bibnamefont
  {Lee}},\ }\bibfield  {title} {\bibinfo {title} {Dimensional transmutation
  from non-hermiticity},\ }\href@noop {} {\bibfield  {journal} {\bibinfo
  {journal} {Physical Review Letters}\ }\textbf {\bibinfo {volume} {131}},\
  \bibinfo {pages} {076401} (\bibinfo {year} {2023})}\BibitemShut {NoStop}%
\bibitem [{\citenamefont {Xiao}\ \emph {et~al.}(2024)\citenamefont {Xiao},
  \citenamefont {Xue}, \citenamefont {Song}, \citenamefont {Hu}, \citenamefont
  {Yi}, \citenamefont {Wang},\ and\ \citenamefont {Xue}}]{XiaoLei2024PRL}%
  \BibitemOpen
  \bibfield  {author} {\bibinfo {author} {\bibfnamefont {L.}~\bibnamefont
  {Xiao}}, \bibinfo {author} {\bibfnamefont {W.-T.}\ \bibnamefont {Xue}},
  \bibinfo {author} {\bibfnamefont {F.}~\bibnamefont {Song}}, \bibinfo {author}
  {\bibfnamefont {Y.-M.}\ \bibnamefont {Hu}}, \bibinfo {author} {\bibfnamefont
  {W.}~\bibnamefont {Yi}}, \bibinfo {author} {\bibfnamefont {Z.}~\bibnamefont
  {Wang}},\ and\ \bibinfo {author} {\bibfnamefont {P.}~\bibnamefont {Xue}},\
  }\bibfield  {title} {\bibinfo {title} {Observation of non-hermitian edge
  burst in quantum dynamics},\ }\href
  {https://doi.org/10.1103/PhysRevLett.133.070801} {\bibfield  {journal}
  {\bibinfo  {journal} {Phys. Rev. Lett.}\ }\textbf {\bibinfo {volume} {133}},\
  \bibinfo {pages} {070801} (\bibinfo {year} {2024})}\BibitemShut {NoStop}%
\bibitem [{\citenamefont {Zhang}\ \emph {et~al.}(2024)\citenamefont {Zhang},
  \citenamefont {Yang},\ and\ \citenamefont {Sun}}]{Zhang2024PRB}%
  \BibitemOpen
  \bibfield  {author} {\bibinfo {author} {\bibfnamefont {K.}~\bibnamefont
  {Zhang}}, \bibinfo {author} {\bibfnamefont {Z.}~\bibnamefont {Yang}},\ and\
  \bibinfo {author} {\bibfnamefont {K.}~\bibnamefont {Sun}},\ }\bibfield
  {title} {\bibinfo {title} {Edge theory of non-hermitian skin modes in higher
  dimensions},\ }\href {https://doi.org/10.1103/PhysRevB.109.165127} {\bibfield
   {journal} {\bibinfo  {journal} {Phys. Rev. B}\ }\textbf {\bibinfo {volume}
  {109}},\ \bibinfo {pages} {165127} (\bibinfo {year} {2024})}\BibitemShut
  {NoStop}%
\bibitem [{\citenamefont {Zhao}\ \emph
  {et~al.}(2025{\natexlab{a}})\citenamefont {Zhao}, \citenamefont {Wang},
  \citenamefont {He}, \citenamefont {Poon}, \citenamefont {Pak}, \citenamefont
  {Liu}, \citenamefont {Ren}, \citenamefont {Liu},\ and\ \citenamefont
  {Jo}}]{ZhaoEntong2025}%
  \BibitemOpen
  \bibfield  {author} {\bibinfo {author} {\bibfnamefont {E.}~\bibnamefont
  {Zhao}}, \bibinfo {author} {\bibfnamefont {Z.}~\bibnamefont {Wang}}, \bibinfo
  {author} {\bibfnamefont {C.}~\bibnamefont {He}}, \bibinfo {author}
  {\bibfnamefont {T.~F.~J.}\ \bibnamefont {Poon}}, \bibinfo {author}
  {\bibfnamefont {K.~K.}\ \bibnamefont {Pak}}, \bibinfo {author} {\bibfnamefont
  {Y.-J.}\ \bibnamefont {Liu}}, \bibinfo {author} {\bibfnamefont
  {P.}~\bibnamefont {Ren}}, \bibinfo {author} {\bibfnamefont {X.-J.}\
  \bibnamefont {Liu}},\ and\ \bibinfo {author} {\bibfnamefont {G.-B.}\
  \bibnamefont {Jo}},\ }\bibfield  {title} {\bibinfo {title} {Two-dimensional
  non-hermitian skin effect in an ultracold fermi gas},\ }\href
  {https://doi.org/10.1038/s41586-024-08347-3} {\bibfield  {journal} {\bibinfo
  {journal} {Nature}\ }\textbf {\bibinfo {volume} {637}},\ \bibinfo {pages}
  {565} (\bibinfo {year} {2025}{\natexlab{a}})}\BibitemShut {NoStop}%
\bibitem [{\citenamefont {Wang}\ and\ \citenamefont
  {Li}(2025{\natexlab{a}})}]{wang2025non}%
  \BibitemOpen
  \bibfield  {author} {\bibinfo {author} {\bibfnamefont {Y.-A.}\ \bibnamefont
  {Wang}}\ and\ \bibinfo {author} {\bibfnamefont {L.}~\bibnamefont {Li}},\
  }\bibfield  {title} {\bibinfo {title} {Non-hermitian skin effects in
  fragmented hilbert spaces of one-dimensional fermionic lattices},\
  }\href@noop {} {\bibfield  {journal} {\bibinfo  {journal} {Chinese Physics
  Letters}\ }\textbf {\bibinfo {volume} {42}},\ \bibinfo {pages} {037301}
  (\bibinfo {year} {2025}{\natexlab{a}})}\BibitemShut {NoStop}%
\bibitem [{\citenamefont {Ammari}\ \emph {et~al.}(2025)\citenamefont {Ammari},
  \citenamefont {Barandun}, \citenamefont {Cao}, \citenamefont {Davies},
  \citenamefont {Hiltunen},\ and\ \citenamefont {Liu}}]{ammari2025non}%
  \BibitemOpen
  \bibfield  {author} {\bibinfo {author} {\bibfnamefont {H.}~\bibnamefont
  {Ammari}}, \bibinfo {author} {\bibfnamefont {S.}~\bibnamefont {Barandun}},
  \bibinfo {author} {\bibfnamefont {J.}~\bibnamefont {Cao}}, \bibinfo {author}
  {\bibfnamefont {B.}~\bibnamefont {Davies}}, \bibinfo {author} {\bibfnamefont
  {E.~O.}\ \bibnamefont {Hiltunen}},\ and\ \bibinfo {author} {\bibfnamefont
  {P.}~\bibnamefont {Liu}},\ }\bibfield  {title} {\bibinfo {title} {The
  non-hermitian skin effect with three-dimensional long-range coupling},\
  }\href@noop {} {\bibfield  {journal} {\bibinfo  {journal} {J. Eur. Math.
  Soc.(JEMS)}\ } (\bibinfo {year} {2025})}\BibitemShut {NoStop}%
\bibitem [{\citenamefont {Li}\ \emph {et~al.}(2025{\natexlab{a}})\citenamefont
  {Li}, \citenamefont {Li},\ and\ \citenamefont {Xu}}]{2025bmq57tf6}%
  \BibitemOpen
  \bibfield  {author} {\bibinfo {author} {\bibfnamefont {Y.}~\bibnamefont
  {Li}}, \bibinfo {author} {\bibfnamefont {L.}~\bibnamefont {Li}},\ and\
  \bibinfo {author} {\bibfnamefont {Z.}~\bibnamefont {Xu}},\ }\bibfield
  {title} {\bibinfo {title} {Size-dependent skin effect transitions in weakly
  coupled nonreciprocal chains},\ }\href {https://doi.org/10.1103/bmq5-7tf6}
  {\bibfield  {journal} {\bibinfo  {journal} {Phys. Rev. B}\ }\textbf {\bibinfo
  {volume} {112}},\ \bibinfo {pages} {235122} (\bibinfo {year}
  {2025}{\natexlab{a}})}\BibitemShut {NoStop}%
\bibitem [{\citenamefont {Yang}\ \emph
  {et~al.}(2025{\natexlab{a}})\citenamefont {Yang}, \citenamefont {Qin},
  \citenamefont {Li},\ and\ \citenamefont {Xu}}]{yang2025conf}%
  \BibitemOpen
  \bibfield  {author} {\bibinfo {author} {\bibfnamefont {J.}~\bibnamefont
  {Yang}}, \bibinfo {author} {\bibfnamefont {Y.}~\bibnamefont {Qin}}, \bibinfo
  {author} {\bibfnamefont {L.}~\bibnamefont {Li}},\ and\ \bibinfo {author}
  {\bibfnamefont {X.}~\bibnamefont {Xu}},\ }\bibfield  {title} {\bibinfo
  {title} {Configurable localized states in non-hermitian extended
  su--schrieffer--heeger model},\ }\href@noop {} {\bibfield  {journal}
  {\bibinfo  {journal} {New Journal of Physics}\ }\textbf {\bibinfo {volume}
  {27}},\ \bibinfo {pages} {113001} (\bibinfo {year}
  {2025}{\natexlab{a}})}\BibitemShut {NoStop}%
\bibitem [{\citenamefont {Ou}\ \emph {et~al.}(2025)\citenamefont {Ou},
  \citenamefont {Liang}, \citenamefont {Xu},\ and\ \citenamefont
  {Li}}]{20254yt24rx4}%
  \BibitemOpen
  \bibfield  {author} {\bibinfo {author} {\bibfnamefont {Z.}~\bibnamefont
  {Ou}}, \bibinfo {author} {\bibfnamefont {H.-Q.}\ \bibnamefont {Liang}},
  \bibinfo {author} {\bibfnamefont {G.-F.}\ \bibnamefont {Xu}},\ and\ \bibinfo
  {author} {\bibfnamefont {L.}~\bibnamefont {Li}},\ }\bibfield  {title}
  {\bibinfo {title} {Anisotropic scaling localization in higher-dimensional
  non-hermitian systems},\ }\href {https://doi.org/10.1103/4yt2-4rx4}
  {\bibfield  {journal} {\bibinfo  {journal} {Phys. Rev. B}\ }\textbf {\bibinfo
  {volume} {112}},\ \bibinfo {pages} {L161109} (\bibinfo {year}
  {2025})}\BibitemShut {NoStop}%
\bibitem [{\citenamefont {Zhang}\ \emph
  {et~al.}(2025{\natexlab{a}})\citenamefont {Zhang}, \citenamefont {Shu},\ and\
  \citenamefont {Sun}}]{2025cwwdbclc}%
  \BibitemOpen
  \bibfield  {author} {\bibinfo {author} {\bibfnamefont {K.}~\bibnamefont
  {Zhang}}, \bibinfo {author} {\bibfnamefont {C.}~\bibnamefont {Shu}},\ and\
  \bibinfo {author} {\bibfnamefont {K.}~\bibnamefont {Sun}},\ }\bibfield
  {title} {\bibinfo {title} {Algebraic non-hermitian skin effect and
  generalized fermi surface formula in arbitrary dimensions},\ }\href
  {https://doi.org/10.1103/cwwd-bclc} {\bibfield  {journal} {\bibinfo
  {journal} {Phys. Rev. X}\ }\textbf {\bibinfo {volume} {15}},\ \bibinfo
  {pages} {031039} (\bibinfo {year} {2025}{\natexlab{a}})}\BibitemShut
  {NoStop}%
\bibitem [{\citenamefont {Shu}\ \emph {et~al.}(2025)\citenamefont {Shu},
  \citenamefont {Zhang},\ and\ \citenamefont {Sun}}]{3927n25r2025}%
  \BibitemOpen
  \bibfield  {author} {\bibinfo {author} {\bibfnamefont {C.}~\bibnamefont
  {Shu}}, \bibinfo {author} {\bibfnamefont {K.}~\bibnamefont {Zhang}},\ and\
  \bibinfo {author} {\bibfnamefont {K.}~\bibnamefont {Sun}},\ }\bibfield
  {title} {\bibinfo {title} {Ultraspectral sensitivity and nonlocal bound
  states in algebraic non-hermitian skin effect},\ }\href
  {https://doi.org/10.1103/3927-n25r} {\bibfield  {journal} {\bibinfo
  {journal} {Phys. Rev. B}\ }\textbf {\bibinfo {volume} {112}},\ \bibinfo
  {pages} {235152} (\bibinfo {year} {2025})}\BibitemShut {NoStop}%
\bibitem [{\citenamefont {Cheng}\ \emph {et~al.}(2025)\citenamefont {Cheng},
  \citenamefont {Jiang}, \citenamefont {Chen}, \citenamefont {Zhang},
  \citenamefont {Ang},\ and\ \citenamefont {Lee}}]{cheng2025stochasticity}%
  \BibitemOpen
  \bibfield  {author} {\bibinfo {author} {\bibfnamefont {X.}~\bibnamefont
  {Cheng}}, \bibinfo {author} {\bibfnamefont {H.}~\bibnamefont {Jiang}},
  \bibinfo {author} {\bibfnamefont {J.}~\bibnamefont {Chen}}, \bibinfo {author}
  {\bibfnamefont {L.}~\bibnamefont {Zhang}}, \bibinfo {author} {\bibfnamefont
  {Y.~S.}\ \bibnamefont {Ang}},\ and\ \bibinfo {author} {\bibfnamefont {C.~H.}\
  \bibnamefont {Lee}},\ }\bibfield  {title} {\bibinfo {title}
  {Stochasticity-induced non-hermitian skin criticality},\ }\href@noop {}
  {\bibfield  {journal} {\bibinfo  {journal} {arXiv preprint arXiv:2511.13176}\
  } (\bibinfo {year} {2025})}\BibitemShut {NoStop}%
\bibitem [{\citenamefont {Zhang}\ \emph
  {et~al.}(2025{\natexlab{b}})\citenamefont {Zhang}, \citenamefont {Su},\ and\
  \citenamefont {Chen}}]{ZhangPRB2025}%
  \BibitemOpen
  \bibfield  {author} {\bibinfo {author} {\bibfnamefont {Y.}~\bibnamefont
  {Zhang}}, \bibinfo {author} {\bibfnamefont {L.}~\bibnamefont {Su}},\ and\
  \bibinfo {author} {\bibfnamefont {S.}~\bibnamefont {Chen}},\ }\bibfield
  {title} {\bibinfo {title} {Scale-free localization versus anderson
  localization in unidirectional quasiperiodic lattices},\ }\href
  {https://doi.org/10.1103/PhysRevB.111.L140201} {\bibfield  {journal}
  {\bibinfo  {journal} {Phys. Rev. B}\ }\textbf {\bibinfo {volume} {111}},\
  \bibinfo {pages} {L140201} (\bibinfo {year}
  {2025}{\natexlab{b}})}\BibitemShut {NoStop}%
\bibitem [{\citenamefont {Yang}\ and\ \citenamefont {Lee}(2025)}]{yang2025rev}%
  \BibitemOpen
  \bibfield  {author} {\bibinfo {author} {\bibfnamefont {M.}~\bibnamefont
  {Yang}}\ and\ \bibinfo {author} {\bibfnamefont {C.~H.}\ \bibnamefont {Lee}},\
  }\bibfield  {title} {\bibinfo {title} {Reversing non-hermitian skin
  accumulation with a non-local transverse switch},\ }\href@noop {} {\bibfield
  {journal} {\bibinfo  {journal} {arXiv preprint arXiv:2509.02686}\ } (\bibinfo
  {year} {2025})}\BibitemShut {NoStop}%
\bibitem [{\citenamefont {Xue}\ \emph {et~al.}(2025)\citenamefont {Xue},
  \citenamefont {Song}, \citenamefont {Hu},\ and\ \citenamefont
  {Wang}}]{xue2025non}%
  \BibitemOpen
  \bibfield  {author} {\bibinfo {author} {\bibfnamefont {W.-T.}\ \bibnamefont
  {Xue}}, \bibinfo {author} {\bibfnamefont {F.}~\bibnamefont {Song}}, \bibinfo
  {author} {\bibfnamefont {Y.-M.}\ \bibnamefont {Hu}},\ and\ \bibinfo {author}
  {\bibfnamefont {Z.}~\bibnamefont {Wang}},\ }\bibfield  {title} {\bibinfo
  {title} {Non-bloch edge dynamics of non-hermitian lattices},\ }\href@noop {}
  {\bibfield  {journal} {\bibinfo  {journal} {arXiv preprint arXiv:2503.13671}\
  } (\bibinfo {year} {2025})}\BibitemShut {NoStop}%
\bibitem [{\citenamefont {Li}\ \emph {et~al.}(2025{\natexlab{b}})\citenamefont
  {Li}, \citenamefont {Chen},\ and\ \citenamefont {Wang}}]{2025z9m13mwb}%
  \BibitemOpen
  \bibfield  {author} {\bibinfo {author} {\bibfnamefont {B.}~\bibnamefont
  {Li}}, \bibinfo {author} {\bibfnamefont {C.}~\bibnamefont {Chen}},\ and\
  \bibinfo {author} {\bibfnamefont {Z.}~\bibnamefont {Wang}},\ }\bibfield
  {title} {\bibinfo {title} {Universal non-hermitian transport in disordered
  systems},\ }\href {https://doi.org/10.1103/z9m1-3mwb} {\bibfield  {journal}
  {\bibinfo  {journal} {Phys. Rev. Lett.}\ }\textbf {\bibinfo {volume} {135}},\
  \bibinfo {pages} {033802} (\bibinfo {year} {2025}{\natexlab{b}})}\BibitemShut
  {NoStop}%
\bibitem [{\citenamefont {Li}\ \emph {et~al.}(2025{\natexlab{c}})\citenamefont
  {Li}, \citenamefont {Li}, \citenamefont {Zhu},\ and\ \citenamefont
  {Li}}]{li2025anderson}%
  \BibitemOpen
  \bibfield  {author} {\bibinfo {author} {\bibfnamefont {S.-Z.}\ \bibnamefont
  {Li}}, \bibinfo {author} {\bibfnamefont {L.}~\bibnamefont {Li}}, \bibinfo
  {author} {\bibfnamefont {S.-L.}\ \bibnamefont {Zhu}},\ and\ \bibinfo {author}
  {\bibfnamefont {Z.}~\bibnamefont {Li}},\ }\bibfield  {title} {\bibinfo
  {title} {Anderson-skin dualism: A boundary-dependent effect in non-hermitian
  disordered coupled systems},\ }\href@noop {} {\bibfield  {journal} {\bibinfo
  {journal} {Physical Review B}\ }\textbf {\bibinfo {volume} {112}},\ \bibinfo
  {pages} {L201108} (\bibinfo {year} {2025}{\natexlab{c}})}\BibitemShut
  {NoStop}%
\bibitem [{\citenamefont {Yang}\ \emph
  {et~al.}(2025{\natexlab{b}})\citenamefont {Yang}, \citenamefont {Yuan},\ and\
  \citenamefont {Lee}}]{yang2025non}%
  \BibitemOpen
  \bibfield  {author} {\bibinfo {author} {\bibfnamefont {M.}~\bibnamefont
  {Yang}}, \bibinfo {author} {\bibfnamefont {L.}~\bibnamefont {Yuan}},\ and\
  \bibinfo {author} {\bibfnamefont {C.~H.}\ \bibnamefont {Lee}},\ }\bibfield
  {title} {\bibinfo {title} {Non-hermitian strong bosonic clustering through
  interaction-induced caging},\ }\href@noop {} {\bibfield  {journal} {\bibinfo
  {journal} {Communications Physics}\ }\textbf {\bibinfo {volume} {8}},\
  \bibinfo {pages} {388} (\bibinfo {year} {2025}{\natexlab{b}})}\BibitemShut
  {NoStop}%
\bibitem [{\citenamefont {Yi}\ and\ \citenamefont {Yang}(2025)}]{d5zcp1sk2025}%
  \BibitemOpen
  \bibfield  {author} {\bibinfo {author} {\bibfnamefont {Y.}~\bibnamefont
  {Yi}}\ and\ \bibinfo {author} {\bibfnamefont {Z.}~\bibnamefont {Yang}},\
  }\bibfield  {title} {\bibinfo {title} {Anomalous scaling behavior of green's
  function in critical skin effects},\ }\href
  {https://doi.org/10.1103/d5zc-p1sk} {\bibfield  {journal} {\bibinfo
  {journal} {Phys. Rev. B}\ }\textbf {\bibinfo {volume} {112}},\ \bibinfo
  {pages} {174303} (\bibinfo {year} {2025})}\BibitemShut {NoStop}%
\bibitem [{\citenamefont {Wu}\ \emph {et~al.}(2025)\citenamefont {Wu},
  \citenamefont {Hu}, \citenamefont {He}, \citenamefont {Deng}, \citenamefont
  {Huang}, \citenamefont {Ke}, \citenamefont {Deng}, \citenamefont {Lu},\ and\
  \citenamefont {Liu}}]{WuPRL2025}%
  \BibitemOpen
  \bibfield  {author} {\bibinfo {author} {\bibfnamefont {J.}~\bibnamefont
  {Wu}}, \bibinfo {author} {\bibfnamefont {Y.}~\bibnamefont {Hu}}, \bibinfo
  {author} {\bibfnamefont {Z.}~\bibnamefont {He}}, \bibinfo {author}
  {\bibfnamefont {K.}~\bibnamefont {Deng}}, \bibinfo {author} {\bibfnamefont
  {X.}~\bibnamefont {Huang}}, \bibinfo {author} {\bibfnamefont
  {M.}~\bibnamefont {Ke}}, \bibinfo {author} {\bibfnamefont {W.}~\bibnamefont
  {Deng}}, \bibinfo {author} {\bibfnamefont {J.}~\bibnamefont {Lu}},\ and\
  \bibinfo {author} {\bibfnamefont {Z.}~\bibnamefont {Liu}},\ }\bibfield
  {title} {\bibinfo {title} {Hybrid-order skin effect from loss-induced
  nonreciprocity},\ }\href {https://doi.org/10.1103/PhysRevLett.134.176601}
  {\bibfield  {journal} {\bibinfo  {journal} {Phys. Rev. Lett.}\ }\textbf
  {\bibinfo {volume} {134}},\ \bibinfo {pages} {176601} (\bibinfo {year}
  {2025})}\BibitemShut {NoStop}%
\bibitem [{\citenamefont {Li}\ \emph {et~al.}(2025{\natexlab{d}})\citenamefont
  {Li}, \citenamefont {Jiang},\ and\ \citenamefont {Lee}}]{li2025phase}%
  \BibitemOpen
  \bibfield  {author} {\bibinfo {author} {\bibfnamefont {Q.}~\bibnamefont
  {Li}}, \bibinfo {author} {\bibfnamefont {H.}~\bibnamefont {Jiang}},\ and\
  \bibinfo {author} {\bibfnamefont {C.~H.}\ \bibnamefont {Lee}},\ }\bibfield
  {title} {\bibinfo {title} {Phase-space generalized brillouin zone for
  spatially inhomogeneous non-hermitian systems},\ }\href@noop {} {\bibfield
  {journal} {\bibinfo  {journal} {Advanced Science}\ }\textbf {\bibinfo
  {volume} {12}},\ \bibinfo {pages} {e08047} (\bibinfo {year}
  {2025}{\natexlab{d}})}\BibitemShut {NoStop}%
\bibitem [{\citenamefont {Gohsrich}\ \emph {et~al.}(2025)\citenamefont
  {Gohsrich}, \citenamefont {Banerjee},\ and\ \citenamefont
  {Kunst}}]{gohsrich2025non}%
  \BibitemOpen
  \bibfield  {author} {\bibinfo {author} {\bibfnamefont {J.~T.}\ \bibnamefont
  {Gohsrich}}, \bibinfo {author} {\bibfnamefont {A.}~\bibnamefont {Banerjee}},\
  and\ \bibinfo {author} {\bibfnamefont {F.~K.}\ \bibnamefont {Kunst}},\
  }\bibfield  {title} {\bibinfo {title} {The non-hermitian skin effect: A
  perspective},\ }\href@noop {} {\bibfield  {journal} {\bibinfo  {journal}
  {Europhysics Letters}\ }\textbf {\bibinfo {volume} {150}},\ \bibinfo {pages}
  {60001} (\bibinfo {year} {2025})}\BibitemShut {NoStop}%
\bibitem [{\citenamefont {Wang}\ \emph {et~al.}(2025)\citenamefont {Wang},
  \citenamefont {Wang}, \citenamefont {Liu}, \citenamefont {Qin}, \citenamefont
  {Zhao}, \citenamefont {Liu}, \citenamefont {Longhi},\ and\ \citenamefont
  {Lu}}]{wang2025nonlinear}%
  \BibitemOpen
  \bibfield  {author} {\bibinfo {author} {\bibfnamefont {S.}~\bibnamefont
  {Wang}}, \bibinfo {author} {\bibfnamefont {B.}~\bibnamefont {Wang}}, \bibinfo
  {author} {\bibfnamefont {C.}~\bibnamefont {Liu}}, \bibinfo {author}
  {\bibfnamefont {C.}~\bibnamefont {Qin}}, \bibinfo {author} {\bibfnamefont
  {L.}~\bibnamefont {Zhao}}, \bibinfo {author} {\bibfnamefont {W.}~\bibnamefont
  {Liu}}, \bibinfo {author} {\bibfnamefont {S.}~\bibnamefont {Longhi}},\ and\
  \bibinfo {author} {\bibfnamefont {P.}~\bibnamefont {Lu}},\ }\bibfield
  {title} {\bibinfo {title} {Nonlinear non-hermitian skin effect and skin
  solitons in temporal photonic feedforward lattices},\ }\href@noop {}
  {\bibfield  {journal} {\bibinfo  {journal} {Physical Review Letters}\
  }\textbf {\bibinfo {volume} {134}},\ \bibinfo {pages} {243805} (\bibinfo
  {year} {2025})}\BibitemShut {NoStop}%
\bibitem [{\citenamefont {Zhao}\ \emph
  {et~al.}(2025{\natexlab{b}})\citenamefont {Zhao}, \citenamefont {Zhang},
  \citenamefont {Xiao}, \citenamefont {Sun},\ and\ \citenamefont
  {Yan}}]{zhao2025magne}%
  \BibitemOpen
  \bibfield  {author} {\bibinfo {author} {\bibfnamefont {Y.}~\bibnamefont
  {Zhao}}, \bibinfo {author} {\bibfnamefont {K.}~\bibnamefont {Zhang}},
  \bibinfo {author} {\bibfnamefont {J.}~\bibnamefont {Xiao}}, \bibinfo {author}
  {\bibfnamefont {K.}~\bibnamefont {Sun}},\ and\ \bibinfo {author}
  {\bibfnamefont {B.}~\bibnamefont {Yan}},\ }\bibfield  {title} {\bibinfo
  {title} {Magnetochiral charge pumping due to charge trapping and skin effect
  in chirality-induced spin selectivity},\ }\href@noop {} {\bibfield  {journal}
  {\bibinfo  {journal} {Nature communications}\ }\textbf {\bibinfo {volume}
  {16}},\ \bibinfo {pages} {37} (\bibinfo {year}
  {2025}{\natexlab{b}})}\BibitemShut {NoStop}%
\bibitem [{\citenamefont {Liu}\ \emph {et~al.}(2025{\natexlab{a}})\citenamefont
  {Liu}, \citenamefont {Liu},\ and\ \citenamefont {Xiao}}]{liu2025anom}%
  \BibitemOpen
  \bibfield  {author} {\bibinfo {author} {\bibfnamefont {T.-R.}\ \bibnamefont
  {Liu}}, \bibinfo {author} {\bibfnamefont {T.}~\bibnamefont {Liu}},\ and\
  \bibinfo {author} {\bibfnamefont {M.}~\bibnamefont {Xiao}},\ }\bibfield
  {title} {\bibinfo {title} {Anomalous non-hermitian skin effect of chiral
  boundary states},\ }\href@noop {} {\bibfield  {journal} {\bibinfo  {journal}
  {Physical Review B}\ }\textbf {\bibinfo {volume} {112}},\ \bibinfo {pages}
  {L081112} (\bibinfo {year} {2025}{\natexlab{a}})}\BibitemShut {NoStop}%
\bibitem [{\citenamefont {Li}\ \emph {et~al.}(2025{\natexlab{e}})\citenamefont
  {Li}, \citenamefont {Lin},\ and\ \citenamefont {Ding}}]{li2025algebraic}%
  \BibitemOpen
  \bibfield  {author} {\bibinfo {author} {\bibfnamefont {M.}~\bibnamefont
  {Li}}, \bibinfo {author} {\bibfnamefont {J.}~\bibnamefont {Lin}},\ and\
  \bibinfo {author} {\bibfnamefont {K.}~\bibnamefont {Ding}},\ }\bibfield
  {title} {\bibinfo {title} {Algebraic skin effect in two-dimensional
  non-hermitian metamaterials},\ }\href@noop {} {\bibfield  {journal} {\bibinfo
   {journal} {arXiv preprint arXiv:2501.13440}\ } (\bibinfo {year}
  {2025}{\natexlab{e}})}\BibitemShut {NoStop}%
\bibitem [{\citenamefont {Hu}(2025)}]{hu2025topological}%
  \BibitemOpen
  \bibfield  {author} {\bibinfo {author} {\bibfnamefont {H.}~\bibnamefont
  {Hu}},\ }\bibfield  {title} {\bibinfo {title} {Topological origin of
  non-hermitian skin effect in higher dimensions and uniform spectra},\
  }\href@noop {} {\bibfield  {journal} {\bibinfo  {journal} {Science Bulletin}\
  }\textbf {\bibinfo {volume} {70}},\ \bibinfo {pages} {51} (\bibinfo {year}
  {2025})}\BibitemShut {NoStop}%
\bibitem [{\citenamefont {Het\'enyi}\ \emph {et~al.}(2025)\citenamefont
  {Het\'enyi}, \citenamefont {L\'aszl\'offy}, \citenamefont {Penc},\ and\
  \citenamefont {\'Ujfalussy}}]{v6ht-jq3l2025}%
  \BibitemOpen
  \bibfield  {author} {\bibinfo {author} {\bibfnamefont {B.}~\bibnamefont
  {Het\'enyi}}, \bibinfo {author} {\bibfnamefont {A.}~\bibnamefont
  {L\'aszl\'offy}}, \bibinfo {author} {\bibfnamefont {K.}~\bibnamefont
  {Penc}},\ and\ \bibinfo {author} {\bibfnamefont {B.}~\bibnamefont
  {\'Ujfalussy}},\ }\bibfield  {title} {\bibinfo {title} {Topological
  indicators for systems with open boundaries: Application to the kitaev
  wire},\ }\href {https://doi.org/10.1103/v6ht-jq3l} {\bibfield  {journal}
  {\bibinfo  {journal} {Phys. Rev. B}\ }\textbf {\bibinfo {volume} {112}},\
  \bibinfo {pages} {075115} (\bibinfo {year} {2025})}\BibitemShut {NoStop}%
\bibitem [{\citenamefont {Het\'enyi}\ and\ \citenamefont
  {D\'ora}(2025)}]{xbj1-hfyf2025}%
  \BibitemOpen
  \bibfield  {author} {\bibinfo {author} {\bibfnamefont {B.}~\bibnamefont
  {Het\'enyi}}\ and\ \bibinfo {author} {\bibfnamefont {B.}~\bibnamefont
  {D\'ora}},\ }\bibfield  {title} {\bibinfo {title} {Localized states and skin
  effect around non-hermitian impurities in tight-binding models},\ }\href
  {https://doi.org/10.1103/xbj1-hfyf} {\bibfield  {journal} {\bibinfo
  {journal} {Phys. Rev. B}\ }\textbf {\bibinfo {volume} {112}},\ \bibinfo
  {pages} {075123} (\bibinfo {year} {2025})}\BibitemShut {NoStop}%
\bibitem [{\citenamefont {Ito}\ and\ \citenamefont
  {Uchino}(2026)}]{Nobuhiro2026}%
  \BibitemOpen
  \bibfield  {author} {\bibinfo {author} {\bibfnamefont {N.}~\bibnamefont
  {Ito}}\ and\ \bibinfo {author} {\bibfnamefont {S.}~\bibnamefont {Uchino}},\
  }\bibfield  {title} {\bibinfo {title} {Edge-controlled non-hermitian skin
  effect in the modified haldane model},\ }\href@noop {} {\bibfield  {journal}
  {\bibinfo  {journal} {arXiv preprint arXiv:2603.01503}\ } (\bibinfo {year}
  {2026})}\BibitemShut {NoStop}%
\bibitem [{\citenamefont {Yi}(2026)}]{yi2026dir}%
  \BibitemOpen
  \bibfield  {author} {\bibinfo {author} {\bibfnamefont {B.}~\bibnamefont
  {Yi}},\ }\bibfield  {title} {\bibinfo {title} {Directional dynamics of the
  non-hermitian skin effect},\ }\href@noop {} {\bibfield  {journal} {\bibinfo
  {journal} {arXiv preprint arXiv:2602.18106}\ } (\bibinfo {year}
  {2026})}\BibitemShut {NoStop}%
\bibitem [{\citenamefont {Lin}\ \emph {et~al.}(2026)\citenamefont {Lin},
  \citenamefont {Qi},\ and\ \citenamefont {Long}}]{lin2026glo}%
  \BibitemOpen
  \bibfield  {author} {\bibinfo {author} {\bibfnamefont {H.}~\bibnamefont
  {Lin}}, \bibinfo {author} {\bibfnamefont {Y.}~\bibnamefont {Qi}},\ and\
  \bibinfo {author} {\bibfnamefont {G.-L.}\ \bibnamefont {Long}},\ }\bibfield
  {title} {\bibinfo {title} {Global bifurcations and basin geometry of the
  nonlinear non-hermitian skin effect},\ }\href@noop {} {\bibfield  {journal}
  {\bibinfo  {journal} {arXiv preprint arXiv:2602.17439}\ } (\bibinfo {year}
  {2026})}\BibitemShut {NoStop}%
\bibitem [{\citenamefont {Rahul}\ and\ \citenamefont
  {Marra}(2026)}]{rahul2026cont}%
  \BibitemOpen
  \bibfield  {author} {\bibinfo {author} {\bibfnamefont {S.}~\bibnamefont
  {Rahul}}\ and\ \bibinfo {author} {\bibfnamefont {P.}~\bibnamefont {Marra}},\
  }\bibfield  {title} {\bibinfo {title} {Controlling energy spectra and skin
  effect via boundary conditions in non-hermitian lattices},\ }\href@noop {}
  {\bibfield  {journal} {\bibinfo  {journal} {arXiv preprint arXiv:2602.16780}\
  } (\bibinfo {year} {2026})}\BibitemShut {NoStop}%
\bibitem [{\citenamefont {Longhi}(2026)}]{longhi2026erra}%
  \BibitemOpen
  \bibfield  {author} {\bibinfo {author} {\bibfnamefont {S.}~\bibnamefont
  {Longhi}},\ }\bibfield  {title} {\bibinfo {title} {Erratic liouvillian skin
  localization and subdiffusive transport},\ }\href@noop {} {\bibfield
  {journal} {\bibinfo  {journal} {arXiv preprint arXiv:2602.14698}\ } (\bibinfo
  {year} {2026})}\BibitemShut {NoStop}%
\bibitem [{\citenamefont {Saito}\ \emph {et~al.}(2026)\citenamefont {Saito},
  \citenamefont {Okugawa}, \citenamefont {Yokomizo}, \citenamefont {Tohyama},\
  and\ \citenamefont {Hsu}}]{saito2026quas}%
  \BibitemOpen
  \bibfield  {author} {\bibinfo {author} {\bibfnamefont {K.}~\bibnamefont
  {Saito}}, \bibinfo {author} {\bibfnamefont {R.}~\bibnamefont {Okugawa}},
  \bibinfo {author} {\bibfnamefont {K.}~\bibnamefont {Yokomizo}}, \bibinfo
  {author} {\bibfnamefont {T.}~\bibnamefont {Tohyama}},\ and\ \bibinfo {author}
  {\bibfnamefont {C.-H.}\ \bibnamefont {Hsu}},\ }\bibfield  {title} {\bibinfo
  {title} {Quasiperiodicity-induced non-hermitian skin effect from the
  breakdown of scale-free localization},\ }\href@noop {} {\bibfield  {journal}
  {\bibinfo  {journal} {arXiv preprint arXiv:2602.11155}\ } (\bibinfo {year}
  {2026})}\BibitemShut {NoStop}%
\bibitem [{\citenamefont {Okuma}(2026)}]{okuma2026ste}%
  \BibitemOpen
  \bibfield  {author} {\bibinfo {author} {\bibfnamefont {N.}~\bibnamefont
  {Okuma}},\ }\bibfield  {title} {\bibinfo {title} {Steady-state skin effect in
  bosonic topological edge states under parametric driving},\ }\href@noop {}
  {\bibfield  {journal} {\bibinfo  {journal} {arXiv preprint arXiv:2602.01625}\
  } (\bibinfo {year} {2026})}\BibitemShut {NoStop}%
\bibitem [{\citenamefont {Bai}\ \emph {et~al.}(2026)\citenamefont {Bai},
  \citenamefont {Yang}, \citenamefont {Yan}, \citenamefont {Li},\ and\
  \citenamefont {Shao}}]{bai2026eng}%
  \BibitemOpen
  \bibfield  {author} {\bibinfo {author} {\bibfnamefont {J.~N.}\ \bibnamefont
  {Bai}}, \bibinfo {author} {\bibfnamefont {F.}~\bibnamefont {Yang}}, \bibinfo
  {author} {\bibfnamefont {D.}~\bibnamefont {Yan}}, \bibinfo {author}
  {\bibfnamefont {W.}~\bibnamefont {Li}},\ and\ \bibinfo {author}
  {\bibfnamefont {X.~Q.}\ \bibnamefont {Shao}},\ }\bibfield  {title} {\bibinfo
  {title} {Engineering the non-hermitian ssh model with skin effects in rydberg
  atom arrays},\ }\href@noop {} {\bibfield  {journal} {\bibinfo  {journal}
  {arXiv preprint arXiv:2601.20114}\ } (\bibinfo {year} {2026})}\BibitemShut
  {NoStop}%
\bibitem [{\citenamefont {Deng}\ \emph {et~al.}(2026)\citenamefont {Deng},
  \citenamefont {Mi}, \citenamefont {Cai}, \citenamefont {Wu},\ and\
  \citenamefont {Chen}}]{deng2026conf}%
  \BibitemOpen
  \bibfield  {author} {\bibinfo {author} {\bibfnamefont {Z.~J.}\ \bibnamefont
  {Deng}}, \bibinfo {author} {\bibfnamefont {X.~Y.}\ \bibnamefont {Mi}},
  \bibinfo {author} {\bibfnamefont {R.~K.}\ \bibnamefont {Cai}}, \bibinfo
  {author} {\bibfnamefont {C.~W.}\ \bibnamefont {Wu}},\ and\ \bibinfo {author}
  {\bibfnamefont {P.~X.}\ \bibnamefont {Chen}},\ }\bibfield  {title} {\bibinfo
  {title} {Confined non-hermitian skin effect in a semi-infinite fock-state
  lattice},\ }\href@noop {} {\bibfield  {journal} {\bibinfo  {journal} {arXiv
  preprint arXiv:2601.13540}\ } (\bibinfo {year} {2026})}\BibitemShut {NoStop}%
\bibitem [{\citenamefont {Yang}\ \emph {et~al.}(2026)\citenamefont {Yang},
  \citenamefont {Feng}, \citenamefont {Wahab},\ and\ \citenamefont
  {Geng}}]{2026s26b8bdl}%
  \BibitemOpen
  \bibfield  {author} {\bibinfo {author} {\bibfnamefont {X.}~\bibnamefont
  {Yang}}, \bibinfo {author} {\bibfnamefont {Y.}~\bibnamefont {Feng}}, \bibinfo
  {author} {\bibfnamefont {A.}~\bibnamefont {Wahab}},\ and\ \bibinfo {author}
  {\bibfnamefont {H.}~\bibnamefont {Geng}},\ }\bibfield  {title} {\bibinfo
  {title} {Non-hermitian second-order topological phases and bipolar skin
  effect in photonic kagome crystals},\ }\href
  {https://doi.org/10.1103/s26b-8bdl} {\bibfield  {journal} {\bibinfo
  {journal} {Phys. Rev. A}\ }\textbf {\bibinfo {volume} {113}},\ \bibinfo
  {pages} {023506} (\bibinfo {year} {2026})}\BibitemShut {NoStop}%
\bibitem [{\citenamefont {Wu}\ \emph {et~al.}(2026)\citenamefont {Wu},
  \citenamefont {Zhang}, \citenamefont {Qi}, \citenamefont {Zhang},
  \citenamefont {Tong},\ and\ \citenamefont {Qiu}}]{wu2026observation}%
  \BibitemOpen
  \bibfield  {author} {\bibinfo {author} {\bibfnamefont {W.}~\bibnamefont
  {Wu}}, \bibinfo {author} {\bibfnamefont {Q.}~\bibnamefont {Zhang}}, \bibinfo
  {author} {\bibfnamefont {L.}~\bibnamefont {Qi}}, \bibinfo {author}
  {\bibfnamefont {K.}~\bibnamefont {Zhang}}, \bibinfo {author} {\bibfnamefont
  {S.}~\bibnamefont {Tong}},\ and\ \bibinfo {author} {\bibfnamefont
  {C.}~\bibnamefont {Qiu}},\ }\bibfield  {title} {\bibinfo {title} {Observation
  of dislocation non-hermitian skin effect in a torus-like acoustic
  metamaterial},\ }\href@noop {} {\bibfield  {journal} {\bibinfo  {journal}
  {Advanced Materials}\ }\textbf {\bibinfo {volume} {38}},\ \bibinfo {pages}
  {e14101} (\bibinfo {year} {2026})}\BibitemShut {NoStop}%
\bibitem [{\citenamefont {Yu}\ \emph {et~al.}(2026)\citenamefont {Yu},
  \citenamefont {Soci}, \citenamefont {Chong},\ and\ \citenamefont
  {Zhang}}]{yu2026sensitivity}%
  \BibitemOpen
  \bibfield  {author} {\bibinfo {author} {\bibfnamefont {L.}~\bibnamefont
  {Yu}}, \bibinfo {author} {\bibfnamefont {C.}~\bibnamefont {Soci}}, \bibinfo
  {author} {\bibfnamefont {Y.}~\bibnamefont {Chong}},\ and\ \bibinfo {author}
  {\bibfnamefont {B.}~\bibnamefont {Zhang}},\ }\bibfield  {title} {\bibinfo
  {title} {Sensitivity evaluation for global perturbations in non-hermitian
  skin-effect sensors},\ }\href@noop {} {\bibfield  {journal} {\bibinfo
  {journal} {Nanophotonics}\ }\textbf {\bibinfo {volume} {15}},\ \bibinfo
  {pages} {e70039} (\bibinfo {year} {2026})}\BibitemShut {NoStop}%
\bibitem [{\citenamefont {Hu}\ \emph {et~al.}(2026)\citenamefont {Hu},
  \citenamefont {Shi}, \citenamefont {Li}, \citenamefont {Teza}, \citenamefont
  {Lee}, \citenamefont {Moessner}, \citenamefont {Zhang},\ and\ \citenamefont
  {Mu}}]{hu2026boundary}%
  \BibitemOpen
  \bibfield  {author} {\bibinfo {author} {\bibfnamefont {Y.-M.}\ \bibnamefont
  {Hu}}, \bibinfo {author} {\bibfnamefont {Y.-B.}\ \bibnamefont {Shi}},
  \bibinfo {author} {\bibfnamefont {L.}~\bibnamefont {Li}}, \bibinfo {author}
  {\bibfnamefont {G.}~\bibnamefont {Teza}}, \bibinfo {author} {\bibfnamefont
  {C.~H.}\ \bibnamefont {Lee}}, \bibinfo {author} {\bibfnamefont
  {R.}~\bibnamefont {Moessner}}, \bibinfo {author} {\bibfnamefont
  {S.}~\bibnamefont {Zhang}},\ and\ \bibinfo {author} {\bibfnamefont
  {S.}~\bibnamefont {Mu}},\ }\bibfield  {title} {\bibinfo {title} {Boundary
  floquet control of bulk non-hermitian systems},\ }\href@noop {} {\bibfield
  {journal} {\bibinfo  {journal} {arXiv preprint arXiv:2603.22396}\ } (\bibinfo
  {year} {2026})}\BibitemShut {NoStop}%
\bibitem [{\citenamefont {Lee}\ \emph {et~al.}(2020)\citenamefont {Lee},
  \citenamefont {Lee},\ and\ \citenamefont {Yang}}]{2020PhysRevB.101.121109}%
  \BibitemOpen
  \bibfield  {author} {\bibinfo {author} {\bibfnamefont {E.}~\bibnamefont
  {Lee}}, \bibinfo {author} {\bibfnamefont {H.}~\bibnamefont {Lee}},\ and\
  \bibinfo {author} {\bibfnamefont {B.-J.}\ \bibnamefont {Yang}},\ }\bibfield
  {title} {\bibinfo {title} {Many-body approach to non-hermitian physics in
  fermionic systems},\ }\href {https://doi.org/10.1103/PhysRevB.101.121109}
  {\bibfield  {journal} {\bibinfo  {journal} {Phys. Rev. B}\ }\textbf {\bibinfo
  {volume} {101}},\ \bibinfo {pages} {121109} (\bibinfo {year}
  {2020})}\BibitemShut {NoStop}%
\bibitem [{\citenamefont {Mu}\ \emph {et~al.}(2020)\citenamefont {Mu},
  \citenamefont {Lee}, \citenamefont {Li},\ and\ \citenamefont
  {Gong}}]{2020PhysRevB.102.081115}%
  \BibitemOpen
  \bibfield  {author} {\bibinfo {author} {\bibfnamefont {S.}~\bibnamefont
  {Mu}}, \bibinfo {author} {\bibfnamefont {C.~H.}\ \bibnamefont {Lee}},
  \bibinfo {author} {\bibfnamefont {L.}~\bibnamefont {Li}},\ and\ \bibinfo
  {author} {\bibfnamefont {J.}~\bibnamefont {Gong}},\ }\bibfield  {title}
  {\bibinfo {title} {Emergent fermi surface in a many-body non-hermitian
  fermionic chain},\ }\href {https://doi.org/10.1103/PhysRevB.102.081115}
  {\bibfield  {journal} {\bibinfo  {journal} {Phys. Rev. B}\ }\textbf {\bibinfo
  {volume} {102}},\ \bibinfo {pages} {081115} (\bibinfo {year}
  {2020})}\BibitemShut {NoStop}%
\bibitem [{\citenamefont {Alsallom}\ \emph {et~al.}(2022)\citenamefont
  {Alsallom}, \citenamefont {Herviou}, \citenamefont {Yazyev},\ and\
  \citenamefont {Brzezi\ifmmode~\acute{n}\else
  \'{n}\fi{}ska}}]{2022PhysRevResearch.4.033122}%
  \BibitemOpen
  \bibfield  {author} {\bibinfo {author} {\bibfnamefont {F.}~\bibnamefont
  {Alsallom}}, \bibinfo {author} {\bibfnamefont {L.}~\bibnamefont {Herviou}},
  \bibinfo {author} {\bibfnamefont {O.~V.}\ \bibnamefont {Yazyev}},\ and\
  \bibinfo {author} {\bibfnamefont {M.}~\bibnamefont
  {Brzezi\ifmmode~\acute{n}\else \'{n}\fi{}ska}},\ }\bibfield  {title}
  {\bibinfo {title} {Fate of the non-hermitian skin effect in many-body
  fermionic systems},\ }\href
  {https://doi.org/10.1103/PhysRevResearch.4.033122} {\bibfield  {journal}
  {\bibinfo  {journal} {Phys. Rev. Res.}\ }\textbf {\bibinfo {volume} {4}},\
  \bibinfo {pages} {033122} (\bibinfo {year} {2022})}\BibitemShut {NoStop}%
\bibitem [{\citenamefont {Zhang}\ \emph
  {et~al.}(2022{\natexlab{b}})\citenamefont {Zhang}, \citenamefont {Denner},
  \citenamefont {Bzdu\ifmmode~\check{s}\else \v{s}\fi{}ek}, \citenamefont
  {Sentef},\ and\ \citenamefont {Neupert}}]{PhysRevB.106.L121102}%
  \BibitemOpen
  \bibfield  {author} {\bibinfo {author} {\bibfnamefont {S.-B.}\ \bibnamefont
  {Zhang}}, \bibinfo {author} {\bibfnamefont {M.~M.}\ \bibnamefont {Denner}},
  \bibinfo {author} {\bibfnamefont {T.~c.~v.}\ \bibnamefont
  {Bzdu\ifmmode~\check{s}\else \v{s}\fi{}ek}}, \bibinfo {author} {\bibfnamefont
  {M.~A.}\ \bibnamefont {Sentef}},\ and\ \bibinfo {author} {\bibfnamefont
  {T.}~\bibnamefont {Neupert}},\ }\bibfield  {title} {\bibinfo {title}
  {Symmetry breaking and spectral structure of the interacting hatano-nelson
  model},\ }\href {https://doi.org/10.1103/PhysRevB.106.L121102} {\bibfield
  {journal} {\bibinfo  {journal} {Phys. Rev. B}\ }\textbf {\bibinfo {volume}
  {106}},\ \bibinfo {pages} {L121102} (\bibinfo {year}
  {2022}{\natexlab{b}})}\BibitemShut {NoStop}%
\bibitem [{\citenamefont {Qin}\ \emph {et~al.}(2023)\citenamefont {Qin},
  \citenamefont {Shen},\ and\ \citenamefont {Lee}}]{qin2023non}%
  \BibitemOpen
  \bibfield  {author} {\bibinfo {author} {\bibfnamefont {F.}~\bibnamefont
  {Qin}}, \bibinfo {author} {\bibfnamefont {R.}~\bibnamefont {Shen}},\ and\
  \bibinfo {author} {\bibfnamefont {C.~H.}\ \bibnamefont {Lee}},\ }\bibfield
  {title} {\bibinfo {title} {Non-hermitian squeezed polarons},\ }\href@noop {}
  {\bibfield  {journal} {\bibinfo  {journal} {Physical Review A}\ }\textbf
  {\bibinfo {volume} {107}},\ \bibinfo {pages} {L010202} (\bibinfo {year}
  {2023})}\BibitemShut {NoStop}%
\bibitem [{\citenamefont {Shen}\ and\ \citenamefont {Lee}(2022)}]{shen2022non}%
  \BibitemOpen
  \bibfield  {author} {\bibinfo {author} {\bibfnamefont {R.}~\bibnamefont
  {Shen}}\ and\ \bibinfo {author} {\bibfnamefont {C.~H.}\ \bibnamefont {Lee}},\
  }\bibfield  {title} {\bibinfo {title} {Non-hermitian skin clusters from
  strong interactions},\ }\href@noop {} {\bibfield  {journal} {\bibinfo
  {journal} {Communications Physics}\ }\textbf {\bibinfo {volume} {5}},\
  \bibinfo {pages} {238} (\bibinfo {year} {2022})}\BibitemShut {NoStop}%
\bibitem [{\citenamefont {Li}\ \emph {et~al.}(2023)\citenamefont {Li},
  \citenamefont {Wu}, \citenamefont {Zheng},\ and\ \citenamefont
  {Yi}}]{2023PhysRevResearch.5.033173}%
  \BibitemOpen
  \bibfield  {author} {\bibinfo {author} {\bibfnamefont {H.}~\bibnamefont
  {Li}}, \bibinfo {author} {\bibfnamefont {H.}~\bibnamefont {Wu}}, \bibinfo
  {author} {\bibfnamefont {W.}~\bibnamefont {Zheng}},\ and\ \bibinfo {author}
  {\bibfnamefont {W.}~\bibnamefont {Yi}},\ }\bibfield  {title} {\bibinfo
  {title} {Many-body non-hermitian skin effect under dynamic gauge coupling},\
  }\href {https://doi.org/10.1103/PhysRevResearch.5.033173} {\bibfield
  {journal} {\bibinfo  {journal} {Phys. Rev. Res.}\ }\textbf {\bibinfo {volume}
  {5}},\ \bibinfo {pages} {033173} (\bibinfo {year} {2023})}\BibitemShut
  {NoStop}%
\bibitem [{\citenamefont {Shen}\ \emph
  {et~al.}(2024{\natexlab{a}})\citenamefont {Shen}, \citenamefont {Qin},
  \citenamefont {Desaules}, \citenamefont {Papi\ifmmode~\acute{c}\else
  \'{c}\fi{}},\ and\ \citenamefont {Lee}}]{2024hysRevLett.133.216601}%
  \BibitemOpen
  \bibfield  {author} {\bibinfo {author} {\bibfnamefont {R.}~\bibnamefont
  {Shen}}, \bibinfo {author} {\bibfnamefont {F.}~\bibnamefont {Qin}}, \bibinfo
  {author} {\bibfnamefont {J.-Y.}\ \bibnamefont {Desaules}}, \bibinfo {author}
  {\bibfnamefont {Z.}~\bibnamefont {Papi\ifmmode~\acute{c}\else \'{c}\fi{}}},\
  and\ \bibinfo {author} {\bibfnamefont {C.~H.}\ \bibnamefont {Lee}},\
  }\bibfield  {title} {\bibinfo {title} {Enhanced many-body quantum scars from
  the non-hermitian fock skin effect},\ }\href
  {https://doi.org/10.1103/PhysRevLett.133.216601} {\bibfield  {journal}
  {\bibinfo  {journal} {Phys. Rev. Lett.}\ }\textbf {\bibinfo {volume} {133}},\
  \bibinfo {pages} {216601} (\bibinfo {year} {2024}{\natexlab{a}})}\BibitemShut
  {NoStop}%
\bibitem [{\citenamefont {Gliozzi}\ \emph {et~al.}(2024)\citenamefont
  {Gliozzi}, \citenamefont {De~Tomasi},\ and\ \citenamefont
  {Hughes}}]{2024PhysRevLett.133.136503}%
  \BibitemOpen
  \bibfield  {author} {\bibinfo {author} {\bibfnamefont {J.}~\bibnamefont
  {Gliozzi}}, \bibinfo {author} {\bibfnamefont {G.}~\bibnamefont {De~Tomasi}},\
  and\ \bibinfo {author} {\bibfnamefont {T.~L.}\ \bibnamefont {Hughes}},\
  }\bibfield  {title} {\bibinfo {title} {Many-body non-hermitian skin effect
  for multipoles},\ }\href {https://doi.org/10.1103/PhysRevLett.133.136503}
  {\bibfield  {journal} {\bibinfo  {journal} {Phys. Rev. Lett.}\ }\textbf
  {\bibinfo {volume} {133}},\ \bibinfo {pages} {136503} (\bibinfo {year}
  {2024})}\BibitemShut {NoStop}%
\bibitem [{\citenamefont {Shimomura}\ and\ \citenamefont
  {Sato}(2024)}]{2024PhysRevLett.133.136502}%
  \BibitemOpen
  \bibfield  {author} {\bibinfo {author} {\bibfnamefont {K.}~\bibnamefont
  {Shimomura}}\ and\ \bibinfo {author} {\bibfnamefont {M.}~\bibnamefont
  {Sato}},\ }\bibfield  {title} {\bibinfo {title} {General criterion for
  non-hermitian skin effects and application: Fock space skin effects in
  many-body systems},\ }\href {https://doi.org/10.1103/PhysRevLett.133.136502}
  {\bibfield  {journal} {\bibinfo  {journal} {Phys. Rev. Lett.}\ }\textbf
  {\bibinfo {volume} {133}},\ \bibinfo {pages} {136502} (\bibinfo {year}
  {2024})}\BibitemShut {NoStop}%
\bibitem [{\citenamefont {Qin}\ and\ \citenamefont
  {Li}(2024)}]{2024PhysRevLett.132.096501}%
  \BibitemOpen
  \bibfield  {author} {\bibinfo {author} {\bibfnamefont {Y.}~\bibnamefont
  {Qin}}\ and\ \bibinfo {author} {\bibfnamefont {L.}~\bibnamefont {Li}},\
  }\bibfield  {title} {\bibinfo {title} {Occupation-dependent particle
  separation in one-dimensional non-hermitian lattices},\ }\href
  {https://doi.org/10.1103/PhysRevLett.132.096501} {\bibfield  {journal}
  {\bibinfo  {journal} {Phys. Rev. Lett.}\ }\textbf {\bibinfo {volume} {132}},\
  \bibinfo {pages} {096501} (\bibinfo {year} {2024})}\BibitemShut {NoStop}%
\bibitem [{\citenamefont {Hamanaka}\ and\ \citenamefont
  {Kawabata}(2025)}]{2025PhysRevB.111.035144}%
  \BibitemOpen
  \bibfield  {author} {\bibinfo {author} {\bibfnamefont {S.}~\bibnamefont
  {Hamanaka}}\ and\ \bibinfo {author} {\bibfnamefont {K.}~\bibnamefont
  {Kawabata}},\ }\bibfield  {title} {\bibinfo {title} {Multifractality of the
  many-body non-hermitian skin effect},\ }\href
  {https://doi.org/10.1103/PhysRevB.111.035144} {\bibfield  {journal} {\bibinfo
   {journal} {Phys. Rev. B}\ }\textbf {\bibinfo {volume} {111}},\ \bibinfo
  {pages} {035144} (\bibinfo {year} {2025})}\BibitemShut {NoStop}%
\bibitem [{\citenamefont {Wang}\ and\ \citenamefont
  {Li}(2025{\natexlab{b}})}]{cpl_42_3_037301}%
  \BibitemOpen
  \bibfield  {author} {\bibinfo {author} {\bibfnamefont {Y.-A.}\ \bibnamefont
  {Wang}}\ and\ \bibinfo {author} {\bibfnamefont {L.}~\bibnamefont {Li}},\
  }\bibfield  {title} {\bibinfo {title} {Non-hermitian skin effects in
  fragmented hilbert spaces of one-dimensional fermionic lattices},\ }\href
  {https://doi.org/10.1088/0256-307X/42/3/037301} {\bibfield  {journal}
  {\bibinfo  {journal} {Chin. Phys. Lett.}\ }\textbf {\bibinfo {volume} {42}},\
  \bibinfo {pages} {037301} (\bibinfo {year} {2025}{\natexlab{b}})}\BibitemShut
  {NoStop}%
\bibitem [{\citenamefont {Hu}\ \emph {et~al.}(2025{\natexlab{a}})\citenamefont
  {Hu}, \citenamefont {Wang}, \citenamefont {Lian},\ and\ \citenamefont
  {Wang}}]{2025wztw-l8wg}%
  \BibitemOpen
  \bibfield  {author} {\bibinfo {author} {\bibfnamefont {Y.-M.}\ \bibnamefont
  {Hu}}, \bibinfo {author} {\bibfnamefont {Z.}~\bibnamefont {Wang}}, \bibinfo
  {author} {\bibfnamefont {B.}~\bibnamefont {Lian}},\ and\ \bibinfo {author}
  {\bibfnamefont {Z.}~\bibnamefont {Wang}},\ }\bibfield  {title} {\bibinfo
  {title} {Many-body non-hermitian skin effect with exact steady states in the
  dissipative quantum link model},\ }\href {https://doi.org/10.1103/wztw-l8wg}
  {\bibfield  {journal} {\bibinfo  {journal} {Phys. Rev. Lett.}\ }\textbf
  {\bibinfo {volume} {135}},\ \bibinfo {pages} {260401} (\bibinfo {year}
  {2025}{\natexlab{a}})}\BibitemShut {NoStop}%
\bibitem [{\citenamefont {Hu}\ \emph {et~al.}(2025{\natexlab{b}})\citenamefont
  {Hu}, \citenamefont {Wang}, \citenamefont {Lian},\ and\ \citenamefont
  {Wang}}]{hu2025many}%
  \BibitemOpen
  \bibfield  {author} {\bibinfo {author} {\bibfnamefont {Y.-M.}\ \bibnamefont
  {Hu}}, \bibinfo {author} {\bibfnamefont {Z.}~\bibnamefont {Wang}}, \bibinfo
  {author} {\bibfnamefont {B.}~\bibnamefont {Lian}},\ and\ \bibinfo {author}
  {\bibfnamefont {Z.}~\bibnamefont {Wang}},\ }\bibfield  {title} {\bibinfo
  {title} {Many-body non-hermitian skin effect with exact steady states in the
  dissipative quantum link model},\ }\href@noop {} {\bibfield  {journal}
  {\bibinfo  {journal} {Physical Review Letters}\ }\textbf {\bibinfo {volume}
  {135}},\ \bibinfo {pages} {260401} (\bibinfo {year}
  {2025}{\natexlab{b}})}\BibitemShut {NoStop}%
\bibitem [{\citenamefont {Wang}\ \emph {et~al.}(2026)\citenamefont {Wang},
  \citenamefont {Zhang}, \citenamefont {Yang},\ and\ \citenamefont
  {Wu}}]{2026bhpz17d2}%
  \BibitemOpen
  \bibfield  {author} {\bibinfo {author} {\bibfnamefont {Y.}~\bibnamefont
  {Wang}}, \bibinfo {author} {\bibfnamefont {X.}~\bibnamefont {Zhang}},
  \bibinfo {author} {\bibfnamefont {Z.}~\bibnamefont {Yang}},\ and\ \bibinfo
  {author} {\bibfnamefont {C.}~\bibnamefont {Wu}},\ }\bibfield  {title}
  {\bibinfo {title} {Explicit wave function of the interacting non-hermitian
  spin-$1/2$ 1d system},\ }\href {https://doi.org/10.1103/bhpz-17d2} {\bibfield
   {journal} {\bibinfo  {journal} {Phys. Rev. Lett.}\ }\textbf {\bibinfo
  {volume} {136}},\ \bibinfo {pages} {036501} (\bibinfo {year}
  {2026})}\BibitemShut {NoStop}%
\bibitem [{\citenamefont {Hao}\ \emph {et~al.}(2025)\citenamefont {Hao},
  \citenamefont {Chan},\ and\ \citenamefont {Lee}}]{hao2025interacting}%
  \BibitemOpen
  \bibfield  {author} {\bibinfo {author} {\bibfnamefont {Z.}~\bibnamefont
  {Hao}}, \bibinfo {author} {\bibfnamefont {W.~J.}\ \bibnamefont {Chan}},\ and\
  \bibinfo {author} {\bibfnamefont {C.~H.}\ \bibnamefont {Lee}},\ }\bibfield
  {title} {\bibinfo {title} {Interacting many-body non-hermitian systems as
  markov chains},\ }\href@noop {} {\bibfield  {journal} {\bibinfo  {journal}
  {arXiv preprint arXiv:2509.05411}\ } (\bibinfo {year} {2025})}\BibitemShut
  {NoStop}%
\bibitem [{\citenamefont {Koh}\ \emph {et~al.}(2025)\citenamefont {Koh},
  \citenamefont {Xue}, \citenamefont {Tai}, \citenamefont {Koh},\ and\
  \citenamefont {Lee}}]{koh2025interacting}%
  \BibitemOpen
  \bibfield  {author} {\bibinfo {author} {\bibfnamefont {J.~M.}\ \bibnamefont
  {Koh}}, \bibinfo {author} {\bibfnamefont {W.-T.}\ \bibnamefont {Xue}},
  \bibinfo {author} {\bibfnamefont {T.}~\bibnamefont {Tai}}, \bibinfo {author}
  {\bibfnamefont {D.~E.}\ \bibnamefont {Koh}},\ and\ \bibinfo {author}
  {\bibfnamefont {C.~H.}\ \bibnamefont {Lee}},\ }\bibfield  {title} {\bibinfo
  {title} {Interacting non-hermitian edge and cluster bursts on a digital
  quantum processor},\ }\href@noop {} {\bibfield  {journal} {\bibinfo
  {journal} {arXiv preprint arXiv:2503.14595}\ } (\bibinfo {year}
  {2025})}\BibitemShut {NoStop}%
\bibitem [{\citenamefont {Qin}\ \emph {et~al.}(2025)\citenamefont {Qin},
  \citenamefont {Lee},\ and\ \citenamefont {Li}}]{Qin2025AnyonNHSE}%
  \BibitemOpen
  \bibfield  {author} {\bibinfo {author} {\bibfnamefont {Y.}~\bibnamefont
  {Qin}}, \bibinfo {author} {\bibfnamefont {C.~H.}\ \bibnamefont {Lee}},\ and\
  \bibinfo {author} {\bibfnamefont {L.}~\bibnamefont {Li}},\ }\bibfield
  {title} {\bibinfo {title} {Dynamical suppression of many-body non-hermitian
  skin effect in anyonic systems},\ }\href
  {https://doi.org/10.1038/s42005-025-01935-3} {\bibfield  {journal} {\bibinfo
  {journal} {Communications Physics}\ }\textbf {\bibinfo {volume} {8}},\
  \bibinfo {pages} {18} (\bibinfo {year} {2025})}\BibitemShut {NoStop}%
\bibitem [{\citenamefont {Ekman}\ \emph {et~al.}(2026)\citenamefont {Ekman},
  \citenamefont {Bergholtz},\ and\ \citenamefont
  {Molignini}}]{Ekman2026symmetry}%
  \BibitemOpen
  \bibfield  {author} {\bibinfo {author} {\bibfnamefont {C.}~\bibnamefont
  {Ekman}}, \bibinfo {author} {\bibfnamefont {E.~J.}\ \bibnamefont
  {Bergholtz}},\ and\ \bibinfo {author} {\bibfnamefont {P.}~\bibnamefont
  {Molignini}},\ }\bibfield  {title} {\bibinfo {title} {Symmetry-fractionalized
  skin effects in non-hermitian luttinger liquids},\ }\href@noop {} {\bibfield
  {journal} {\bibinfo  {journal} {arXiv preprint arXiv:2603.28849}\ } (\bibinfo
  {year} {2026})}\BibitemShut {NoStop}%
\bibitem [{\citenamefont {Qin}\ \emph {et~al.}(2026)\citenamefont {Qin},
  \citenamefont {Ang}, \citenamefont {Lee},\ and\ \citenamefont
  {Li}}]{Qin2026MBNHSE}%
  \BibitemOpen
  \bibfield  {author} {\bibinfo {author} {\bibfnamefont {Y.}~\bibnamefont
  {Qin}}, \bibinfo {author} {\bibfnamefont {Y.~S.}\ \bibnamefont {Ang}},
  \bibinfo {author} {\bibfnamefont {C.~H.}\ \bibnamefont {Lee}},\ and\ \bibinfo
  {author} {\bibfnamefont {L.}~\bibnamefont {Li}},\ }\bibfield  {title}
  {\bibinfo {title} {Many-body critical non-hermitian skin effect},\ }\href
  {https://doi.org/10.1038/s42005-025-02448-9} {\bibfield  {journal} {\bibinfo
  {journal} {Communications Physics}\ }\textbf {\bibinfo {volume} {9}},\
  \bibinfo {pages} {16} (\bibinfo {year} {2026})}\BibitemShut {NoStop}%
\bibitem [{\citenamefont {Pu}\ \emph {et~al.}(2026)\citenamefont {Pu},
  \citenamefont {Xi}, \citenamefont {Tang}, \citenamefont {Lu}, \citenamefont
  {Deng}, \citenamefont {Cheng}, \citenamefont {Ke}, \citenamefont {Chen},\
  and\ \citenamefont {Liu}}]{xr3h-j9pv2026}%
  \BibitemOpen
  \bibfield  {author} {\bibinfo {author} {\bibfnamefont {Z.}~\bibnamefont
  {Pu}}, \bibinfo {author} {\bibfnamefont {Y.}~\bibnamefont {Xi}}, \bibinfo
  {author} {\bibfnamefont {Y.}~\bibnamefont {Tang}}, \bibinfo {author}
  {\bibfnamefont {J.}~\bibnamefont {Lu}}, \bibinfo {author} {\bibfnamefont
  {W.}~\bibnamefont {Deng}}, \bibinfo {author} {\bibfnamefont {H.}~\bibnamefont
  {Cheng}}, \bibinfo {author} {\bibfnamefont {M.}~\bibnamefont {Ke}}, \bibinfo
  {author} {\bibfnamefont {S.}~\bibnamefont {Chen}},\ and\ \bibinfo {author}
  {\bibfnamefont {Z.}~\bibnamefont {Liu}},\ }\bibfield  {title} {\bibinfo
  {title} {Acoustic bound pairs under nonreciprocal two-body interactions},\
  }\href {https://doi.org/10.1103/xr3h-j9pv} {\bibfield  {journal} {\bibinfo
  {journal} {Phys. Rev. Lett.}\ }\textbf {\bibinfo {volume} {136}},\ \bibinfo
  {pages} {146603} (\bibinfo {year} {2026})}\BibitemShut {NoStop}%
\bibitem [{\citenamefont {Lee}(2021)}]{2021PhysRevB.104.195102}%
  \BibitemOpen
  \bibfield  {author} {\bibinfo {author} {\bibfnamefont {C.~H.}\ \bibnamefont
  {Lee}},\ }\bibfield  {title} {\bibinfo {title} {Many-body topological and
  skin states without open boundaries},\ }\href
  {https://doi.org/10.1103/PhysRevB.104.195102} {\bibfield  {journal} {\bibinfo
   {journal} {Phys. Rev. B}\ }\textbf {\bibinfo {volume} {104}},\ \bibinfo
  {pages} {195102} (\bibinfo {year} {2021})}\BibitemShut {NoStop}%
\bibitem [{\citenamefont {Xu}\ \emph {et~al.}(2021)\citenamefont {Xu},
  \citenamefont {Xu}, \citenamefont {Mandal}, \citenamefont {Banerjee},
  \citenamefont {Ghosh},\ and\ \citenamefont {Liew}}]{2021PhysRevB.103.235306}%
  \BibitemOpen
  \bibfield  {author} {\bibinfo {author} {\bibfnamefont {X.}~\bibnamefont
  {Xu}}, \bibinfo {author} {\bibfnamefont {H.}~\bibnamefont {Xu}}, \bibinfo
  {author} {\bibfnamefont {S.}~\bibnamefont {Mandal}}, \bibinfo {author}
  {\bibfnamefont {R.}~\bibnamefont {Banerjee}}, \bibinfo {author}
  {\bibfnamefont {S.}~\bibnamefont {Ghosh}},\ and\ \bibinfo {author}
  {\bibfnamefont {T.~C.~H.}\ \bibnamefont {Liew}},\ }\bibfield  {title}
  {\bibinfo {title} {Interaction-induced double-sided skin effect in an
  exciton-polariton system},\ }\href
  {https://doi.org/10.1103/PhysRevB.103.235306} {\bibfield  {journal} {\bibinfo
   {journal} {Phys. Rev. B}\ }\textbf {\bibinfo {volume} {103}},\ \bibinfo
  {pages} {235306} (\bibinfo {year} {2021})}\BibitemShut {NoStop}%
\bibitem [{\citenamefont {Faugno}\ and\ \citenamefont
  {Ozawa}(2022)}]{2022PhysRevLett.129.180401}%
  \BibitemOpen
  \bibfield  {author} {\bibinfo {author} {\bibfnamefont {W.~N.}\ \bibnamefont
  {Faugno}}\ and\ \bibinfo {author} {\bibfnamefont {T.}~\bibnamefont {Ozawa}},\
  }\bibfield  {title} {\bibinfo {title} {Interaction-induced non-hermitian
  topological phases from a dynamical gauge field},\ }\href
  {https://doi.org/10.1103/PhysRevLett.129.180401} {\bibfield  {journal}
  {\bibinfo  {journal} {Phys. Rev. Lett.}\ }\textbf {\bibinfo {volume} {129}},\
  \bibinfo {pages} {180401} (\bibinfo {year} {2022})}\BibitemShut {NoStop}%
\bibitem [{\citenamefont {Kawabata}\ \emph {et~al.}(2022)\citenamefont
  {Kawabata}, \citenamefont {Shiozaki},\ and\ \citenamefont
  {Ryu}}]{2022PhysRevB.105.165137}%
  \BibitemOpen
  \bibfield  {author} {\bibinfo {author} {\bibfnamefont {K.}~\bibnamefont
  {Kawabata}}, \bibinfo {author} {\bibfnamefont {K.}~\bibnamefont {Shiozaki}},\
  and\ \bibinfo {author} {\bibfnamefont {S.}~\bibnamefont {Ryu}},\ }\bibfield
  {title} {\bibinfo {title} {Many-body topology of non-hermitian systems},\
  }\href {https://doi.org/10.1103/PhysRevB.105.165137} {\bibfield  {journal}
  {\bibinfo  {journal} {Phys. Rev. B}\ }\textbf {\bibinfo {volume} {105}},\
  \bibinfo {pages} {165137} (\bibinfo {year} {2022})}\BibitemShut {NoStop}%
\bibitem [{\citenamefont {Hamanaka}\ \emph {et~al.}(2023)\citenamefont
  {Hamanaka}, \citenamefont {Yamamoto},\ and\ \citenamefont
  {Yoshida}}]{2023PhysRevB.108.155114}%
  \BibitemOpen
  \bibfield  {author} {\bibinfo {author} {\bibfnamefont {S.}~\bibnamefont
  {Hamanaka}}, \bibinfo {author} {\bibfnamefont {K.}~\bibnamefont {Yamamoto}},\
  and\ \bibinfo {author} {\bibfnamefont {T.}~\bibnamefont {Yoshida}},\
  }\bibfield  {title} {\bibinfo {title} {Interaction-induced liouvillian skin
  effect in a fermionic chain with a two-body loss},\ }\href
  {https://doi.org/10.1103/PhysRevB.108.155114} {\bibfield  {journal} {\bibinfo
   {journal} {Phys. Rev. B}\ }\textbf {\bibinfo {volume} {108}},\ \bibinfo
  {pages} {155114} (\bibinfo {year} {2023})}\BibitemShut {NoStop}%
\bibitem [{\citenamefont {Poddubny}(2023)}]{2023PhysRevB.107.045131}%
  \BibitemOpen
  \bibfield  {author} {\bibinfo {author} {\bibfnamefont {A.~N.}\ \bibnamefont
  {Poddubny}},\ }\bibfield  {title} {\bibinfo {title} {Interaction-induced
  analog of a non-hermitian skin effect in a lattice two-body problem},\ }\href
  {https://doi.org/10.1103/PhysRevB.107.045131} {\bibfield  {journal} {\bibinfo
   {journal} {Phys. Rev. B}\ }\textbf {\bibinfo {volume} {107}},\ \bibinfo
  {pages} {045131} (\bibinfo {year} {2023})}\BibitemShut {NoStop}%
\bibitem [{\citenamefont {Kawabata}\ \emph {et~al.}(2023)\citenamefont
  {Kawabata}, \citenamefont {Numasawa},\ and\ \citenamefont
  {Ryu}}]{2023PhysRevX.13.021007}%
  \BibitemOpen
  \bibfield  {author} {\bibinfo {author} {\bibfnamefont {K.}~\bibnamefont
  {Kawabata}}, \bibinfo {author} {\bibfnamefont {T.}~\bibnamefont {Numasawa}},\
  and\ \bibinfo {author} {\bibfnamefont {S.}~\bibnamefont {Ryu}},\ }\bibfield
  {title} {\bibinfo {title} {Entanglement phase transition induced by the
  non-hermitian skin effect},\ }\href
  {https://doi.org/10.1103/PhysRevX.13.021007} {\bibfield  {journal} {\bibinfo
  {journal} {Phys. Rev. X}\ }\textbf {\bibinfo {volume} {13}},\ \bibinfo
  {pages} {021007} (\bibinfo {year} {2023})}\BibitemShut {NoStop}%
\bibitem [{\citenamefont {Wang}\ \emph {et~al.}(2024)\citenamefont {Wang},
  \citenamefont {Fang},\ and\ \citenamefont {Ren}}]{2024PhysRevB.110.035113}%
  \BibitemOpen
  \bibfield  {author} {\bibinfo {author} {\bibfnamefont {Y.-P.}\ \bibnamefont
  {Wang}}, \bibinfo {author} {\bibfnamefont {C.}~\bibnamefont {Fang}},\ and\
  \bibinfo {author} {\bibfnamefont {J.}~\bibnamefont {Ren}},\ }\bibfield
  {title} {\bibinfo {title} {Absence of measurement-induced entanglement
  transition due to feedback-induced skin effect},\ }\href
  {https://doi.org/10.1103/PhysRevB.110.035113} {\bibfield  {journal} {\bibinfo
   {journal} {Phys. Rev. B}\ }\textbf {\bibinfo {volume} {110}},\ \bibinfo
  {pages} {035113} (\bibinfo {year} {2024})}\BibitemShut {NoStop}%
\bibitem [{\citenamefont {Liu}\ \emph {et~al.}(2025{\natexlab{b}})\citenamefont
  {Liu}, \citenamefont {Jiang}, \citenamefont {Xue}, \citenamefont {Li},
  \citenamefont {Gong}, \citenamefont {Liu},\ and\ \citenamefont
  {Lee}}]{liu2025non}%
  \BibitemOpen
  \bibfield  {author} {\bibinfo {author} {\bibfnamefont {S.}~\bibnamefont
  {Liu}}, \bibinfo {author} {\bibfnamefont {H.}~\bibnamefont {Jiang}}, \bibinfo
  {author} {\bibfnamefont {W.-T.}\ \bibnamefont {Xue}}, \bibinfo {author}
  {\bibfnamefont {Q.}~\bibnamefont {Li}}, \bibinfo {author} {\bibfnamefont
  {J.}~\bibnamefont {Gong}}, \bibinfo {author} {\bibfnamefont {X.}~\bibnamefont
  {Liu}},\ and\ \bibinfo {author} {\bibfnamefont {C.~H.}\ \bibnamefont {Lee}},\
  }\bibfield  {title} {\bibinfo {title} {Non-hermitian entanglement dip from
  scaling-induced exceptional criticality},\ }\href@noop {} {\bibfield
  {journal} {\bibinfo  {journal} {Science Bulletin}\ }\textbf {\bibinfo
  {volume} {70}},\ \bibinfo {pages} {2929} (\bibinfo {year}
  {2025}{\natexlab{b}})}\BibitemShut {NoStop}%
\bibitem [{\citenamefont {Xue}\ and\ \citenamefont
  {Lee}(2026)}]{xue2026topologically}%
  \BibitemOpen
  \bibfield  {author} {\bibinfo {author} {\bibfnamefont {W.-T.}\ \bibnamefont
  {Xue}}\ and\ \bibinfo {author} {\bibfnamefont {C.~H.}\ \bibnamefont {Lee}},\
  }\bibfield  {title} {\bibinfo {title} {Topologically protected negative
  entanglement},\ }\href@noop {} {\bibfield  {journal} {\bibinfo  {journal}
  {Advanced Science}\ }\textbf {\bibinfo {volume} {13}},\ \bibinfo {pages}
  {e13868} (\bibinfo {year} {2026})}\BibitemShut {NoStop}%
\bibitem [{\citenamefont {Greiner}\ \emph {et~al.}(2002)\citenamefont
  {Greiner}, \citenamefont {Mandel}, \citenamefont {Esslinger}, \citenamefont
  {H{\"a}nsch},\ and\ \citenamefont {Bloch}}]{Greiner2002}%
  \BibitemOpen
  \bibfield  {author} {\bibinfo {author} {\bibfnamefont {M.}~\bibnamefont
  {Greiner}}, \bibinfo {author} {\bibfnamefont {O.}~\bibnamefont {Mandel}},
  \bibinfo {author} {\bibfnamefont {T.}~\bibnamefont {Esslinger}}, \bibinfo
  {author} {\bibfnamefont {T.~W.}\ \bibnamefont {H{\"a}nsch}},\ and\ \bibinfo
  {author} {\bibfnamefont {I.}~\bibnamefont {Bloch}},\ }\bibfield  {title}
  {\bibinfo {title} {Quantum phase transition from a superfluid to a mott
  insulator in a gas of ultracold atoms},\ }\href
  {https://doi.org/10.1038/415039a} {\bibfield  {journal} {\bibinfo  {journal}
  {Nature}\ }\textbf {\bibinfo {volume} {415}},\ \bibinfo {pages} {39}
  (\bibinfo {year} {2002})}\BibitemShut {NoStop}%
\bibitem [{\citenamefont {Lewenstein}\ \emph {et~al.}(2007)\citenamefont
  {Lewenstein}, \citenamefont {Sanpera}, \citenamefont {Ahufinger},
  \citenamefont {Damski}, \citenamefont {Sen},\ and\ \citenamefont
  {Sen}}]{Lewenstein2007}%
  \BibitemOpen
  \bibfield  {author} {\bibinfo {author} {\bibfnamefont {M.}~\bibnamefont
  {Lewenstein}}, \bibinfo {author} {\bibfnamefont {A.}~\bibnamefont {Sanpera}},
  \bibinfo {author} {\bibfnamefont {V.}~\bibnamefont {Ahufinger}}, \bibinfo
  {author} {\bibfnamefont {B.}~\bibnamefont {Damski}}, \bibinfo {author}
  {\bibfnamefont {A.}~\bibnamefont {Sen}},\ and\ \bibinfo {author}
  {\bibfnamefont {U.}~\bibnamefont {Sen}},\ }\bibfield  {title} {\bibinfo
  {title} {Ultracold atomic gases in optical lattices: Mimicking condensed
  matter physics and beyond},\ }\href
  {https://doi.org/10.1080/00018730701223200} {\bibfield  {journal} {\bibinfo
  {journal} {Advances in Physics}\ }\textbf {\bibinfo {volume} {56}},\ \bibinfo
  {pages} {243} (\bibinfo {year} {2007})}\BibitemShut {NoStop}%
\bibitem [{\citenamefont {Bloch}\ \emph {et~al.}(2008)\citenamefont {Bloch},
  \citenamefont {Dalibard},\ and\ \citenamefont {Zwerger}}]{Bloch2008RMP}%
  \BibitemOpen
  \bibfield  {author} {\bibinfo {author} {\bibfnamefont {I.}~\bibnamefont
  {Bloch}}, \bibinfo {author} {\bibfnamefont {J.}~\bibnamefont {Dalibard}},\
  and\ \bibinfo {author} {\bibfnamefont {W.}~\bibnamefont {Zwerger}},\
  }\bibfield  {title} {\bibinfo {title} {Many-body physics with ultracold
  gases},\ }\href {https://doi.org/10.1103/RevModPhys.80.885} {\bibfield
  {journal} {\bibinfo  {journal} {Rev. Mod. Phys.}\ }\textbf {\bibinfo {volume}
  {80}},\ \bibinfo {pages} {885} (\bibinfo {year} {2008})}\BibitemShut
  {NoStop}%
\bibitem [{\citenamefont {Schreck}\ \emph
  {et~al.}(2001{\natexlab{a}})\citenamefont {Schreck}, \citenamefont {Ferrari},
  \citenamefont {Corwin}, \citenamefont {Cubizolles}, \citenamefont
  {Khaykovich}, \citenamefont {Mewes},\ and\ \citenamefont
  {Salomon}}]{2001PhysRevA.64.011402}%
  \BibitemOpen
  \bibfield  {author} {\bibinfo {author} {\bibfnamefont {F.}~\bibnamefont
  {Schreck}}, \bibinfo {author} {\bibfnamefont {G.}~\bibnamefont {Ferrari}},
  \bibinfo {author} {\bibfnamefont {K.~L.}\ \bibnamefont {Corwin}}, \bibinfo
  {author} {\bibfnamefont {J.}~\bibnamefont {Cubizolles}}, \bibinfo {author}
  {\bibfnamefont {L.}~\bibnamefont {Khaykovich}}, \bibinfo {author}
  {\bibfnamefont {M.-O.}\ \bibnamefont {Mewes}},\ and\ \bibinfo {author}
  {\bibfnamefont {C.}~\bibnamefont {Salomon}},\ }\bibfield  {title} {\bibinfo
  {title} {Sympathetic cooling of bosonic and fermionic lithium gases towards
  quantum degeneracy},\ }\href {https://doi.org/10.1103/PhysRevA.64.011402}
  {\bibfield  {journal} {\bibinfo  {journal} {Phys. Rev. A}\ }\textbf {\bibinfo
  {volume} {64}},\ \bibinfo {pages} {011402} (\bibinfo {year}
  {2001}{\natexlab{a}})}\BibitemShut {NoStop}%
\bibitem [{\citenamefont {Schreck}\ \emph
  {et~al.}(2001{\natexlab{b}})\citenamefont {Schreck}, \citenamefont
  {Khaykovich}, \citenamefont {Corwin}, \citenamefont {Ferrari}, \citenamefont
  {Bourdel}, \citenamefont {Cubizolles},\ and\ \citenamefont
  {Salomon}}]{2001PhysRevLett.87.080403}%
  \BibitemOpen
  \bibfield  {author} {\bibinfo {author} {\bibfnamefont {F.}~\bibnamefont
  {Schreck}}, \bibinfo {author} {\bibfnamefont {L.}~\bibnamefont {Khaykovich}},
  \bibinfo {author} {\bibfnamefont {K.~L.}\ \bibnamefont {Corwin}}, \bibinfo
  {author} {\bibfnamefont {G.}~\bibnamefont {Ferrari}}, \bibinfo {author}
  {\bibfnamefont {T.}~\bibnamefont {Bourdel}}, \bibinfo {author} {\bibfnamefont
  {J.}~\bibnamefont {Cubizolles}},\ and\ \bibinfo {author} {\bibfnamefont
  {C.}~\bibnamefont {Salomon}},\ }\bibfield  {title} {\bibinfo {title}
  {Quasipure bose-einstein condensate immersed in a fermi sea},\ }\href
  {https://doi.org/10.1103/PhysRevLett.87.080403} {\bibfield  {journal}
  {\bibinfo  {journal} {Phys. Rev. Lett.}\ }\textbf {\bibinfo {volume} {87}},\
  \bibinfo {pages} {080403} (\bibinfo {year} {2001}{\natexlab{b}})}\BibitemShut
  {NoStop}%
\bibitem [{\citenamefont {Albus}\ \emph {et~al.}(2002)\citenamefont {Albus},
  \citenamefont {Gardiner}, \citenamefont {Illuminati},\ and\ \citenamefont
  {Wilkens}}]{2002PhysRevA.65.053607}%
  \BibitemOpen
  \bibfield  {author} {\bibinfo {author} {\bibfnamefont {A.~P.}\ \bibnamefont
  {Albus}}, \bibinfo {author} {\bibfnamefont {S.~A.}\ \bibnamefont {Gardiner}},
  \bibinfo {author} {\bibfnamefont {F.}~\bibnamefont {Illuminati}},\ and\
  \bibinfo {author} {\bibfnamefont {M.}~\bibnamefont {Wilkens}},\ }\bibfield
  {title} {\bibinfo {title} {Quantum field theory of dilute homogeneous
  bose-fermi mixtures at zero temperature: General formalism and beyond
  mean-field corrections},\ }\href {https://doi.org/10.1103/PhysRevA.65.053607}
  {\bibfield  {journal} {\bibinfo  {journal} {Phys. Rev. A}\ }\textbf {\bibinfo
  {volume} {65}},\ \bibinfo {pages} {053607} (\bibinfo {year}
  {2002})}\BibitemShut {NoStop}%
\bibitem [{\citenamefont {Santamore}\ and\ \citenamefont
  {Timmermans}(2008)}]{2008PhysRevA.78.013619}%
  \BibitemOpen
  \bibfield  {author} {\bibinfo {author} {\bibfnamefont {D.~H.}\ \bibnamefont
  {Santamore}}\ and\ \bibinfo {author} {\bibfnamefont {E.}~\bibnamefont
  {Timmermans}},\ }\bibfield  {title} {\bibinfo {title} {Fermion-mediated
  interactions in a dilute bose-einstein condensate},\ }\href
  {https://doi.org/10.1103/PhysRevA.78.013619} {\bibfield  {journal} {\bibinfo
  {journal} {Phys. Rev. A}\ }\textbf {\bibinfo {volume} {78}},\ \bibinfo
  {pages} {013619} (\bibinfo {year} {2008})}\BibitemShut {NoStop}%
\bibitem [{\citenamefont {Nishida}(2010)}]{2010PhysRevA.82.011605}%
  \BibitemOpen
  \bibfield  {author} {\bibinfo {author} {\bibfnamefont {Y.}~\bibnamefont
  {Nishida}},\ }\bibfield  {title} {\bibinfo {title} {Phases of a bilayer fermi
  gas},\ }\href {https://doi.org/10.1103/PhysRevA.82.011605} {\bibfield
  {journal} {\bibinfo  {journal} {Phys. Rev. A}\ }\textbf {\bibinfo {volume}
  {82}},\ \bibinfo {pages} {011605} (\bibinfo {year} {2010})}\BibitemShut
  {NoStop}%
\bibitem [{\citenamefont {Yu}\ and\ \citenamefont
  {Pethick}(2012)}]{2012PhysRevA.85.063616}%
  \BibitemOpen
  \bibfield  {author} {\bibinfo {author} {\bibfnamefont {Z.}~\bibnamefont
  {Yu}}\ and\ \bibinfo {author} {\bibfnamefont {C.~J.}\ \bibnamefont
  {Pethick}},\ }\bibfield  {title} {\bibinfo {title} {Induced interactions in
  dilute atomic gases and liquid helium mixtures},\ }\href
  {https://doi.org/10.1103/PhysRevA.85.063616} {\bibfield  {journal} {\bibinfo
  {journal} {Phys. Rev. A}\ }\textbf {\bibinfo {volume} {85}},\ \bibinfo
  {pages} {063616} (\bibinfo {year} {2012})}\BibitemShut {NoStop}%
\bibitem [{\citenamefont {Li}\ \emph {et~al.}(2020)\citenamefont {Li},
  \citenamefont {Lee},\ and\ \citenamefont {Gong}}]{li2020topological}%
  \BibitemOpen
  \bibfield  {author} {\bibinfo {author} {\bibfnamefont {L.}~\bibnamefont
  {Li}}, \bibinfo {author} {\bibfnamefont {C.~H.}\ \bibnamefont {Lee}},\ and\
  \bibinfo {author} {\bibfnamefont {J.}~\bibnamefont {Gong}},\ }\bibfield
  {title} {\bibinfo {title} {Topological switch for non-hermitian skin effect
  in cold-atom systems with loss},\ }\href@noop {} {\bibfield  {journal}
  {\bibinfo  {journal} {Physical review letters}\ }\textbf {\bibinfo {volume}
  {124}},\ \bibinfo {pages} {250402} (\bibinfo {year} {2020})}\BibitemShut
  {NoStop}%
\bibitem [{\citenamefont {Ferrier-Barbut}\ \emph {et~al.}(2014)\citenamefont
  {Ferrier-Barbut}, \citenamefont {Delehaye}, \citenamefont {Laurent},
  \citenamefont {Grier}, \citenamefont {Pierce}, \citenamefont {Rem},
  \citenamefont {Chevy},\ and\ \citenamefont {Salomon}}]{Ferrier-Barbut2014}%
  \BibitemOpen
  \bibfield  {author} {\bibinfo {author} {\bibfnamefont {I.}~\bibnamefont
  {Ferrier-Barbut}}, \bibinfo {author} {\bibfnamefont {M.}~\bibnamefont
  {Delehaye}}, \bibinfo {author} {\bibfnamefont {S.}~\bibnamefont {Laurent}},
  \bibinfo {author} {\bibfnamefont {A.~T.}\ \bibnamefont {Grier}}, \bibinfo
  {author} {\bibfnamefont {M.}~\bibnamefont {Pierce}}, \bibinfo {author}
  {\bibfnamefont {B.~S.}\ \bibnamefont {Rem}}, \bibinfo {author} {\bibfnamefont
  {F.}~\bibnamefont {Chevy}},\ and\ \bibinfo {author} {\bibfnamefont
  {C.}~\bibnamefont {Salomon}},\ }\bibfield  {title} {\bibinfo {title} {A
  mixture of bose and fermi superfluids},\ }\href
  {https://doi.org/10.1126/science.1255380} {\bibfield  {journal} {\bibinfo
  {journal} {Science}\ }\textbf {\bibinfo {volume} {345}},\ \bibinfo {pages}
  {1035} (\bibinfo {year} {2014})}\BibitemShut {NoStop}%
\bibitem [{\citenamefont {Kinnunen}\ and\ \citenamefont
  {Bruun}(2015)}]{2015PhysRevA.91.041605}%
  \BibitemOpen
  \bibfield  {author} {\bibinfo {author} {\bibfnamefont {J.~J.}\ \bibnamefont
  {Kinnunen}}\ and\ \bibinfo {author} {\bibfnamefont {G.~M.}\ \bibnamefont
  {Bruun}},\ }\bibfield  {title} {\bibinfo {title} {Induced interactions in a
  superfluid bose-fermi mixture},\ }\href
  {https://doi.org/10.1103/PhysRevA.91.041605} {\bibfield  {journal} {\bibinfo
  {journal} {Phys. Rev. A}\ }\textbf {\bibinfo {volume} {91}},\ \bibinfo
  {pages} {041605} (\bibinfo {year} {2015})}\BibitemShut {NoStop}%
\bibitem [{\citenamefont {Onofrio}(2016)}]{Onofrio2016Cooling}%
  \BibitemOpen
  \bibfield  {author} {\bibinfo {author} {\bibfnamefont {R.}~\bibnamefont
  {Onofrio}},\ }\bibfield  {title} {\bibinfo {title} {Cooling and thermometry
  of atomic fermi gases},\ }\href {https://doi.org/10.3367/UFNe.2016.07.037873}
  {\bibfield  {journal} {\bibinfo  {journal} {Physics-Uspekhi}\ }\textbf
  {\bibinfo {volume} {59}},\ \bibinfo {pages} {1129} (\bibinfo {year}
  {2016})}\BibitemShut {NoStop}%
\bibitem [{\citenamefont {Yao}\ \emph {et~al.}(2016)\citenamefont {Yao},
  \citenamefont {Chen}, \citenamefont {Wu}, \citenamefont {Liu}, \citenamefont
  {Wang}, \citenamefont {Jiang}, \citenamefont {Deng}, \citenamefont {Chen},\
  and\ \citenamefont {Pan}}]{2016PhysRevLett.117.145301}%
  \BibitemOpen
  \bibfield  {author} {\bibinfo {author} {\bibfnamefont {X.-C.}\ \bibnamefont
  {Yao}}, \bibinfo {author} {\bibfnamefont {H.-Z.}\ \bibnamefont {Chen}},
  \bibinfo {author} {\bibfnamefont {Y.-P.}\ \bibnamefont {Wu}}, \bibinfo
  {author} {\bibfnamefont {X.-P.}\ \bibnamefont {Liu}}, \bibinfo {author}
  {\bibfnamefont {X.-Q.}\ \bibnamefont {Wang}}, \bibinfo {author}
  {\bibfnamefont {X.}~\bibnamefont {Jiang}}, \bibinfo {author} {\bibfnamefont
  {Y.}~\bibnamefont {Deng}}, \bibinfo {author} {\bibfnamefont {Y.-A.}\
  \bibnamefont {Chen}},\ and\ \bibinfo {author} {\bibfnamefont {J.-W.}\
  \bibnamefont {Pan}},\ }\bibfield  {title} {\bibinfo {title} {Observation of
  coupled vortex lattices in a mass-imbalance bose and fermi superfluid
  mixture},\ }\href {https://doi.org/10.1103/PhysRevLett.117.145301} {\bibfield
   {journal} {\bibinfo  {journal} {Phys. Rev. Lett.}\ }\textbf {\bibinfo
  {volume} {117}},\ \bibinfo {pages} {145301} (\bibinfo {year}
  {2016})}\BibitemShut {NoStop}%
\bibitem [{\citenamefont {Wu}\ and\ \citenamefont
  {Bruun}(2016)}]{2016PhysRevLett.117.245302}%
  \BibitemOpen
  \bibfield  {author} {\bibinfo {author} {\bibfnamefont {Z.}~\bibnamefont
  {Wu}}\ and\ \bibinfo {author} {\bibfnamefont {G.~M.}\ \bibnamefont {Bruun}},\
  }\bibfield  {title} {\bibinfo {title} {Topological superfluid in a fermi-bose
  mixture with a high critical temperature},\ }\href
  {https://doi.org/10.1103/PhysRevLett.117.245302} {\bibfield  {journal}
  {\bibinfo  {journal} {Phys. Rev. Lett.}\ }\textbf {\bibinfo {volume} {117}},\
  \bibinfo {pages} {245302} (\bibinfo {year} {2016})}\BibitemShut {NoStop}%
\bibitem [{\citenamefont {Roy}\ \emph {et~al.}(2017)\citenamefont {Roy},
  \citenamefont {Green}, \citenamefont {Bowler},\ and\ \citenamefont
  {Gupta}}]{2017PhysRevLett.118.055301}%
  \BibitemOpen
  \bibfield  {author} {\bibinfo {author} {\bibfnamefont {R.}~\bibnamefont
  {Roy}}, \bibinfo {author} {\bibfnamefont {A.}~\bibnamefont {Green}}, \bibinfo
  {author} {\bibfnamefont {R.}~\bibnamefont {Bowler}},\ and\ \bibinfo {author}
  {\bibfnamefont {S.}~\bibnamefont {Gupta}},\ }\bibfield  {title} {\bibinfo
  {title} {Two-element mixture of bose and fermi superfluids},\ }\href
  {https://doi.org/10.1103/PhysRevLett.118.055301} {\bibfield  {journal}
  {\bibinfo  {journal} {Phys. Rev. Lett.}\ }\textbf {\bibinfo {volume} {118}},\
  \bibinfo {pages} {055301} (\bibinfo {year} {2017})}\BibitemShut {NoStop}%
\bibitem [{\citenamefont {Suchet}\ \emph {et~al.}(2017)\citenamefont {Suchet},
  \citenamefont {Wu}, \citenamefont {Chevy},\ and\ \citenamefont
  {Bruun}}]{2017PhysRevA.95.043643}%
  \BibitemOpen
  \bibfield  {author} {\bibinfo {author} {\bibfnamefont {D.}~\bibnamefont
  {Suchet}}, \bibinfo {author} {\bibfnamefont {Z.}~\bibnamefont {Wu}}, \bibinfo
  {author} {\bibfnamefont {F.}~\bibnamefont {Chevy}},\ and\ \bibinfo {author}
  {\bibfnamefont {G.~M.}\ \bibnamefont {Bruun}},\ }\bibfield  {title} {\bibinfo
  {title} {Long-range mediated interactions in a mixed-dimensional system},\
  }\href {https://doi.org/10.1103/PhysRevA.95.043643} {\bibfield  {journal}
  {\bibinfo  {journal} {Phys. Rev. A}\ }\textbf {\bibinfo {volume} {95}},\
  \bibinfo {pages} {043643} (\bibinfo {year} {2017})}\BibitemShut {NoStop}%
\bibitem [{\citenamefont {Camacho-Guardian}\ and\ \citenamefont
  {Bruun}(2018)}]{2018PhysRevX.8.031042}%
  \BibitemOpen
  \bibfield  {author} {\bibinfo {author} {\bibfnamefont {A.}~\bibnamefont
  {Camacho-Guardian}}\ and\ \bibinfo {author} {\bibfnamefont {G.~M.}\
  \bibnamefont {Bruun}},\ }\bibfield  {title} {\bibinfo {title} {Landau
  effective interaction between quasiparticles in a bose-einstein condensate},\
  }\href {https://doi.org/10.1103/PhysRevX.8.031042} {\bibfield  {journal}
  {\bibinfo  {journal} {Phys. Rev. X}\ }\textbf {\bibinfo {volume} {8}},\
  \bibinfo {pages} {031042} (\bibinfo {year} {2018})}\BibitemShut {NoStop}%
\bibitem [{\citenamefont {Kinnunen}\ \emph {et~al.}(2018)\citenamefont
  {Kinnunen}, \citenamefont {Wu},\ and\ \citenamefont
  {Bruun}}]{2018PhysRevLett.121.253402}%
  \BibitemOpen
  \bibfield  {author} {\bibinfo {author} {\bibfnamefont {J.~J.}\ \bibnamefont
  {Kinnunen}}, \bibinfo {author} {\bibfnamefont {Z.}~\bibnamefont {Wu}},\ and\
  \bibinfo {author} {\bibfnamefont {G.~M.}\ \bibnamefont {Bruun}},\ }\bibfield
  {title} {\bibinfo {title} {Induced $p$-wave pairing in bose-fermi mixtures},\
  }\href {https://doi.org/10.1103/PhysRevLett.121.253402} {\bibfield  {journal}
  {\bibinfo  {journal} {Phys. Rev. Lett.}\ }\textbf {\bibinfo {volume} {121}},\
  \bibinfo {pages} {253402} (\bibinfo {year} {2018})}\BibitemShut {NoStop}%
\bibitem [{\citenamefont {DeSalvo}\ \emph {et~al.}(2019)\citenamefont
  {DeSalvo}, \citenamefont {Patel}, \citenamefont {Cai},\ and\ \citenamefont
  {Chin}}]{DeSalvo2019}%
  \BibitemOpen
  \bibfield  {author} {\bibinfo {author} {\bibfnamefont {B.~J.}\ \bibnamefont
  {DeSalvo}}, \bibinfo {author} {\bibfnamefont {K.}~\bibnamefont {Patel}},
  \bibinfo {author} {\bibfnamefont {G.}~\bibnamefont {Cai}},\ and\ \bibinfo
  {author} {\bibfnamefont {C.}~\bibnamefont {Chin}},\ }\bibfield  {title}
  {\bibinfo {title} {Observation of fermion-mediated interactions between
  bosonic atoms},\ }\href {https://doi.org/10.1038/s41586-019-1055-0}
  {\bibfield  {journal} {\bibinfo  {journal} {Nature}\ }\textbf {\bibinfo
  {volume} {568}},\ \bibinfo {pages} {61} (\bibinfo {year} {2019})}\BibitemShut
  {NoStop}%
\bibitem [{\citenamefont {Edri}\ \emph {et~al.}(2020)\citenamefont {Edri},
  \citenamefont {Raz}, \citenamefont {Matzliah}, \citenamefont {Davidson},\
  and\ \citenamefont {Ozeri}}]{2020PhysRevLett.124.163401}%
  \BibitemOpen
  \bibfield  {author} {\bibinfo {author} {\bibfnamefont {H.}~\bibnamefont
  {Edri}}, \bibinfo {author} {\bibfnamefont {B.}~\bibnamefont {Raz}}, \bibinfo
  {author} {\bibfnamefont {N.}~\bibnamefont {Matzliah}}, \bibinfo {author}
  {\bibfnamefont {N.}~\bibnamefont {Davidson}},\ and\ \bibinfo {author}
  {\bibfnamefont {R.}~\bibnamefont {Ozeri}},\ }\bibfield  {title} {\bibinfo
  {title} {Observation of spin-spin fermion-mediated interactions between
  ultracold bosons},\ }\href {https://doi.org/10.1103/PhysRevLett.124.163401}
  {\bibfield  {journal} {\bibinfo  {journal} {Phys. Rev. Lett.}\ }\textbf
  {\bibinfo {volume} {124}},\ \bibinfo {pages} {163401} (\bibinfo {year}
  {2020})}\BibitemShut {NoStop}%
\bibitem [{\citenamefont {Qin}\ \emph {et~al.}(2024)\citenamefont {Qin},
  \citenamefont {Shen}, \citenamefont {Li},\ and\ \citenamefont
  {Lee}}]{qin2024kinked}%
  \BibitemOpen
  \bibfield  {author} {\bibinfo {author} {\bibfnamefont {F.}~\bibnamefont
  {Qin}}, \bibinfo {author} {\bibfnamefont {R.}~\bibnamefont {Shen}}, \bibinfo
  {author} {\bibfnamefont {L.}~\bibnamefont {Li}},\ and\ \bibinfo {author}
  {\bibfnamefont {C.~H.}\ \bibnamefont {Lee}},\ }\bibfield  {title} {\bibinfo
  {title} {Kinked linear response from non-hermitian cold-atom pumping},\
  }\href@noop {} {\bibfield  {journal} {\bibinfo  {journal} {Physical Review
  A}\ }\textbf {\bibinfo {volume} {109}},\ \bibinfo {pages} {053311} (\bibinfo
  {year} {2024})}\BibitemShut {NoStop}%
\bibitem [{\citenamefont {M\o{}lmer}(1998)}]{1998PhysRevLett.80.1804}%
  \BibitemOpen
  \bibfield  {author} {\bibinfo {author} {\bibfnamefont {K.}~\bibnamefont
  {M\o{}lmer}},\ }\bibfield  {title} {\bibinfo {title} {Bose condensates and
  fermi gases at zero temperature},\ }\href
  {https://doi.org/10.1103/PhysRevLett.80.1804} {\bibfield  {journal} {\bibinfo
   {journal} {Phys. Rev. Lett.}\ }\textbf {\bibinfo {volume} {80}},\ \bibinfo
  {pages} {1804} (\bibinfo {year} {1998})}\BibitemShut {NoStop}%
\bibitem [{\citenamefont {Viverit}\ \emph {et~al.}(2000)\citenamefont
  {Viverit}, \citenamefont {Pethick},\ and\ \citenamefont
  {Smith}}]{2000PhysRevA.61.053605}%
  \BibitemOpen
  \bibfield  {author} {\bibinfo {author} {\bibfnamefont {L.}~\bibnamefont
  {Viverit}}, \bibinfo {author} {\bibfnamefont {C.~J.}\ \bibnamefont
  {Pethick}},\ and\ \bibinfo {author} {\bibfnamefont {H.}~\bibnamefont
  {Smith}},\ }\bibfield  {title} {\bibinfo {title} {Zero-temperature phase
  diagram of binary boson-fermion mixtures},\ }\href
  {https://doi.org/10.1103/PhysRevA.61.053605} {\bibfield  {journal} {\bibinfo
  {journal} {Phys. Rev. A}\ }\textbf {\bibinfo {volume} {61}},\ \bibinfo
  {pages} {053605} (\bibinfo {year} {2000})}\BibitemShut {NoStop}%
\bibitem [{\citenamefont {Yi}\ and\ \citenamefont
  {Sun}(2001)}]{2001PhysRevA.64.043608}%
  \BibitemOpen
  \bibfield  {author} {\bibinfo {author} {\bibfnamefont {X.~X.}\ \bibnamefont
  {Yi}}\ and\ \bibinfo {author} {\bibfnamefont {C.~P.}\ \bibnamefont {Sun}},\
  }\bibfield  {title} {\bibinfo {title} {Phase separation of a trapped
  bose-fermi gas mixture: Beyond the thomas-fermi approximation},\ }\href
  {https://doi.org/10.1103/PhysRevA.64.043608} {\bibfield  {journal} {\bibinfo
  {journal} {Phys. Rev. A}\ }\textbf {\bibinfo {volume} {64}},\ \bibinfo
  {pages} {043608} (\bibinfo {year} {2001})}\BibitemShut {NoStop}%
\bibitem [{\citenamefont {Viverit}\ and\ \citenamefont
  {Giorgini}(2002)}]{2002PhysRevA.66.063604}%
  \BibitemOpen
  \bibfield  {author} {\bibinfo {author} {\bibfnamefont {L.}~\bibnamefont
  {Viverit}}\ and\ \bibinfo {author} {\bibfnamefont {S.}~\bibnamefont
  {Giorgini}},\ }\bibfield  {title} {\bibinfo {title} {Ground-state properties
  of a dilute bose-fermi mixture},\ }\href
  {https://doi.org/10.1103/PhysRevA.66.063604} {\bibfield  {journal} {\bibinfo
  {journal} {Phys. Rev. A}\ }\textbf {\bibinfo {volume} {66}},\ \bibinfo
  {pages} {063604} (\bibinfo {year} {2002})}\BibitemShut {NoStop}%
\bibitem [{\citenamefont {Roth}\ and\ \citenamefont
  {Feldmeier}(2002)}]{2002PhysRevA.65.021603}%
  \BibitemOpen
  \bibfield  {author} {\bibinfo {author} {\bibfnamefont {R.}~\bibnamefont
  {Roth}}\ and\ \bibinfo {author} {\bibfnamefont {H.}~\bibnamefont
  {Feldmeier}},\ }\bibfield  {title} {\bibinfo {title} {Mean-field instability
  of trapped dilute boson-fermion mixtures},\ }\href
  {https://doi.org/10.1103/PhysRevA.65.021603} {\bibfield  {journal} {\bibinfo
  {journal} {Phys. Rev. A}\ }\textbf {\bibinfo {volume} {65}},\ \bibinfo
  {pages} {021603} (\bibinfo {year} {2002})}\BibitemShut {NoStop}%
\bibitem [{\citenamefont {Capuzzi}\ \emph {et~al.}(2003)\citenamefont
  {Capuzzi}, \citenamefont {Minguzzi},\ and\ \citenamefont
  {Tosi}}]{2003PhysRevA.68.033605}%
  \BibitemOpen
  \bibfield  {author} {\bibinfo {author} {\bibfnamefont {P.}~\bibnamefont
  {Capuzzi}}, \bibinfo {author} {\bibfnamefont {A.}~\bibnamefont {Minguzzi}},\
  and\ \bibinfo {author} {\bibfnamefont {M.~P.}\ \bibnamefont {Tosi}},\
  }\bibfield  {title} {\bibinfo {title} {Collective excitations in trapped
  boson-fermion mixtures: From demixing to collapse},\ }\href
  {https://doi.org/10.1103/PhysRevA.68.033605} {\bibfield  {journal} {\bibinfo
  {journal} {Phys. Rev. A}\ }\textbf {\bibinfo {volume} {68}},\ \bibinfo
  {pages} {033605} (\bibinfo {year} {2003})}\BibitemShut {NoStop}%
\bibitem [{\citenamefont {Chui}\ and\ \citenamefont
  {Ryzhov}(2004)}]{2004PhysRevA.69.043607}%
  \BibitemOpen
  \bibfield  {author} {\bibinfo {author} {\bibfnamefont {S.~T.}\ \bibnamefont
  {Chui}}\ and\ \bibinfo {author} {\bibfnamefont {V.~N.}\ \bibnamefont
  {Ryzhov}},\ }\bibfield  {title} {\bibinfo {title} {Collapse transition in
  mixtures of bosons and fermions},\ }\href
  {https://doi.org/10.1103/PhysRevA.69.043607} {\bibfield  {journal} {\bibinfo
  {journal} {Phys. Rev. A}\ }\textbf {\bibinfo {volume} {69}},\ \bibinfo
  {pages} {043607} (\bibinfo {year} {2004})}\BibitemShut {NoStop}%
\bibitem [{\citenamefont {Salasnich}\ and\ \citenamefont
  {Toigo}(2007)}]{2007PhysRevA.75.013623}%
  \BibitemOpen
  \bibfield  {author} {\bibinfo {author} {\bibfnamefont {L.}~\bibnamefont
  {Salasnich}}\ and\ \bibinfo {author} {\bibfnamefont {F.}~\bibnamefont
  {Toigo}},\ }\bibfield  {title} {\bibinfo {title} {Fermi-bose mixture across a
  feshbach resonance},\ }\href {https://doi.org/10.1103/PhysRevA.75.013623}
  {\bibfield  {journal} {\bibinfo  {journal} {Phys. Rev. A}\ }\textbf {\bibinfo
  {volume} {75}},\ \bibinfo {pages} {013623} (\bibinfo {year}
  {2007})}\BibitemShut {NoStop}%
\bibitem [{\citenamefont {Marchetti}\ \emph {et~al.}(2008)\citenamefont
  {Marchetti}, \citenamefont {Mathy}, \citenamefont {Huse},\ and\ \citenamefont
  {Parish}}]{2008PhysRevB.78.134517}%
  \BibitemOpen
  \bibfield  {author} {\bibinfo {author} {\bibfnamefont {F.~M.}\ \bibnamefont
  {Marchetti}}, \bibinfo {author} {\bibfnamefont {C.~J.~M.}\ \bibnamefont
  {Mathy}}, \bibinfo {author} {\bibfnamefont {D.~A.}\ \bibnamefont {Huse}},\
  and\ \bibinfo {author} {\bibfnamefont {M.~M.}\ \bibnamefont {Parish}},\
  }\bibfield  {title} {\bibinfo {title} {Phase separation and collapse in
  bose-fermi mixtures with a feshbach resonance},\ }\href
  {https://doi.org/10.1103/PhysRevB.78.134517} {\bibfield  {journal} {\bibinfo
  {journal} {Phys. Rev. B}\ }\textbf {\bibinfo {volume} {78}},\ \bibinfo
  {pages} {134517} (\bibinfo {year} {2008})}\BibitemShut {NoStop}%
\bibitem [{\citenamefont {B\"uchler}\ and\ \citenamefont
  {Blatter}(2004)}]{2004PhysRevA.69.063603}%
  \BibitemOpen
  \bibfield  {author} {\bibinfo {author} {\bibfnamefont {H.~P.}\ \bibnamefont
  {B\"uchler}}\ and\ \bibinfo {author} {\bibfnamefont {G.}~\bibnamefont
  {Blatter}},\ }\bibfield  {title} {\bibinfo {title} {Phase separation of
  atomic bose-fermi mixtures in an optical lattice},\ }\href
  {https://doi.org/10.1103/PhysRevA.69.063603} {\bibfield  {journal} {\bibinfo
  {journal} {Phys. Rev. A}\ }\textbf {\bibinfo {volume} {69}},\ \bibinfo
  {pages} {063603} (\bibinfo {year} {2004})}\BibitemShut {NoStop}%
\bibitem [{\citenamefont {Yu}\ \emph {et~al.}(2011)\citenamefont {Yu},
  \citenamefont {Zhang},\ and\ \citenamefont {Zhai}}]{2011hysRevA.83.041603}%
  \BibitemOpen
  \bibfield  {author} {\bibinfo {author} {\bibfnamefont {Z.-Q.}\ \bibnamefont
  {Yu}}, \bibinfo {author} {\bibfnamefont {S.}~\bibnamefont {Zhang}},\ and\
  \bibinfo {author} {\bibfnamefont {H.}~\bibnamefont {Zhai}},\ }\bibfield
  {title} {\bibinfo {title} {Stability condition of a strongly interacting
  boson-fermion mixture across an interspecies feshbach resonance},\ }\href
  {https://doi.org/10.1103/PhysRevA.83.041603} {\bibfield  {journal} {\bibinfo
  {journal} {Phys. Rev. A}\ }\textbf {\bibinfo {volume} {83}},\ \bibinfo
  {pages} {041603} (\bibinfo {year} {2011})}\BibitemShut {NoStop}%
\bibitem [{\citenamefont {Bertaina}\ \emph {et~al.}(2013)\citenamefont
  {Bertaina}, \citenamefont {Fratini}, \citenamefont {Giorgini},\ and\
  \citenamefont {Pieri}}]{2013PhysRevLett.110.115303}%
  \BibitemOpen
  \bibfield  {author} {\bibinfo {author} {\bibfnamefont {G.}~\bibnamefont
  {Bertaina}}, \bibinfo {author} {\bibfnamefont {E.}~\bibnamefont {Fratini}},
  \bibinfo {author} {\bibfnamefont {S.}~\bibnamefont {Giorgini}},\ and\
  \bibinfo {author} {\bibfnamefont {P.}~\bibnamefont {Pieri}},\ }\bibfield
  {title} {\bibinfo {title} {Quantum monte carlo study of a resonant bose-fermi
  mixture},\ }\href {https://doi.org/10.1103/PhysRevLett.110.115303} {\bibfield
   {journal} {\bibinfo  {journal} {Phys. Rev. Lett.}\ }\textbf {\bibinfo
  {volume} {110}},\ \bibinfo {pages} {115303} (\bibinfo {year}
  {2013})}\BibitemShut {NoStop}%
\bibitem [{\citenamefont {DeSalvo}\ \emph {et~al.}(2017)\citenamefont
  {DeSalvo}, \citenamefont {Patel}, \citenamefont {Johansen},\ and\
  \citenamefont {Chin}}]{2017PhysRevLett.119.233401}%
  \BibitemOpen
  \bibfield  {author} {\bibinfo {author} {\bibfnamefont {B.~J.}\ \bibnamefont
  {DeSalvo}}, \bibinfo {author} {\bibfnamefont {K.}~\bibnamefont {Patel}},
  \bibinfo {author} {\bibfnamefont {J.}~\bibnamefont {Johansen}},\ and\
  \bibinfo {author} {\bibfnamefont {C.}~\bibnamefont {Chin}},\ }\bibfield
  {title} {\bibinfo {title} {Observation of a degenerate fermi gas trapped by a
  bose-einstein condensate},\ }\href
  {https://doi.org/10.1103/PhysRevLett.119.233401} {\bibfield  {journal}
  {\bibinfo  {journal} {Phys. Rev. Lett.}\ }\textbf {\bibinfo {volume} {119}},\
  \bibinfo {pages} {233401} (\bibinfo {year} {2017})}\BibitemShut {NoStop}%
\bibitem [{\citenamefont {Wu}\ \emph {et~al.}(2018)\citenamefont {Wu},
  \citenamefont {Yao}, \citenamefont {Liu}, \citenamefont {Wang}, \citenamefont
  {Wang}, \citenamefont {Chen}, \citenamefont {Deng}, \citenamefont {Chen},\
  and\ \citenamefont {Pan}}]{2018PhysRevB.97.020506}%
  \BibitemOpen
  \bibfield  {author} {\bibinfo {author} {\bibfnamefont {Y.-P.}\ \bibnamefont
  {Wu}}, \bibinfo {author} {\bibfnamefont {X.-C.}\ \bibnamefont {Yao}},
  \bibinfo {author} {\bibfnamefont {X.-P.}\ \bibnamefont {Liu}}, \bibinfo
  {author} {\bibfnamefont {X.-Q.}\ \bibnamefont {Wang}}, \bibinfo {author}
  {\bibfnamefont {Y.-X.}\ \bibnamefont {Wang}}, \bibinfo {author}
  {\bibfnamefont {H.-Z.}\ \bibnamefont {Chen}}, \bibinfo {author}
  {\bibfnamefont {Y.}~\bibnamefont {Deng}}, \bibinfo {author} {\bibfnamefont
  {Y.-A.}\ \bibnamefont {Chen}},\ and\ \bibinfo {author} {\bibfnamefont
  {J.-W.}\ \bibnamefont {Pan}},\ }\bibfield  {title} {\bibinfo {title} {Coupled
  dipole oscillations of a mass-imbalanced bose-fermi superfluid mixture},\
  }\href {https://doi.org/10.1103/PhysRevB.97.020506} {\bibfield  {journal}
  {\bibinfo  {journal} {Phys. Rev. B}\ }\textbf {\bibinfo {volume} {97}},\
  \bibinfo {pages} {020506} (\bibinfo {year} {2018})}\BibitemShut {NoStop}%
\bibitem [{\citenamefont {Modugno}\ \emph {et~al.}(2002)\citenamefont
  {Modugno}, \citenamefont {Roati}, \citenamefont {Riboli}, \citenamefont
  {Ferlaino}, \citenamefont {Brecha},\ and\ \citenamefont
  {Inguscio}}]{Modugno2002}%
  \BibitemOpen
  \bibfield  {author} {\bibinfo {author} {\bibfnamefont {G.}~\bibnamefont
  {Modugno}}, \bibinfo {author} {\bibfnamefont {G.}~\bibnamefont {Roati}},
  \bibinfo {author} {\bibfnamefont {F.}~\bibnamefont {Riboli}}, \bibinfo
  {author} {\bibfnamefont {F.}~\bibnamefont {Ferlaino}}, \bibinfo {author}
  {\bibfnamefont {R.~J.}\ \bibnamefont {Brecha}},\ and\ \bibinfo {author}
  {\bibfnamefont {M.}~\bibnamefont {Inguscio}},\ }\bibfield  {title} {\bibinfo
  {title} {Collapse of a degenerate fermi gas},\ }\href
  {https://doi.org/10.1126/science.1077386} {\bibfield  {journal} {\bibinfo
  {journal} {Science}\ }\textbf {\bibinfo {volume} {297}},\ \bibinfo {pages}
  {2240} (\bibinfo {year} {2002})}\BibitemShut {NoStop}%
\bibitem [{\citenamefont {Lous}\ \emph {et~al.}(2018)\citenamefont {Lous},
  \citenamefont {Fritsche}, \citenamefont {Jag}, \citenamefont {Lehmann},
  \citenamefont {Kirilov}, \citenamefont {Huang},\ and\ \citenamefont
  {Grimm}}]{2018PhysRevLett.120.243403}%
  \BibitemOpen
  \bibfield  {author} {\bibinfo {author} {\bibfnamefont {R.~S.}\ \bibnamefont
  {Lous}}, \bibinfo {author} {\bibfnamefont {I.}~\bibnamefont {Fritsche}},
  \bibinfo {author} {\bibfnamefont {M.}~\bibnamefont {Jag}}, \bibinfo {author}
  {\bibfnamefont {F.}~\bibnamefont {Lehmann}}, \bibinfo {author} {\bibfnamefont
  {E.}~\bibnamefont {Kirilov}}, \bibinfo {author} {\bibfnamefont
  {B.}~\bibnamefont {Huang}},\ and\ \bibinfo {author} {\bibfnamefont
  {R.}~\bibnamefont {Grimm}},\ }\bibfield  {title} {\bibinfo {title} {Probing
  the interface of a phase-separated state in a repulsive bose-fermi mixture},\
  }\href {https://doi.org/10.1103/PhysRevLett.120.243403} {\bibfield  {journal}
  {\bibinfo  {journal} {Phys. Rev. Lett.}\ }\textbf {\bibinfo {volume} {120}},\
  \bibinfo {pages} {243403} (\bibinfo {year} {2018})}\BibitemShut {NoStop}%
\bibitem [{\citenamefont {Kim}\ and\ \citenamefont
  {Chien}(2018)}]{2018PhysRevA.97.033628}%
  \BibitemOpen
  \bibfield  {author} {\bibinfo {author} {\bibfnamefont {T.}~\bibnamefont
  {Kim}}\ and\ \bibinfo {author} {\bibfnamefont {C.-C.}\ \bibnamefont
  {Chien}},\ }\bibfield  {title} {\bibinfo {title} {Thermodynamics and
  structural transition of binary atomic bose-fermi mixtures in box or harmonic
  potentials: A path-integral study},\ }\href
  {https://doi.org/10.1103/PhysRevA.97.033628} {\bibfield  {journal} {\bibinfo
  {journal} {Phys. Rev. A}\ }\textbf {\bibinfo {volume} {97}},\ \bibinfo
  {pages} {033628} (\bibinfo {year} {2018})}\BibitemShut {NoStop}%
\bibitem [{\citenamefont {Shen}\ \emph {et~al.}(2023)\citenamefont {Shen},
  \citenamefont {Chen}, \citenamefont {Aliyu}, \citenamefont {Qin},
  \citenamefont {Zhong}, \citenamefont {Loh},\ and\ \citenamefont
  {Lee}}]{shen2023proposal}%
  \BibitemOpen
  \bibfield  {author} {\bibinfo {author} {\bibfnamefont {R.}~\bibnamefont
  {Shen}}, \bibinfo {author} {\bibfnamefont {T.}~\bibnamefont {Chen}}, \bibinfo
  {author} {\bibfnamefont {M.~M.}\ \bibnamefont {Aliyu}}, \bibinfo {author}
  {\bibfnamefont {F.}~\bibnamefont {Qin}}, \bibinfo {author} {\bibfnamefont
  {Y.}~\bibnamefont {Zhong}}, \bibinfo {author} {\bibfnamefont
  {H.}~\bibnamefont {Loh}},\ and\ \bibinfo {author} {\bibfnamefont {C.~H.}\
  \bibnamefont {Lee}},\ }\bibfield  {title} {\bibinfo {title} {Proposal for
  observing yang-lee criticality in rydberg atomic arrays},\ }\href@noop {}
  {\bibfield  {journal} {\bibinfo  {journal} {Physical Review Letters}\
  }\textbf {\bibinfo {volume} {131}},\ \bibinfo {pages} {080403} (\bibinfo
  {year} {2023})}\BibitemShut {NoStop}%
\bibitem [{\citenamefont {Manabe}\ and\ \citenamefont
  {Ohashi}(2021)}]{2021PhysRevA.103.063317}%
  \BibitemOpen
  \bibfield  {author} {\bibinfo {author} {\bibfnamefont {K.}~\bibnamefont
  {Manabe}}\ and\ \bibinfo {author} {\bibfnamefont {Y.}~\bibnamefont
  {Ohashi}},\ }\bibfield  {title} {\bibinfo {title} {Thermodynamic stability,
  compressibility matrices, and effects of mediated interactions in a strongly
  interacting bose-fermi mixture},\ }\href
  {https://doi.org/10.1103/PhysRevA.103.063317} {\bibfield  {journal} {\bibinfo
   {journal} {Phys. Rev. A}\ }\textbf {\bibinfo {volume} {103}},\ \bibinfo
  {pages} {063317} (\bibinfo {year} {2021})}\BibitemShut {NoStop}%
\bibitem [{\citenamefont {Patel}\ \emph {et~al.}(2023)\citenamefont {Patel},
  \citenamefont {Cai}, \citenamefont {Ando},\ and\ \citenamefont
  {Chin}}]{2023PhysRevLett.131.083003}%
  \BibitemOpen
  \bibfield  {author} {\bibinfo {author} {\bibfnamefont {K.}~\bibnamefont
  {Patel}}, \bibinfo {author} {\bibfnamefont {G.}~\bibnamefont {Cai}}, \bibinfo
  {author} {\bibfnamefont {H.}~\bibnamefont {Ando}},\ and\ \bibinfo {author}
  {\bibfnamefont {C.}~\bibnamefont {Chin}},\ }\bibfield  {title} {\bibinfo
  {title} {Sound propagation in a bose-fermi mixture: From weak to strong
  interactions},\ }\href {https://doi.org/10.1103/PhysRevLett.131.083003}
  {\bibfield  {journal} {\bibinfo  {journal} {Phys. Rev. Lett.}\ }\textbf
  {\bibinfo {volume} {131}},\ \bibinfo {pages} {083003} (\bibinfo {year}
  {2023})}\BibitemShut {NoStop}%
\bibitem [{\citenamefont {Shen}\ \emph
  {et~al.}(2024{\natexlab{b}})\citenamefont {Shen}, \citenamefont {Davidson},
  \citenamefont {Bruun}, \citenamefont {Sun},\ and\ \citenamefont
  {Wu}}]{PhysRevLett.132.033401}%
  \BibitemOpen
  \bibfield  {author} {\bibinfo {author} {\bibfnamefont {X.}~\bibnamefont
  {Shen}}, \bibinfo {author} {\bibfnamefont {N.}~\bibnamefont {Davidson}},
  \bibinfo {author} {\bibfnamefont {G.~M.}\ \bibnamefont {Bruun}}, \bibinfo
  {author} {\bibfnamefont {M.}~\bibnamefont {Sun}},\ and\ \bibinfo {author}
  {\bibfnamefont {Z.}~\bibnamefont {Wu}},\ }\bibfield  {title} {\bibinfo
  {title} {Strongly interacting bose-fermi mixtures: Mediated interaction,
  phase diagram, and sound propagation},\ }\href
  {https://doi.org/10.1103/PhysRevLett.132.033401} {\bibfield  {journal}
  {\bibinfo  {journal} {Phys. Rev. Lett.}\ }\textbf {\bibinfo {volume} {132}},\
  \bibinfo {pages} {033401} (\bibinfo {year} {2024}{\natexlab{b}})}\BibitemShut
  {NoStop}%
\bibitem [{\citenamefont {Mostaan}\ \emph {et~al.}(2022)\citenamefont
  {Mostaan}, \citenamefont {Grusdt},\ and\ \citenamefont
  {Goldman}}]{Mostaan2022}%
  \BibitemOpen
  \bibfield  {author} {\bibinfo {author} {\bibfnamefont {N.}~\bibnamefont
  {Mostaan}}, \bibinfo {author} {\bibfnamefont {F.}~\bibnamefont {Grusdt}},\
  and\ \bibinfo {author} {\bibfnamefont {N.}~\bibnamefont {Goldman}},\
  }\bibfield  {title} {\bibinfo {title} {Quantized topological pumping of
  solitons in nonlinear photonics and ultracold atomic mixtures},\ }\href
  {https://doi.org/10.1038/s41467-022-33478-4} {\bibfield  {journal} {\bibinfo
  {journal} {Nature Communications}\ }\textbf {\bibinfo {volume} {13}},\
  \bibinfo {pages} {5997} (\bibinfo {year} {2022})}\BibitemShut {NoStop}%
\bibitem [{\citenamefont {Padhan}\ \emph {et~al.}(2025)\citenamefont {Padhan},
  \citenamefont {Barbiero},\ and\ \citenamefont {Mishra}}]{2025lw8k-7h6p}%
  \BibitemOpen
  \bibfield  {author} {\bibinfo {author} {\bibfnamefont {A.}~\bibnamefont
  {Padhan}}, \bibinfo {author} {\bibfnamefont {L.}~\bibnamefont {Barbiero}},\
  and\ \bibinfo {author} {\bibfnamefont {T.}~\bibnamefont {Mishra}},\
  }\bibfield  {title} {\bibinfo {title} {Correlated-hopping-induced topological
  order in an atomic mixture},\ }\href {https://doi.org/10.1103/lw8k-7h6p}
  {\bibfield  {journal} {\bibinfo  {journal} {Phys. Rev. A}\ }\textbf {\bibinfo
  {volume} {112}},\ \bibinfo {pages} {L011305} (\bibinfo {year}
  {2025})}\BibitemShut {NoStop}%
\bibitem [{\citenamefont {Yan}\ \emph {et~al.}(2024)\citenamefont {Yan},
  \citenamefont {Ni}, \citenamefont {Chuang}, \citenamefont {Dolgirev},
  \citenamefont {Seetharam}, \citenamefont {Demle}, \citenamefont {Robens},\
  and\ \citenamefont {Zwierlein}}]{YanZoeZ2024}%
  \BibitemOpen
  \bibfield  {author} {\bibinfo {author} {\bibfnamefont {Z.~Z.}\ \bibnamefont
  {Yan}}, \bibinfo {author} {\bibfnamefont {Y.}~\bibnamefont {Ni}}, \bibinfo
  {author} {\bibfnamefont {A.}~\bibnamefont {Chuang}}, \bibinfo {author}
  {\bibfnamefont {P.~E.}\ \bibnamefont {Dolgirev}}, \bibinfo {author}
  {\bibfnamefont {K.}~\bibnamefont {Seetharam}}, \bibinfo {author}
  {\bibfnamefont {E.}~\bibnamefont {Demle}}, \bibinfo {author} {\bibfnamefont
  {C.}~\bibnamefont {Robens}},\ and\ \bibinfo {author} {\bibfnamefont
  {M.}~\bibnamefont {Zwierlein}},\ }\bibfield  {title} {\bibinfo {title}
  {Collective flow of fermionic impurities immersed in a bose–einstein
  condensate},\ }\href {https://doi.org/10.1038/s41567-024-02541-w} {\bibfield
  {journal} {\bibinfo  {journal} {Nat. Phys.}\ }\textbf {\bibinfo {volume}
  {20}},\ \bibinfo {pages} {1395} (\bibinfo {year} {2024})}\BibitemShut
  {NoStop}%
\bibitem [{202()}]{2026supplemental}%
  \BibitemOpen
  \href@noop {} {}\bibinfo {note} {See Supplemental Materials for details of
  (S1) Persistence of the drag-induced skin effect and interaction blockade in
  a uniform asymmetric-hopping model; (S2) Real spectrum from a non-unitary
  similarity transformation; (S3) Drag-induced skin effect beyond the
  strong-coupling limit; (S4) Effective single-particle model of a
  boson-fermion pair; (S5) Criterion for distinguishing the bound and
  scattering sectors; (S6) Interaction dependence of the drag-induced skin
  effect; (S7) Robustness of the drag-induced skin effect against fermionic
  hopping; (S8) Eigenstate-resolved density distributions within each energy
  band; (S9) Dynamics in the absence of boson-fermion interaction; (S10)
  Floquet engineering of nonreciprocal tunneling; and (S11) Drag-induced skin
  effect at higher particle numbers.}\BibitemShut {Stop}%
\bibitem [{\citenamefont {Inouye}\ \emph {et~al.}(2004)\citenamefont {Inouye},
  \citenamefont {Goldwin}, \citenamefont {Olsen}, \citenamefont {Ticknor},
  \citenamefont {Bohn},\ and\ \citenamefont {Jin}}]{PhysRevLett.93.183201}%
  \BibitemOpen
  \bibfield  {author} {\bibinfo {author} {\bibfnamefont {S.}~\bibnamefont
  {Inouye}}, \bibinfo {author} {\bibfnamefont {J.}~\bibnamefont {Goldwin}},
  \bibinfo {author} {\bibfnamefont {M.~L.}\ \bibnamefont {Olsen}}, \bibinfo
  {author} {\bibfnamefont {C.}~\bibnamefont {Ticknor}}, \bibinfo {author}
  {\bibfnamefont {J.~L.}\ \bibnamefont {Bohn}},\ and\ \bibinfo {author}
  {\bibfnamefont {D.~S.}\ \bibnamefont {Jin}},\ }\bibfield  {title} {\bibinfo
  {title} {Observation of heteronuclear feshbach resonances in a mixture of
  bosons and fermions},\ }\href {https://doi.org/10.1103/PhysRevLett.93.183201}
  {\bibfield  {journal} {\bibinfo  {journal} {Phys. Rev. Lett.}\ }\textbf
  {\bibinfo {volume} {93}},\ \bibinfo {pages} {183201} (\bibinfo {year}
  {2004})}\BibitemShut {NoStop}%
\bibitem [{\citenamefont {Wang}\ \emph {et~al.}(2011)\citenamefont {Wang},
  \citenamefont {Fu}, \citenamefont {Chai},\ and\ \citenamefont
  {Zhang}}]{Wang_2011}%
  \BibitemOpen
  \bibfield  {author} {\bibinfo {author} {\bibfnamefont {P.-J.}\ \bibnamefont
  {Wang}}, \bibinfo {author} {\bibfnamefont {Z.-K.}\ \bibnamefont {Fu}},
  \bibinfo {author} {\bibfnamefont {S.-J.}\ \bibnamefont {Chai}},\ and\
  \bibinfo {author} {\bibfnamefont {J.}~\bibnamefont {Zhang}},\ }\bibfield
  {title} {\bibinfo {title} {Feshbach resonances in an ultracold mixture of
  87rb and 40k},\ }\href {https://doi.org/10.1088/1674-1056/20/10/103401}
  {\bibfield  {journal} {\bibinfo  {journal} {Chinese Physics B}\ }\textbf
  {\bibinfo {volume} {20}},\ \bibinfo {pages} {103401} (\bibinfo {year}
  {2011})}\BibitemShut {NoStop}%
\bibitem [{\citenamefont {Chin}\ \emph {et~al.}(2010)\citenamefont {Chin},
  \citenamefont {Grimm}, \citenamefont {Julienne},\ and\ \citenamefont
  {Tiesinga}}]{Chin2010}%
  \BibitemOpen
  \bibfield  {author} {\bibinfo {author} {\bibfnamefont {C.}~\bibnamefont
  {Chin}}, \bibinfo {author} {\bibfnamefont {R.}~\bibnamefont {Grimm}},
  \bibinfo {author} {\bibfnamefont {P.}~\bibnamefont {Julienne}},\ and\
  \bibinfo {author} {\bibfnamefont {E.}~\bibnamefont {Tiesinga}},\ }\bibfield
  {title} {\bibinfo {title} {Feshbach resonances in ultracold gases},\ }\href
  {https://doi.org/10.1103/RevModPhys.82.1225} {\bibfield  {journal} {\bibinfo
  {journal} {Rev. Mod. Phys.}\ }\textbf {\bibinfo {volume} {82}},\ \bibinfo
  {pages} {1225} (\bibinfo {year} {2010})}\BibitemShut {NoStop}%
\bibitem [{\citenamefont {Ozawa}\ \emph {et~al.}(2019)\citenamefont {Ozawa},
  \citenamefont {Price}, \citenamefont {Amo}, \citenamefont {Goldman},
  \citenamefont {Hafezi}, \citenamefont {Lu}, \citenamefont {Rechtsman},
  \citenamefont {Schuster}, \citenamefont {Simon}, \citenamefont {Zilberberg},\
  and\ \citenamefont {Carusotto}}]{Ozawa2019}%
  \BibitemOpen
  \bibfield  {author} {\bibinfo {author} {\bibfnamefont {T.}~\bibnamefont
  {Ozawa}}, \bibinfo {author} {\bibfnamefont {H.~M.}\ \bibnamefont {Price}},
  \bibinfo {author} {\bibfnamefont {A.}~\bibnamefont {Amo}}, \bibinfo {author}
  {\bibfnamefont {N.}~\bibnamefont {Goldman}}, \bibinfo {author} {\bibfnamefont
  {M.}~\bibnamefont {Hafezi}}, \bibinfo {author} {\bibfnamefont
  {L.}~\bibnamefont {Lu}}, \bibinfo {author} {\bibfnamefont {M.~C.}\
  \bibnamefont {Rechtsman}}, \bibinfo {author} {\bibfnamefont {D.}~\bibnamefont
  {Schuster}}, \bibinfo {author} {\bibfnamefont {J.}~\bibnamefont {Simon}},
  \bibinfo {author} {\bibfnamefont {O.}~\bibnamefont {Zilberberg}},\ and\
  \bibinfo {author} {\bibfnamefont {I.}~\bibnamefont {Carusotto}},\ }\bibfield
  {title} {\bibinfo {title} {Topological photonics},\ }\href
  {https://doi.org/10.1103/RevModPhys.91.015006} {\bibfield  {journal}
  {\bibinfo  {journal} {Rev. Mod. Phys.}\ }\textbf {\bibinfo {volume} {91}},\
  \bibinfo {pages} {015006} (\bibinfo {year} {2019})}\BibitemShut {NoStop}%
\bibitem [{\citenamefont {Bakr}\ \emph {et~al.}(2009)\citenamefont {Bakr},
  \citenamefont {Gillen}, \citenamefont {Peng}, \citenamefont {F{\"o}lling},\
  and\ \citenamefont {Markus}}]{BakrWS2009}%
  \BibitemOpen
  \bibfield  {author} {\bibinfo {author} {\bibfnamefont {W.~S.}\ \bibnamefont
  {Bakr}}, \bibinfo {author} {\bibfnamefont {J.~I.}\ \bibnamefont {Gillen}},
  \bibinfo {author} {\bibfnamefont {A.}~\bibnamefont {Peng}}, \bibinfo {author}
  {\bibfnamefont {S.}~\bibnamefont {F{\"o}lling}},\ and\ \bibinfo {author}
  {\bibfnamefont {G.}~\bibnamefont {Markus}},\ }\bibfield  {title} {\bibinfo
  {title} {A quantum gas microscope for detecting single atoms in a
  hubbard-regime optical lattice},\ }\href
  {https://doi.org/10.1038/nature08482} {\bibfield  {journal} {\bibinfo
  {journal} {Nature}\ }\textbf {\bibinfo {volume} {462}},\ \bibinfo {pages}
  {74} (\bibinfo {year} {2009})}\BibitemShut {NoStop}%
\bibitem [{\citenamefont {Sherson}\ \emph {et~al.}(2010)\citenamefont
  {Sherson}, \citenamefont {Weitenberg}, \citenamefont {Engelsen},
  \citenamefont {Guzman}, \citenamefont {Reichs{\"o}llner}, \citenamefont
  {Bloch},\ and\ \citenamefont {Kuhr}}]{Sherson2010}%
  \BibitemOpen
  \bibfield  {author} {\bibinfo {author} {\bibfnamefont {J.~F.}\ \bibnamefont
  {Sherson}}, \bibinfo {author} {\bibfnamefont {C.}~\bibnamefont {Weitenberg}},
  \bibinfo {author} {\bibfnamefont {N.~J.}\ \bibnamefont {Engelsen}}, \bibinfo
  {author} {\bibfnamefont {S.}~\bibnamefont {Guzman}}, \bibinfo {author}
  {\bibfnamefont {L.}~\bibnamefont {Reichs{\"o}llner}}, \bibinfo {author}
  {\bibfnamefont {I.}~\bibnamefont {Bloch}},\ and\ \bibinfo {author}
  {\bibfnamefont {S.}~\bibnamefont {Kuhr}},\ }\bibfield  {title} {\bibinfo
  {title} {Single-atom-resolved fluorescence imaging for quantum gas
  microscopy},\ }\href {https://doi.org/10.1038/nature09378} {\bibfield
  {journal} {\bibinfo  {journal} {Nature}\ }\textbf {\bibinfo {volume} {467}},\
  \bibinfo {pages} {68} (\bibinfo {year} {2010})}\BibitemShut {NoStop}%
\bibitem [{\citenamefont {Mostafazadeh}(2002)}]{MostafazadehA2002}%
  \BibitemOpen
  \bibfield  {author} {\bibinfo {author} {\bibfnamefont {A.}~\bibnamefont
  {Mostafazadeh}},\ }\bibfield  {title} {\bibinfo {title} {Pseudo-hermiticity
  versus pt-symmetry. ii. a complete characterization of non-hermitian
  hamiltonians with a real spectrum},\ }\href
  {https://doi.org/10.1063/1.1461427} {\bibfield  {journal} {\bibinfo
  {journal} {J. Math. Phys.}\ }\textbf {\bibinfo {volume} {43}},\ \bibinfo
  {pages} {2814} (\bibinfo {year} {2002})}\BibitemShut {NoStop}%
\bibitem [{\citenamefont {Liu}\ \emph {et~al.}(2023)\citenamefont {Liu},
  \citenamefont {Ke}, \citenamefont {Lei},\ and\ \citenamefont
  {Lee}}]{Liu2023NJP}%
  \BibitemOpen
  \bibfield  {author} {\bibinfo {author} {\bibfnamefont {W.}~\bibnamefont
  {Liu}}, \bibinfo {author} {\bibfnamefont {Y.}~\bibnamefont {Ke}}, \bibinfo
  {author} {\bibfnamefont {Z.}~\bibnamefont {Lei}},\ and\ \bibinfo {author}
  {\bibfnamefont {C.}~\bibnamefont {Lee}},\ }\bibfield  {title} {\bibinfo
  {title} {Magnon boundary states tailored by longitudinal spin--spin
  interactions and topology},\ }\href
  {https://doi.org/10.1088/1367-2630/acf8ea} {\bibfield  {journal} {\bibinfo
  {journal} {New Journal of Physics}\ }\textbf {\bibinfo {volume} {25}},\
  \bibinfo {pages} {093042} (\bibinfo {year} {2023})}\BibitemShut {NoStop}%
\bibitem [{\citenamefont {Bukov}\ \emph {et~al.}(2015)\citenamefont {Bukov},
  \citenamefont {D'Alessio},\ and\ \citenamefont {Polkovnikov}}]{Bukov2015}%
  \BibitemOpen
  \bibfield  {author} {\bibinfo {author} {\bibfnamefont {M.}~\bibnamefont
  {Bukov}}, \bibinfo {author} {\bibfnamefont {L.}~\bibnamefont {D'Alessio}},\
  and\ \bibinfo {author} {\bibfnamefont {A.}~\bibnamefont {Polkovnikov}},\
  }\bibfield  {title} {\bibinfo {title} {Universal high-frequency behavior of
  periodically driven systems: from dynamical stabilization to floquet
  engineering},\ }\href {https://doi.org/10.1080/00018732.2015.1055918}
  {\bibfield  {journal} {\bibinfo  {journal} {Adv. Phys.}\ }\textbf {\bibinfo
  {volume} {64}},\ \bibinfo {pages} {139} (\bibinfo {year} {2015})}\BibitemShut
  {NoStop}%
\bibitem [{\citenamefont {Lee}\ \emph {et~al.}(2018)\citenamefont {Lee},
  \citenamefont {Ho}, \citenamefont {Yang}, \citenamefont {Gong},\ and\
  \citenamefont {Papi{\'c}}}]{lee2018floquet}%
  \BibitemOpen
  \bibfield  {author} {\bibinfo {author} {\bibfnamefont {C.~H.}\ \bibnamefont
  {Lee}}, \bibinfo {author} {\bibfnamefont {W.~W.}\ \bibnamefont {Ho}},
  \bibinfo {author} {\bibfnamefont {B.}~\bibnamefont {Yang}}, \bibinfo {author}
  {\bibfnamefont {J.}~\bibnamefont {Gong}},\ and\ \bibinfo {author}
  {\bibfnamefont {Z.}~\bibnamefont {Papi{\'c}}},\ }\bibfield  {title} {\bibinfo
  {title} {Floquet mechanism for non-abelian fractional quantum hall states},\
  }\href@noop {} {\bibfield  {journal} {\bibinfo  {journal} {Physical review
  letters}\ }\textbf {\bibinfo {volume} {121}},\ \bibinfo {pages} {237401}
  (\bibinfo {year} {2018})}\BibitemShut {NoStop}%
\end{thebibliography}
\end{document}